\newcount\mgnf  %ingrandimento
\mgnf=2

\ifnum\mgnf=0
\magnification=1000   
\hsize=15truecm\vsize=20.2truecm%\voffset2.truecm\hoffset.5truecm
   \parindent=0.3cm\baselineskip=0.45cm\fi
\ifnum\mgnf=1
   \magnification=\magstephalf
\voffset=.5truecm
 %  \hoffset=0.truecm
 %  \hsize=15truecm\vsize=20.2truecm
   \baselineskip=18truept plus0.1pt minus0.1pt \parindent=0.9truecm
 %  \lineskip=0.5truecm\lineskiplimit=0.1pt      \parskip=0.1pt plus1pt
\fi

%\ifnum\mgnf=2%
%   \magnification=1200%
%   \hsize=15truecm\vsize=20.2truecm%
%   \baselineskip=18truept plus0.1pt minus0.1pt \parindent=0.9truecm%
%   \lineskip=0.5truecm\lineskiplimit=0.1pt      \parskip=0.1pt plus1pt%
%\fi
\ifnum\mgnf=2\magnification=1200\fi

\ifnum\mgnf=0
\def\openone{\leavevmode\hbox{\ninerm 1\kern-3.3pt\tenrm1}}%
\def\*{\vglue0.2truecm}\fi
\ifnum\mgnf=1
\def\openone{\leavevmode\hbox{\ninerm 1\kern-3.63pt\tenrm1}}%
\def\*{\vglue0.3truecm}\fi
\ifnum\mgnf=2\def\*{\vglue0.7truecm}\fi
\openout15=\jobname.aux

%%%%%%%%%%%%%%%%%%%%%%%%%%%%%%%%%%%%%%%%%%%%%%%%%%%%%%%%%%%%%%%%%%%%%%%%%%%%%
%%%%%%%%%%%%%%%%%%%%%%%%%  DEFINIZIONI DI FONT    %%%%%%%%%%%%%%%%%%%%%%%%%%%
%%%%%%%%%%%%%%%%%%%%%%%%%%%%%%%%%%%%%%%%%%%%%%%%%%%%%%%%%%%%%%%%%%%%%%%%%%%%%
\font\titolo=cmbx12\font\titolone=cmbx10 scaled\magstep 2%
\font\cs=cmcsc10\font\sc=cmcsc10\font\css=cmcsc8%
\font\indbf=cmbx10 scaled\magstep2
\font\ottorm=cmr8\font\ninerm=cmr9%
\font\msytw=msbm9 scaled\magstep1%
\font\msytww=msbm7 scaled\magstep1%
\font\msytwww=msbm5 scaled\magstep1%
%\font\msytwwww=msbm4 scaled\magstep1%
%
%
%
\def\st{\scriptstyle}%
%

%%%%%%%%%%%%%%%%%%%%%%%%%%%%%%%%%%%%%%%%%%%%%%%%%%%%%%%%%%%%%%%%%%%%%%%%%%%%%
%%%%%%%%%%%%%%%%%    LETTERE GRECHE E LATINE IN NERETTO     %%%%%%%%%%%%%%%%%
%%%%%%%%%%%%%%%%%%%%%%%%%%%%%%%%%%%%%%%%%%%%%%%%%%%%%%%%%%%%%%%%%%%%%%%%%%%%%

% lettere greche e latine in neretto italico - pag.430 del manuale
\font\tenmib=cmmib10 \font\eightmib=cmmib8
\font\sevenmib=cmmib7\font\fivemib=cmmib5 
\font\ottoit=cmti8
\font\fiveit=cmti5\font\sixit=cmti6%%
%!!!@@@\font\fiveit=cmti7\font\sixit=cmti7%%
\font\fivei=cmmi5\font\sixi=cmmi6\font\ottoi=cmmi8
\font\ottorm=cmr8
\font\ottosy=cmsy8\font\sixsy=cmsy6\font\fivesy=cmsy5%%
\font\ottobf=cmbx8\font\sixbf=cmbx6\font\fivebf=cmbx5%
\font\ottocss=cmcsc8%

\def\ottopunti{\def\rm{\fam0\ottorm}\def\it{\fam6\ottoit}%
\def\bf{\fam7\ottobf}%
\textfont1=\ottoi\scriptfont1=\sixi\scriptscriptfont1=\fivei%
\textfont2=\ottosy\scriptfont2=\sixsy\scriptscriptfont2=\fivesy%
%\textfont3=\tenex\scriptfont3=\tenex\scriptscriptfont3=\tenex%
\textfont4=\ottocss\scriptfont4=\sc\scriptscriptfont4=\sc%
%\scriptfont4=\ottocss\scriptscriptfont4=\ottocss%
\textfont5=\eightmib\scriptfont5=\sevenmib\scriptscriptfont5=\fivemib%
\textfont6=\ottoit\scriptfont6=\sixit\scriptscriptfont6=\fiveit%
\textfont7=\ottobf\scriptfont7=\sixbf\scriptscriptfont7=\fivebf%
%\textfont\bffam=\eightmib\scriptfont\bffam=\sevenmib%
%\scriptscriptfont\bffam=\fivemib%
\setbox\strutbox=\hbox{\vrule height7pt depth2pt width0pt}%
\normalbaselineskip=9pt\rm}
\let\nota=\ottopunti%

\textfont5=\tenmib\scriptfont5=\sevenmib\scriptscriptfont5=\fivemib
\mathchardef\Ba   = "050B  %alfa
\mathchardef\Bb   = "050C  %beta
\mathchardef\Bg   = "050D  %gamma
\mathchardef\Bd   = "050E  %delta
\mathchardef\Be   = "0522  %varepsilon
\mathchardef\Bee  = "050F  %epsilon
\mathchardef\Bz   = "0510  %zeta
\mathchardef\Bh   = "0511  %eta
\mathchardef\Bthh = "0512  %teta
\mathchardef\Bth  = "0523  %varteta
\mathchardef\Bi   = "0513  %iota
\mathchardef\Bk   = "0514  %kappa
\mathchardef\Bl   = "0515  %lambda
\mathchardef\Bm   = "0516  %mu
\mathchardef\Bn   = "0517  %nu
\mathchardef\Bx   = "0518  %xi
\mathchardef\Bom  = "0530  %omi
\mathchardef\Bp   = "0519  %pi
\mathchardef\Br   = "0525  %ro
\mathchardef\Bro  = "051A  %varrho
\mathchardef\Bs   = "051B  %sigma
\mathchardef\Bsi  = "0526  %varsigma
\mathchardef\Bt   = "051C  %tau
\mathchardef\Bu   = "051D  %upsilon
\mathchardef\Bf   = "0527  %phi
\mathchardef\Bff  = "051E  %varphi
\mathchardef\Bch  = "051F  %chi
\mathchardef\Bps  = "0520  %psi
\mathchardef\Bo   = "0521  %omega
\mathchardef\Bome = "0524  %varomega
\mathchardef\BG   = "0500  %Gamma
\mathchardef\BD   = "0501  %Delta
\mathchardef\BTh  = "0502  %Theta
\mathchardef\BL   = "0503  %Lambda
\mathchardef\BX   = "0504  %Xi
\mathchardef\BP   = "0505  %Pi
\mathchardef\BS   = "0506  %Sigma
\mathchardef\BU   = "0507  %Upsilon
\mathchardef\BF   = "0508  %Fi
\mathchardef\BPs  = "0509  %Psi
\mathchardef\BO   = "050A  %Omega
\mathchardef\BDpr = "0540  %Dpr
\mathchardef\Bstl = "053F  %*
\def\BK{\bf K}
%%%%%%%%%%%%%%%%%%%%%%%%%%%%%%%%%%%%%%%%%%%%%%%%%%%%%%%%%%%%%%%%%%%%%%%%%%%%%
%%%%%%%%%   RIFERIMENTI SIMBOLICI A FORMULE, PARAGRAFI E FIGURE    %%%%%%%%%%
%%%%%%%%%%%%%%%%%%%%%%%%%%%%%%%%%%%%%%%%%%%%%%%%%%%%%%%%%%%%%%%%%%%%%%%%%%%%%
%
% Ogni paragrafo deve iniziare con il comando \section(#1,#2), dove #1
% e' il simbolo associato al paragrafo e #2 e' il titolo. Per le
% appendici bisogna pero' usare \appendix(#1,#2).
%
% Se nel titolo compaiono riferimenti ad altri simboli, questi vanno
% racchiusi fra parentesi graffe, per es. {\equ(1.2)}; in caso contrario
% si provoca un errore.
%
% Ogni sottoparagrafo deve iniziare con il comando \sub(#1) o \asub(#1),
% nelle appendici.
%
% I riferimenti a paragrafi e sottoparagrafi si realizzano con il comando
% \sec(#1), che produce il numero effettivo preceduto dal simbolo di
% paragrafo, o \secc(#1), che produce solo il numero (serve nel caso si
% faccia riferimento ad un sottoparagrafo, che e' un Lemma, un Teorema o
% altro oggetto suscettibile di una denominazione speciale).
%
% Le formule sono contrassegnate con \Eq(#1), eccetto che all'interno
% del comando \eqalignno, dove si deve usare \eq(#1). Nelle appendici
% i comandi corrispondenti sono \Eqa(#1) e \eqa(#1).
% I riferimenti alle formule si realizzano con \equ(#1).
%
% La numerazione delle figure utilizza il comando \eqg(#1), per
% contrassegnarle, e \graf(#1) per citarle.
%

\global\newcount\numsec\global\newcount\numapp
\global\newcount\numfor\global\newcount\numfig
\global\newcount\numsub
\numsec=0\numapp=0\numfig=0
\def\veroparagrafo{\number\numsec}\def\veraformula{\number\numfor}
\def\veraappendice{\number\numapp}\def\verasub{\number\numsub}
\def\verafigura{\number\numfig}

\def\Section(#1,#2){\advance\numsec by 1\numfor=1\numsub=1\numfig=1%
\SIA p,#1,{\veroparagrafo} %
\write15{\string\Fp (#1){\secc(#1)}}%
\write16{ sec. #1 ==> \secc(#1)  }%
\0\hbox%to \hsize
{\titolo\hfill
\number\numsec. #2\hfill%
\expandafter{\hglue-1truecm\alato(sec. #1)}}}

\def\appendix(#1,#2){\advance\numapp by 1\numfor=1\numsub=1\numfig=1%
\SIA p,#1,{A\veraappendice} %
\write15{\string\Fp (#1){\secc(#1)}}%
\write16{ app. #1 ==> \secc(#1)  }%
\hbox to \hsize{\titolo Appendix A\number\numapp. #2\hfill%
\expandafter{\alato(app. #1)}}\*%
}

\def\senondefinito#1{\expandafter\ifx\csname#1\endcsname\relax}

\def\SIA #1,#2,#3 {\senondefinito{#1#2}%
\expandafter\xdef\csname #1#2\endcsname{#3}\else
\write16{???? ma #1#2 e' gia' stato definito !!!!} \fi}

\def \Fe(#1)#2{\SIA fe,#1,#2 }
\def \Fp(#1)#2{\SIA fp,#1,#2 }
\def \Fg(#1)#2{\SIA fg,#1,#2 }

\def\etichetta(#1){(\veroparagrafo.\veraformula)%
\SIA e,#1,(\veroparagrafo.\veraformula) %
\global\advance\numfor by 1%
\write15{\string\Fe (#1){\equ(#1)}}%
\write16{ EQ #1 ==> \equ(#1)  }}

\def\etichettaa(#1){(A\veraappendice.\veraformula)%
\SIA e,#1,(A\veraappendice.\veraformula) %
\global\advance\numfor by 1%
\write15{\string\Fe (#1){\equ(#1)}}%
\write16{ EQ #1 ==> \equ(#1) }}

\def\getichetta(#1){\veroparagrafo.\verafigura%
\SIA g,#1,{\veroparagrafo.\verafigura} %
\global\advance\numfig by 1%
\write15{\string\Fg (#1){\graf(#1)}}%
\write16{ Fig. #1 ==> \graf(#1) }}

\def\etichettap(#1){\veroparagrafo.\verasub%
\SIA p,#1,{\veroparagrafo.\verasub} %
\global\advance\numsub by 1%
\write15{\string\Fp (#1){\secc(#1)}}%
\write16{ par #1 ==> \secc(#1)  }}

\def\Eq(#1){\eqno{\etichetta(#1)\alato(#1)}}
\def\eq(#1){\etichetta(#1)\alato(#1)}
\def\Eqa(#1){\eqno{\etichettaa(#1)\alato(#1)}}
\def\eqa(#1){\etichettaa(#1)\alato(#1)}
\def\eqg(#1){\getichetta(#1)\alato(fig. #1)}
\def\sub(#1){\0\palato(p. #1){\bf \etichettap(#1).}}
\def\asub(#1){\0\palato(p. #1){\bf \etichettapa(#1).}}
\def\apprif(#1){\senondefinito{e#1}%
\eqv(#1)\else\csname e#1\endcsname\fi}

\def\equv(#1){\senondefinito{fe#1}$\clubsuit$#1%
\write16{eq. #1 non e' (ancora) definita}%
\else\csname fe#1\endcsname\fi}
\def\grafv(#1){\senondefinito{fg#1}$\clubsuit$#1%
\write16{fig. #1 non e' (ancora) definito}%
\else\csname fg#1\endcsname\fi}
\def\secv(#1){\senondefinito{fp#1}$\clubsuit$#1%
\write16{par. #1 non e' (ancora) definito}%
\else\csname fp#1\endcsname\fi}

\def\eqo{{\global\advance\numfor by 1}}
\def\equ(#1){\senondefinito{e#1}\equv(#1)\else\csname e#1\endcsname\fi}
\def\graf(#1){\senondefinito{g#1}\grafv(#1)\else\csname g#1\endcsname\fi}
\def\figura(#1){{\css Figura} \getichetta(#1)}
\def\secc(#1){\senondefinito{p#1}\secv(#1)\else\csname p#1\endcsname\fi}
\def\sec(#1){{\secc(#1)}}
\def\refe(#1){{[\secc(#1)]}}

\def\BOZZA{%\bz=1
\def\alato(##1){\rlap{\kern-\hsize\kern-.5truecm{$\scriptstyle##1$}}}
\def\palato(##1){\rlap{\kern-.5truecm{$\scriptstyle##1$}}}
}

\def\alato(#1){}
\def\galato(#1){}
\def\palato(#1){}

%%%%%%%%%%%%%%%%%%%%%%%%%%%%%%%%%%%%%%%%%%%%%%%%%%%%%%%%%%%%%%%%%%%%%%%%%%%%%
%%%%%%%%%%%%%%%%%%%%      DATA E PIE' DI PAGINA        %%%%%%%%%%%%%%%%%%%%%%
%%%%%%%%%%%%%%%%%%%%%%%%%%%%%%%%%%%%%%%%%%%%%%%%%%%%%%%%%%%%%%%%%%%%%%%%%%%%%

{\count255=\time\divide\count255 by 60 \xdef\hourmin{\number\count255}
        \multiply\count255 by-60\advance\count255 by\time
   \xdef\hourmin{\hourmin:\ifnum\count255<10 0\fi\the\count255}}

\def\oramin{\hourmin }

\def\data{\number\day/\ifcase\month\or gennaio \or febbraio \or marzo \or
aprile \or maggio \or giugno \or luglio \or agosto \or settembre
\or ottobre \or novembre \or dicembre \fi/\number\year;\ \oramin}
\setbox200\hbox{$\scriptscriptstyle \data $}

%%%%%%%%%%%%%%%%%%%%%%%%%%%%%%%%%%%%%%%%%%%%%%%%%%%%%%%%%%%%%%%%%%%%%%%%%%%%%
%%%%%%%%%%%%%%%      INSERIMENTO FIGURE ( se si usa DVIPS )    %%%%%%%%%%%%%%
%%%%%%%%%%%%%%%%%%%%%%%%%%%%%%%%%%%%%%%%%%%%%%%%%%%%%%%%%%%%%%%%%%%%%%%%%%%%%
\newdimen\xshift \newdimen\xwidth \newdimen\yshift \newdimen\ywidth

\def\ins#1#2#3{\vbox to0pt{\kern-#2\hbox{\kern#1 #3}\vss}\nointerlineskip}

\def\eqfig#1#2#3#4#5{
\par\xwidth=#1 \xshift=\hsize \advance\xshift
by-\xwidth \divide\xshift by 2
\yshift=#2 \divide\yshift by 2%
%\line
{\hglue\xshift \vbox to #2{\vfil
#3 \includegraphics{#4.ps}
}\hfill\raise\yshift\hbox{#5}}}

\def\8{\write12}

\openin13=#1.aux \ifeof13 \relax \else
\input #1.aux \closein13\fi
\openin14=\jobname.aux \ifeof14 \relax \else
\input \jobname.aux \closein14 \fi
\immediate\openout15=\jobname.aux

%%%%%%%%%%%%%%%%%%%%%%%%%%%%%%%%%%%%%%%%%%%%%%%%%%%%%%%%%%%%%%%%%%%%%%%%%%%%%
%%%%%%%%%%%%%%%%%%%%%%         SIMBOLI VARI           %%%%%%%%%%%%%%%%%%%%%%%
%%%%%%%%%%%%%%%%%%%%%%%%%%%%%%%%%%%%%%%%%%%%%%%%%%%%%%%%%%%%%%%%%%%%%%%%%%%%%

\let\a=\alpha \let\b=\beta  \let\g=\gamma  \let\d=\delta \let\e=\varepsilon
\let\z=\zeta  \let\h=\eta   \let\th=\theta \let\k=\kappa \let\l=\lambda
\let\m=\mu    \let\n=\nu    \let\x=\xi     \let\p=\pi    \let\r=\rho
 \let\t=\tau   \let\f=\varphi 
\let\ch=\chi  \let\ps=\psi   \let\o=\omega
\let\G=\Gamma \let\D=\Delta  \let\Th=\Theta 
     \let\F=\Phi

\def\\{\hfill\break} \let\==\equiv
\let\txt=\textstyle

\let\io=\infty 
\def\Dpr{\BDpr\,}%\def\Dpr{\V\dpr\,}
\def\ap{{\it a priori\ }}
\let\0=\noindent\def\pagina{{\vfill\eject}}

\def\ie{\hbox{\it i.e.\ }}\def\eg{\hbox{\it e.g.\ }}
\let\dpr=\partial

\def\tende#1{\,\vtop{\ialign{##\crcr\rightarrowfill\crcr
 \noalign{\kern-1pt\nointerlineskip} \hskip3.pt${\scriptstyle
 #1}$\hskip3.pt\crcr}}\,}
\def\circage{\lower2pt\hbox{$\,\buildrel > \over {\scriptstyle \sim}\,$}}
\def\otto{\,{\kern-1.truept\leftarrow\kern-5.truept\to\kern-1.truept}\,}
\def\fra#1#2{{#1\over#2}}

\def\PP{{\cal P}}\def\EE{{\cal E}} \def\MM{{\cal M}} 
\def\CC{{\cal C}} \def\HH{{\cal H}} 
\def\TT{{\cal T}}  \def\II{{\cal I}}
\def\LL{{\cal L}}  
\def\AA{{\cal A}}

\def\T#1{{#1_{\kern-3pt\lower7pt\hbox{$\widetilde{}$}}\kern3pt}}
\def\VVV#1{{\underline #1}_{\kern-3pt
\lower7pt\hbox{$\widetilde{}$}}\kern3pt\,}
\def\W#1{#1_{\kern-3pt\lower7.5pt\hbox{$\widetilde{}$}}\kern2pt\,}

\def\lis{\overline}\def\tto{\Rightarrow}

\def\indica{\leaders \hbox to 0.5cm{\hss.\hss}\hfill}
\def\guida{\leaders\hbox to 1em{\hss.\hss}\hfill}

\def\qed{\raise1pt\hbox{\vrule height5pt width5pt depth0pt}}

\def\Val{{\rm Val}}
\def\indic{\hbox{\raise-2pt \hbox{\indbf 1}}}

\def\RRR{\hbox{\msytw R}}

 \def\ZZZ{\hbox{\msytw Z}}
 \def\zzz{\hbox{\msytwww Z}}
\def\TTT{\hbox{\msytw T}} \def\tttt{\hbox{\msytww T}}

\def\defi{\,{\buildrel def\over=}\,}

\def\rhs{{\it r.h.s.}\ }

\def\sqr#1#2{{\vcenter{\vbox{\hrule height.#2pt%
        \hbox{\vrule width.#2pt height#1pt \kern#1pt%
          \vrule width.#2pt}%
        \hrule height.#2pt}}}}

\def\ig{\int}

\footline={\rlap{\hbox{\copy200}}\tenrm\hss \number\pageno\hss}
\def\V#1{{\bf#1}}

\def\atan{{\,\mathop{\rm atan}\,}}
\def\tb{{\th}_0}\def\psb{{\ps}_0}
\def\fb{{\f}_0}

\vglue1.truecm
\centerline{\titolone Classical Mechanics}
\*\*

\centerline
{\it Giovanni Gallavotti}

\centerline
{\it I.N.F.N. Roma 1, Fisica Roma1}

\*\*
\Section(1, General principles)

\*
Classical Mechanics is a theory of point particles motions. If $\V
X=(\V x_1,\ldots$, $\V x_n)$ are the particles positions in a Cartesian
inertial system of coordinates, the equations of motion are determined
by their masses $(m_1,\ldots,m_n)$, $m_j>0$, and by the {\it potential
energy} of interaction $V(\V x_1,\ldots,\V x_n)$ as

$$m_i {\ddot{\V x}}_i=-\dpr_{\V x_i} V(\V x_1,\ldots,\V
x_n),\qquad i=1,\ldots,n\Eq(1.1)$$
here $\V x_i=(x_{i1},\ldots,x_{id})$ are coordinates of the $i$-th
particle and $\dpr_{\V x_i}$ is the gradient $(\dpr_{x_{i1}},\ldots,$
$\dpr_{x_{id}})$; $d$ is the space dimension (\ie $d=3$, usually). The
potential energy function will be supposed ``smooth'', \ie
{\it analytic} except, possibly, when two positions coincide. The latter
exception is necessary to include the important cases of gravitational
attraction or, when dealing with electrically charged particles, of
Coulomb interaction. A basic result is that if $V$ is bounded below
the equation \equ(1.1) admits, given initial data $\V X_0=\V X(0),\V
{\dot X}_0=\V{\dot X}(0)$, a unique global solution $t\to\V X(t),\,
t\in(-\io,\io)$; otherwise a solution can fail to be global if and
only if, in a finite time, it reaches infinity or a singularity point
(\ie a configuration in which two or more particles occupy the same
point: an event called a {\it collision}).

In Eq. \equ(1.1) $-\dpr_{\V x_i} V(\V x_1,\ldots,\V x_n)$ is the {\it
force} acting on the points. More general forces are often
admitted. For instance velocity dependent {\it friction forces}: they
are not considered here because of their phenomenological nature as
models for microscopic phenomena which should also, in principle, be
explained in terms of conservative forces (furthermore, even from a
macroscopic viewpoint, they are rather incomplete models as they
should be considered together with the important heat generation
phenomena that accompany them). Another interesting example of forces
not corresponding to a potential are certain velocity dependent forces
like the {\it Coriolis force} (which however appears only in non
inertial frames of reference) and the closely related {\it Lorentz
force} (in electromagnetism): they could be easily accomodated in the
upcoming Hamiltonian formulation of mechanics, see Appendix A2.

The {\it action principle} states that an equivalent formulation of
the equations \equ(1.1) is that a motion $t\to \V X_0(t)$
satisfying \equ(1.1) during a time interval $[t_1,t_2]$ and leading
from $\V X^1=\V X_0(t_1)$ to $\V X^2=\V X_0(t_2)$, renders stationary the
{\it action}

$$\AA(\{\V X\})= \ig_{t_1}^{t_2} \Big(\sum_{i=1}^n \fra12
\,m_i\,{\dot{\V X}_i}(t)^2 -V(\V X(t))\Big)\,dt\Eq(1.2)$$
within the class $\MM_{t_1,t_2}{(\V X^1,\V X^2)}$ of smooth (\ie
analytic) ``motions'' $t\to\V X(t)$ defined for $t\in[t_1,t_2]$ and
leading from $\V X^1$ to $\V X^2$.

The function $\LL(\V Y,\V X)=\fra12\sum_{i=1}^n m_i \V y^2_i-V(\V
X)\defi K(\V Y)-V(\V X)$, $\V Y=(\V y_1,\ldots,\V y_n)$, is called the
{\it Lagrangian} function and the action can be written as
$\ig_{t_1}^{t_2}\LL(\dot{\V X}(t),\V X(t))\,dt$. The quantity
$K(\dot{\V X}(t))$ is called {\it kinetic energy} and motions
satisfying \equ(1.1) conserve {\it energy} as time $t$ varies, \ie

$$K(\dot{\V X}(t))+V(\V X(t))=E=const\Eq(1.3)$$
Hence the action principle can be intuitively thought of as saying
that motions proceed by keeping constant the energy, sum of the kinetic and
potential energies, while trying to share as evenly as possible
their (average over time) contribution to the energy.

In the special case in which $V$ is translation invariant motions
conserve {\it linear momentum} $\V Q\defi\sum_i m_i \dot{\V x}_i$; if
$V$ is rotation invariant around the origin $O$ motions conserve {\it
angular momentum} $\V M\defi$ $\sum_i m_i\, \V x_i\wedge\V{\dot x}_i$,
where $\wedge$ denotes the vector product in $\RRR^d$, \ie it is the
tensor $(\V a\wedge \V b)_{ij}=a_ib_j-b_ia_j,\, i,j=1,\ldots,d$: if
the dimension $d=3$ the $\V a\wedge\V b$ will be naturally
regarded as a vector. More generally to any continuous symmetry group
of the Lagrangian correspond conserved quantities: this is formalized
in the {\it Noether theorem}.

It is convenient to think that the scalar product in $\RRR^{dn}$ is
defined in terms of the ordinary scalar product in $\RRR^d$, $\V
a\cdot\V b=\sum_{j=1}^d a_j b_j$, by $(\V v,\V w)=\sum_{i=1}^n m_i \V
v_i\cdot\V w_i$: so that kinetic
energy and line element $ds$ can be written as $K(\V{\dot X})=\fra12
(\V{\dot X},\V{\dot X})$ and, respectively, $ds^2=\sum_{i=1}^n m_i d\V
x_i^2$. Therefore the metric generated by the latter scalar product
can be called {\it kinetic energy metric}.

The interest of the kinetic metric appears from the {\it Maupertuis'
principle} ({\it equivalent} to \equ(1.1)): the principle allows us to
identify the trajectory traced in $\RRR^d$ by a motion that leads from
$\V X^1$ to $\V X^2$ moving with energy $E$. Parameterizing such
trajectories as $\t\to \V X(\t)$ by a parameter $\t$ varying in
$[0,1]$ so that the line element is $ds^2=(\V{\dpr_\t X},\V{\dpr_\t
X})d\t^2$, the principle states that the trajectory of a motion with
energy $E$ which leads from $\V X^1$ to $\V X^2$ makes stationary,
among the analytic curves $\Bx\in\MM_{0,1}(\V X^1,\V X^2)$, the
function

$$L(\Bx)=\ig_{\Bx} \sqrt{E-V(\Bx(s))}\,ds\Eq(1.4)$$
so that the possible trajectories traced by the solutions of \equ(1.1)
in $\RRR^{nd}$ and with energy $E$ can be identified with the {\it
geodesics} of the metric $d m^2\defi (E-V(\V X))\cdot d s^2$.
\*

\0{\it References:} [LL68], [Ga83]
\*

\Section(2,Constraints)
\*

Often particles are subject to {\it constraints} which force the
motion to take place on a surface $M\subset \RRR^{nd}$: \ie $\V X(t)$
is forced to be a point on the manifold $M$.  A typical example is
provided by {\it rigid systems} in which motions are subject to forces
which keep the mutual distances of the particles constant: $|\V x_i-\V
x_j|=\r_{ij}$ with $\r_{ij}$ time independent positive quantities. In
essentially all cases the forces that imply constraints,
called {\it constraints reactions}, are velocity dependent and,
therefore, {\it are not} in the class of {\it conservative} forces
considered here, cf. \equ(1.1).  Hence, from a fundamental viewpoint
admitting only conservative forces, constrained systems should be
regarded as idealizations of systems subject to conservative forces
which {\it approximately} imply the constraints.

In general the $\ell$-dimensional manifold $M$ will not admit a global
system of coordinates: however it will be possible to describe points
in the vicinity of any $\V X^0\in M$ by using $N=nd$
coordinates $\V q=(q_1,\ldots,q_\ell, q_{\ell+1},\ldots,q_{N})$
varying in an open ball $B_{\V X^0}$: $\V X=\V X(q_1,\ldots,q_\ell,
q_{\ell+1},$ $\ldots,q_{N})$.

The $q$-coordinates can be chosen {\it well adapted} to the surface
$M$ and to the kinetic metric, \ie so that the points of $M$ are
identified by $q_{\ell+1}=\ldots=q_N=0$ (which is the meaning of
``adapted'') and furthermore infinitesimal displacements
$(0,\ldots,0,d\e_{\ell+1},\ldots,d\e_N)$ out of a point $\V X^0\in M$
are orthogonal to $M$ (in the kinetic metric) and have a length
independent of the position of $\V X^0$ on $M$ (which is the meaning
of `well adapted'' to the kinetic metric).

Motions constrained on $M$ arise when the potential $V$ has
the form

$$V(\V X)= V_a(\V X)+\l\,W(\V X)\Eq(2.1)$$
where $W$ is a smooth function which reaches its minimum value, say
equal to $0$, precisely on the manifold $M$ while $V_a$ is another
smooth potential. The factor $\l>0$ is a parameter called the {\it
rigidity} of the constraint.

A particularly interesting case arises when, furthermore, the level
surfaces of $W$ have the geometric property of being ``parallel'' to
the surface $M$: in the precise sense that the matrix
$\dpr^2_{q_iq_j}W(\V X), \, i,j>\ell$ is positive definite and $\V
X$-independent for all $\V X\in M$ in a system of coordinates well
adapted to the kinetic metric.

A potential $W$ with the latter properties can be
called an {\it approximately ideal constraint reaction}.
In fact it can be proved that, given an initial datum $\V X^0\in M$ with
velocity $\V{\dot X}^0$ tangent to $M$, \ie given an initial datum
whose coordinates in a local system of coordinates are $(\V q_0,\V 0)$
and $(\dot{\V q}_0,\V0)$ with $\V q_0=(q_{01},\ldots,q_{0\ell})$ and
$\dot{\V q}_0=(\dot q_{01},\ldots,\dot q_{0\ell})$, the motion
generated by \equ(1.1) with $V$ given by \equ(2.1) is a motion $t\to
\V X_\l(t)$ which
\*

\0(1) as $\l\to\io$ tends to a motion $t\to \V X_\io(t)$,

\0(2) as long as $\V X_\io(t)$ stays in the vicinity of the initial data,
say for $0\le t\le t_1$, so that it can be described in the above local
adapted coordinates, its coordinates have the form $t\to(\V q(t),\V
0)=(q_{1}(t),\ldots, q_\ell(t),0,\ldots,0)$: \ie it is a motion
developing on the constraint surface $M$,

\0(3) the curve $t\to \V X_\io(t),\, t\in[0,t_1]$, as an element of
the space $\MM_{0,t_1}{(\V X^0,\V X_{\io}(t_1))}$ of analytic curves
on $M$ connecting $\V X^0$ to $\V X_{\io}(t_1)$,
renders the action

$$A(\V X)=\ig_{0}^{t_1} \big(K(\dot{\V X}(t))-V_a(\V X(t))\big)\,dt
\Eq(2.2)$$
 stationary.
\*

The latter property can be formulated ``intrinsically'', \ie referring
only to $M$ as a surface, via the restriction of the metric $d
s^2$ to line elements $d \V s=(d q_1,\ldots,d q_\ell,0,\ldots,0)$
tangent to $M$ at the point $\V X=(\V q_0,0,\ldots,0)\in M$; we
write $d s^2=\sum_{i,j}^{1,\ell} g_{ij}(\V q) dq_i\,dq_j$. The
$\ell\times\ell$ symmetric positive definite matrix $g$ is called the
{\it metric} on $M$ induced by the kinetic metric. Then the action in
\equ(2.2) can be written as

$$\AA(\V q)=\ig_0^{t_1}\Big(\fra12 \sum_{i,j}^{1,\ell} g_{ij}(\V q(t))
\dot q_i(t)\dot q_j(t)-\lis V_a(\V q(t))\Big)\,dt\Eq(2.3)$$
where $\lis V_a(\V q)\defi V_a(\V X(q_1,\ldots,q_\ell,0,\ldots,0))$:
the function

$$\LL(\V \Bh,\V q)\defi\fra12 \sum_{i,j}^{1,\ell}
g_{ij}(\V q)\h_i\h_j-\lis V_a(\V q)\=\fra12 g(\V q)\Bh\cdot\Bh-
\lis V_a(\V q)\Eq(2.4)$$
is called the {\it constrained Lagrangian} of the system.

An important property is that the constrained
motions conserve the energy defined as $E=\fra12 (g(\V q)\dot{\V
q},\dot{\V q})+\lis V_a(\V q)$, see Sec. \sec(3).

The constrained motion $\V X_\io(t)$ of energy $E$ satisfies the
Maupertuis' principle in the sense that the curve on $M$ on which the
motion develops renders

$$L(\Bx)=\ig_\Bx \sqrt{E- V_a(\Bx(s))} \,d s\Eq(2.5)$$
stationary among the (smooth) curves that develop on $M$ connecting
two fixed values $\V X_1$ and $\V X_2$. In the particular case in
which $\ell=n$ this is again Maupertuis' principle for unconstrained
motions under the potential $V(\V X)$. In general $\ell$ is called the
{\it number of degrees of freedom} because a complete description of
the initial data requires $2\ell$ coordinates $\V q(0),\dot{\V q}(0)$.

If $W$ is minimal on $M$ but the condition on $W$ of having level
surfaces parallel to $M$ is not satisfied, \ie if $W$ is not an
approximate ideal constraint reaction, it remains still true that the
limit motion $\V X_\io(t)$ takes place on $M$. However, {\it in general},
it will not satisfy the above variational principles. For this reason
motions arising as limits as $\l\to\io$ of motions developing under
the potential \equ(2.1) with $W$ having minimum on $M$ and level
curves parallel (in the above sense) to $M$ are called {\it ideally
constrained motions} or motions subject by ideal constraints to the
surface $M$.

As an example suppose that $W$ has the form $W(\V X)=\sum_{i, j\in
\PP} w_{ij}(|\V x_i-\V x_j|)$ with $w_{ij}(|\Bx|)\ge0$ an analytic
function vanishing only when $|\Bx|=\r_{ij}$ for $i,j$ in some set of
pairs $\PP$ and for some given distances $\r_{ij}$ (for instance
$w_{ij}(\Bx)=(\Bx^2-\r_{ij}^2)^2\,\g$, $\g>0$). Therefore {\it the so
constrained motions $\V X_\io(t)$ of the body satisfy the
variational principles} mentioned in connection with \equ(2.3) and
\equ(2.5): in other words the above natural way of realizing a rather general
rigidity constraint is ideal.  \*

This is the modern viewpoint on the physical meaning of the
constraints reactions: looking at motions in an inertial cartesian
system it will appear that the system is subject to the {\it applied
forces} with potential $V_a(\V X)$ and to {\it constraint forces}
which are {\it defined} as the differences $\V R_i=m_i\ddot{\V
x}_i+\Dpr_{\V x_i} V_a(\V X)$. The latter reflect the action of the
forces with potential $\l W(\V X)$ in the limit of infinite rigidity
($\l\to\io$).

In applications {\it sometimes} the action of a constraint can be
regarded as ideal: so that the motion will verify the mentioned
variational principles and the $\V R$ can be computed as the
differences between the $m_i\ddot{\V x}_i$ and the active forces
$-\Dpr_{\V x_i} V_a(\V X)$.  In dynamics problems it is, {\it
however}, a very difficult and important matter, particularly in
engineering, to judge whether a system of particles can be considered
as subject to ideal constraints: this leads to important decisions in
the construction of machines. It simplifies the calculations of the
efforts of the materials but a misjudgement can have serious
consequences about stability and safety. For statics problems the
difficulty is of lower order: usually assuming that the constraints
reaction is ideal leads to an overestimate of the requirements for
stability of equilibria. Hence employing the action principle to
statics problems, where it constitutes the {\it virtual works
principle}, generally leads to economic problems rather than to safety
issues. Its discovery even predates Newtonian mechanics.

\*

\0{\it References:} [Ar68], [Ga83].
\*

\Section(3,Lagrange and Hamilton form of the equations of motion)
\*

The stationarity condition for the action $\AA(\V q)$,
cf. \equ(2.3),\equ(2.4), is formulated in terms of the Lagrangian
$\LL(\Bh,\Bx)$, see \equ(2.4), by

$$\fra{d}{dt}\dpr_{\h_i} \LL(\dot {\V q}(t),{\V q}(t)) =
\dpr_{q_i} \LL(\dot {\V q}(t),{\V q}(t)), \qquad i=1,\ldots,\ell\Eq(3.1)$$
which is a second order differential equation called {\it Lagrangian
equation of motion}. It can be cast in ``normal form'': for this
purpose, adopting the convention of ``summation over repeated
indices'', introduce the ``generalized momenta''

$$p_i\defi g(\V q)_{ij} \dot q_j,\qquad i=1,\ldots,\ell\Eq(3.2)$$
Since $g(\V q)>0$ the motions $t\to\V q(t)$ and the corresponding
velocities $t\to\dot{\V q}(t)$ can be described equivalently by
$t\to(\V q(t),\V p(t))$: and the equations of motion \equ(3.1) become
the first order equations

$$\dot{q}_i=\dpr_{p_i} \HH(\V p,\V q),\qquad \dot{p}_i=-\dpr_{q_i} H(\V
p,\V q)\Eq(3.3)$$
where the function $\HH$, called {\it Hamiltonian} of the system, is
defined by

$$\HH(\V p,\V q)\defi \fra12 (g(\V q)^{-1}\V p,\V p)+ \lis V_a(\V
q)\Eq(3.4)$$
Eq. \equ(3.3), regarded as equations of motion for phase space
points $(\V p,\V q)$, are called {\it Hamilton equations}.
In general $\V q$ are local coordinates on $M$ and motions are
specified by giving $\V q,\dot{\V q}$ or $\V p,\V q$.

Looking for a coordinate free
representation of motions consider the pairs $\V
X,\V Y$ with $\V X \in M$ and $\V Y$ a vector $\V Y\in T_{\V X}$
tangent to $M$ at the point $\V X$.
The collection of pairs $(\V Y,\V X)$ is denoted
$T(M)=\cup_{\V X\in M}\big(T_{\V X}\times \{\V X\}\big)$ and a motion $t\to
(\dot{\V X}(t),\V X(t))\in T(M)$ in local coordinates is
represented by $(\dot{\V q}(t),\V q(t))$. The space $T(M)$ can be called the
space of initial data for Lagrange's equations of motion: it has
$2\ell$ dimensions (more fancily known as the {\it tangent bundle} of $M$).

Likewise the space of initial data for the Hamilton equations will
be denoted $T^*(M)$ and it consists in pairs $\V X,\V P$
with $\V X\in M$ and $\V P= g(\V X) \V Y$ with $\V Y$ a vector
tangent to $M$ at $\V X$. The space $T^*(M)$ is called the {\it phase
space} of the system: it has $2\ell$ dimensions (and it is occasionally
called the {\it cotangent bundle} of $M$).

Immediate consequence of \equ(3.3) is $\fra{d}{dt} \HH(\V p(t),\V
q(t))\=0$ and it means that $\HH(\V p(t),\V q(t))$ is constant along the
solutions of the \equ(3.3).  Remarking that $\HH(\V p,\V q)=\fra12 (g(\V
q)\dot{\V q},\dot{\V q})+\lis V_a(\V q)$ is the sum of the kinetic and
potential energies it follows that the conservation of $\HH$ along
solutions has the meaning of {\it energy conservation} in presence of
ideal constraints.

Let $S_t$ be the {\it flow} generated on the phase space variables $(\V p,\V
q)$ by the solutions of the equations of motion \equ(3.3), \ie let
$t\to S_t(\V p,\V q)\=(\V p(t),\V q(t))$ denote a solution of
\equ(3.3) with initial data $(\V p,\V q)$. Then a (measurable) set $\D$ in phase
space evolves in time $t$ into a new set $S_t\D$ {\it with the same
volume}: this is obvious because the Hamilton equations \equ(3.3) have
manifestly $0$ divergence (``{\it Liouville's theorem}'').

The Hamilton equations also satisfy a variational principle, called
{\it Hamilton action principle}: \ie if $\MM_{t_1,t_2}{((\V p_1,\V q_1),(\V
p_2,\V q_2);M)}$ denotes the space of the analytic functions
$\Bf:\,t\to (\Bp(t),\Bk(t))$ which in the time interval $[t_1,t_2]$
lead from $(\V p_1,\V q_1)$ to $(\V p_2,\V q_2)$, then the condition
that $\Bf_0(t)=(\V p(t),\V q(t))$ satisfies \equ(3.3) can be
equivalently formulated by requiring that the function

$$\AA_\HH(\Bf)\defi\ig_{t_1}^{t_2} \Big(\Bp(t)\cdot\dot\Bk(t)-
\HH(\Bp(t),\Bk(t))\Big)\,dt \Eq(3.5)$$
be stationary for $\Bf=\Bf_0$: in fact the \equ(3.3) are the
stationarity conditions for the Hamilton action \equ(3.5) on
$\MM_{t_0,t_1} {((\V p_1,\V q_1),(\V p_2,\V q_2);M)}$. And since the
derivatives of $\Bp(t)$ do not appear in \equ(3.5) stationarity is
even achieved in the larger space $\MM_{t_1,t_2}{(\V q_1,\V q_2;M)}$
of the motions $\Bf:\,t\to (\Bp(t),\Bk(t))$ leading from $\V q_1$ to
$\V q_2$ {\it without} any restriction on the initial and final
momenta $\V p_1,\V p_2$ (which therefore cannot be prescribed \ap
independently of $\V q_1,\V q_2$). If the prescribed data $\V p_1,\V
q_1,\V p_2,\V q_2$ are not compatible with the equations of motion,
\eg $H(\V p_1,\V q_2)\ne H(\V p_2,\V q_2)$, then the action functional
has no stationary trajectory in $\MM_{t_1,t_2}{((\V p_1,\V q_1),(\V
p_2\V q_2);M)}$.  \*

\0{\it References:} [LL68], [Ar68], [Ga83]
\*

\Section(4, Canonical transformations of phase space coordinates)
\*

The Hamiltonian form, \equ(3.4), of the equations of motion turns out
to be quite useful in several problems. It is therefore important to
remark that it is {\it invariant} under a special class of
transformations of coordinates, called {\it canonical
transformations}.

Consider a local change of coordinates on phase space, \ie a smooth
smoothly invertible map $\CC(\Bp,\Bk)=(\Bp',\Bk')$ between an open set
$U$ in the phase space of a $\ell$ degrees of freedom Hamiltonian
system, into an open set $U'$ in a $2\ell$ dimensional space.  The
change of coordinates is said {\it canonical} if for any solution
$t\to(\Bp(t),\Bk(t))$ of equations like \equ(3.3), for any Hamiltonian
$\HH(\Bp,\Bk)$ defined on $U$, the $\CC$--image $t\to
(\Bp'(t),\Bk'(t))=\CC(\Bp(t),\Bk(t))$ are solutions of the \equ(3.3)
with the ``same'' Hamiltonian, \ie with Hamiltonian
$\HH'(\Bp',\Bk')\defi \HH(\CC^{-1}(\Bp',\Bk'))$.

The condition that a transformation of coordinates is canonical is
obtained by using the arbitrariness of the function $\HH$ and is simply
expressed as a necessary and sufficient property of the Jacobian $L$

$$\txt L=\pmatrix{A&B\cr C&D\cr},\quad
A_{ij}=\dpr_{\p_j} \p'_i, \ B_{ij}=\dpr_{\k_j}\p'_i, \
C_{ij}=\dpr_{\p_j} \k'_i, \ D_{ij}=\dpr_{\k_j}\k'_i, \Eq(4.1)$$
where $i,j=1,\ldots,\ell$.  Let $E=\pmatrix{0&1\cr-1&0}$ denote the
$2\ell\times2\ell$ matrix formed by four $\ell\times\ell$ blocks, equal
to the $0$ matrix or, as indicated, to the $\pm$ identity matrix;
then, if a superscript $T$ denotes matrix transposition, the condition
that the map be canonical is that

$$L^{-1}=E L^T E^T \qquad{\rm or}\quad L^{-1}=\pmatrix{D^T&-B^T\cr
-C^T& A^T\cr}\Eq(4.2)$$
which immediately implies that $\det L=\pm1$. In fact it is possible
to show that \equ(4.2) implies $\det L=1$.  The \equ(4.2) is
equivalent to the four relations $AD^T-B C^T=1$, $-A B^T+B A^T=0$,
$CD^T-D C^T=0$, $-C B^T+D A^T=1$. More explicitly, since the first and
the fourth coincide, the relations are

$$\big\{\p'_i,\k'_j\big\}=\d_{ij},\quad
\big\{\p'_i,\p'_j\big\}=0,\quad \big\{\k'_i,\k'_j\big\}=0\Eq(4.3) $$
where, for any two functions
$F(\Bp,\Bk),G(\Bp,\Bk)$, the {\it Poisson bracket} is

$$\big\{F,G\big\}(\Bp,\Bk)\defi \sum_{k=1}^\ell \Big(\dpr_{\p_k} F(\Bp,\Bk)\,
\dpr_{\k_k} G(\Bp,\Bk)-\dpr_{\k_k} F(\Bp,\Bk)\,
\dpr_{\p_k} G(\Bp,\Bk)\Big)\Eq(4.4)$$
The latter satisfies {\it Jacobi's identity}:
$\big\{\big\{F,G\big\},Q\big\}+
\big\{\big\{G,Q\big\},F\big\}+\big\{\big\{Q,F\big\},G\big\}$ $=0$,
for any three functions $F,G,Q$ on phase space. It is quite
useful to remark that if $t\to (\V p(t),\V q(t))=S_t(\V p,\V q)$ is a
solution to Hamilton equations with Hamiltonian $\HH$ then, given
any {\it observable} $F(\V p,\V q)$, it ``evolves'' as $F(t)\defi F(\V
p(t),\V q(t))$ satisfying $\dpr_t F(\V p(t),\V
q(t))=\,\big\{\HH,F\big\}(\V p(t),\V q(t))$. Requiring the latter
identity to hold for all observables $F$ is {\it equivalent} to
requiring that the $t\to(\V p(t),\V q(t))$ be a solution of Hamilton's
equations for $\HH$.

Let $\CC: U\otto U'$ be a smooth, smoothly invertible, transformation
between two $2\ell$ dimensional sets: $\CC(\Bp,\Bk)=(\Bp',\Bk')$.
Suppose that there is a function $\F(\Bp',\Bk)$ defined on a
suitable domain $W$ and such that

$$\CC(\Bp,\Bk)=(\Bp',\Bk')\ \tto\ \cases{\Bp=\dpr_{\Bk}\F(\Bp',\Bk)&\cr
\Bk'=\dpr_{\Bp'}\F(\Bp',\Bk)&\cr}\Eq(4.5)$$
then $\CC$ is canonical. This is because \equ(4.5) implies that if
$\Bk,\Bp'$ are varied and if $\Bp,\Bk',\Bp',\Bk$ are related by
$\CC(\Bp,\Bk)=(\Bp',\Bk')$, then $\Bp\cdot d\Bk + \Bk'\cdot d\Bp'=
d\F(\Bp',\Bk)$: which implies

$$\Bp\cdot d\Bk-\HH(\Bp,\Bk)dt\= \Bp'\cdot d\Bk'-\HH(\CC^{-1}(\Bp',\Bk'))dt
+d \F(\Bp',\Bk)-d(\Bp'\cdot\Bk')\Eq(4.6)$$
and means that the Hamiltonians $\HH(\V p,\V q))$ and $\HH'(\V p',\V
q'))\defi \HH(\CC^{-1}(\V p',\V q'))$ have Hamilton actions $\AA_\HH$ and
$\AA_{\HH'}$ differing by a constant, if evaluated on corresponding
motions $(\V p(t),\V q(t))$ and $(\V p'(t),\V q'(t))=\CC(\V p(t),\V
q(t))$.

The constant depends only on the initial and final values
$(\V p(t_1),\V q(t_1))$ and $(\V p(t_2),$ $\V q(t_2))$ and, respectively,
$(\V p'(t_1),\V q'(t_1))$ and $(\V p'(t_2),\V q'(t_2))$ so that if
$(\V p(t)$, $\V q(t))$ makes $\AA_\HH$ extreme also $(\V p'(t),\V
q'(t))=\CC(\V p(t),\V q(t))$ makes $\AA_{\HH'}$ extreme.

Hence if $t\to (\V p(t),\V q(t))$ solves the Hamilton equations with
Hamiltonian $\HH(\V p,\V q)$ then the motion $t\to (\V p'(t),\V q'(t))=\CC(\V
p(t),\V q(t))$ solves the Hamilton equations with Hamiltonian $\HH'(\V
p',\V q')= \HH(\CC^{-1} (\V p',\V q'))$ {\it no matter which is the function
$\HH$}: therefore the transformation is canonical. The function $\F$ is
called its {\it generating function}.

Eq. \equ(4.5) provides a way to construct canonical maps. Suppose that a
function $\F(\Bp',\Bk)$ is given and defined on some domain $W$; then
setting $\cases{\Bp=\dpr_{\Bk}\F(\Bp',\Bk)&\cr
\Bk'=\dpr_{\Bp'}\F(\Bp',\Bk)&\cr}$ and inverting the first equation in
the form $\Bp'=\BX(\Bp,\Bk)$ and substituting the value for $\Bp'$,
thus obtained, in the second equation, a map
$\CC(\Bp,\Bk)=(\Bp',\Bk')$ is defined on some domain (where the mentioned
operations can be performed) and if such domain is open and not empty
then $\CC$ is a canonical map.

For similar reasons if $\G(\Bk,\Bk')$ is a function defined on some
domain then setting $\Bp=\dpr_\Bk \G(\Bk,\Bk')$, $\Bp'=-\dpr_{\Bk'}
\G(\Bk,\Bk')$ and solving the first relation to express
$\Bk'=\BD(\Bp,\Bk)$ and substituting in the second relation a map
$(\Bp',\Bk')=\CC(\Bp,\Bk)$ is defined on some domain (where the mentioned
operations can be performed) and if such domain is open and not empty
then $\CC$ is a canonical map.

Likewise canonical transformations can be constructed starting from
\ap given functions $F(\Bp,\Bk')$ or $G(\Bp,\Bp')$. And the most
general canonical map can be generated locally (\ie near a given point
in phase space) by a single one of the above four ways, possibly
composed with a few ``trivial'' canonical maps in which one pair of
coordinates $(\p_i,\k_i)$ is transformed into $(-\k_i,\p_i)$. The
necessity of including also the trivial maps can be traced to the
existence of {\it homogeneous} canonical maps, \ie maps such that
$\Bp\cdot d\Bk=\Bp'\cdot d\Bk'$ (\eg the identity map, see below or
\equ(9.7) for nontrivial examples) which are action preserving hence
canonical, but which evidently cannot be generated by a function
$\F(\Bk,\Bk')$ although they can be generated by a function depending
on $\Bp',\Bk$.

Simple examples of homogeneous canonical maps are maps in which the coordinates
$\V q$ are changed into $\V q'=\V R(\V q)$ and, correspondingly, the
$\V p$'s are transformed as $\V p'= \big(\dpr_{\V q}
\V R(\V q)\big)^{-1\,{T}}\V p$, linearly:
indeed this map is generated by the function
$F(\V p',\V q)\defi \V p'\cdot\V R(\V q)$.

For instance consider the map ``Cartesian--polar'' coordinates
$(q_1,q_2)\otto$ $(\r,\th)$ with $(\r,\th)$ the polar coordinates of
$\V q$ (namely $\r=\sqrt{q_1^2+q_2^2},\th=\atan q_2/q_1$) and let $\V
n\defi {\V q}/\,|\V q|$ $=(n_1,n_2)$ and $\V t=(-n_2,n_1)$.  Setting
$p_\r\defi\V p\cdot\V n,\,p_\th\defi \r\,\V p\cdot\V t$, the map
$(p_1,p_2,q_1,q_2)\otto $ $(p_\r,p_\th,\r,\th)$ is homogeneous
canonical (because $\V p\cdot d\V q=\V p\cdot \V n \,d\r+\V p\cdot\V t
\,\r\, d\th=p_\r\, d\r+p_\th\, d\th$).

As a further example any area preserving map
$(p,q)\otto(p',q')$ defined on an open region of the
plane $\RRR^2$ is canonical: because in this case the matrices $A,B,C,D$ are
just numbers which satisfy $AD-BC=1$ and, therefore, the \equ(4.2) holds.
\*

\0{\it References:} [LL68], [Ga83]
\*

\Section(5,Quadratures)
\*

The simplest mechanical systems are {\it integrable by
quadratures}. For instance the Hamiltonian on
$\RRR^2$

$$\HH(p,q)=\fra1{2m} p^2+ V(q)\Eq(5.1)$$
generates a motion
$t\to q(t)$ with initial data $q_0,\dot q_0$ such that $\HH(p_0,q_0)=E$,
\ie $\fra12m \dot q_0^2+V(q_0)=E$, satisfying
$\dot q(t)=\pm\sqrt{\fra2m(E-V(q(t)))}$. If the equation
$E=V(q)$ has only two solutions $q_-(E)<q_+(E)$ and
$|\dpr_q V(q_\pm(E))|>0$ the motion is periodic with period

$$\txt T(E)=2\ig_{q_-(E)}^{q_+(E)} \fra{dx}{\sqrt{\fra2m (E-V(x))}}.\Eq(5.2)$$
The special solution with initial data $q_0=q_-(E),\dot q_0=0$ will be
denoted $Q(t)$ and it is an analytic function (by the general
regularity theorem on ordinary differential equations). For $0\le t\le
\fra{T}2$ or for $\fra{T}2\le t\le T$ it is given, respectively, by

$$\txt t=\ig_{q_-(E)}^{Q(t)}
\fra{dx}{\sqrt{\fra2m(E-V(x))}},\qquad{\rm or}\qquad 
t=\fra{T}2-\ig^{Q(t)}_{q_+(E)} \fra{dx}{\sqrt{\fra2m(E-V(x))}}.\Eq(5.3)
$$
The most general solution with energy $E$ has the form $q(t)=Q(t_0+t)$
where $t_0$ is defined by $q_0=Q(t_0), \dot q_0=\dot Q(t_0)$, \ie it
is the time needed to the ``standard solution'' $Q(t)$ to reach the
initial data for the new motion.

If the derivative of $V$ vanishes in one of the extremes or if one at
least of the two solutions $q_\pm(E)$ does not exist the motion is not
periodic and it may be unbounded: nevertheless it is still expressible
via integrals of the type \equ(5.2). If the potential $V$ is periodic
in $q$ and the variable $q$ is considered varying on a circle then
essentially {\it all solutions are periodic}: exceptions can occurr if
the energy $E$ has a value such that $V(q)=E$ admits a solution where
$V$ has zero derivative,

Typical examples are the {\it harmonic oscillator}, the {\it
pendulum}, the {\it Kepler oscillator}: whose Hamiltonians, if
$m,\o,g,h,G,k$ are positive constants, are respectively

$$\fra{p^2}{2m}+\fra12 m \o^2 q^2,\qquad
\fra{p^2}{2m}+m g(1-\cos \fra{q}h),\qquad
\fra{p^2}{2m}-m k \fra1{|q|}+ m \fra{ G^2}{2q^2}\Eq(5.4)$$
the latter has a potential which is singular at $q=0$ but if $G\ne0$
the energy conservation forbids too close an approach to $q=0$ and the
singularity becomes irrelevant.

The integral in \equ(5.3) is called a {\it quadrature} and the systems
in \equ(5.1) are therefore {\it integrable by quadratures}.  Such
systems, at least in the cases in which the motion is periodic, are
best described in new coordinates in which periodicity is more
manifest. Namely when $V(q)=E$ has only two roots $q_\pm(E)$ and $\mp
V'(q_\pm(E))>0$ the {\it energy--time} coordinates can be used by
replacing $q,\dot q$ or $p,q$ by $E,\t$ where $\t$ is the time needed
to the standard solution $t\to Q(t)$ to reach the given data,
\ie $Q(\t)=q,\dot Q(\t)=\dot q$. In such coordinates the motion is
simply $(E,\t)\to(E,\t+t)$ and, of course, the variable $\t$ has to be
regarded as varying on a circle of radius $T/2\p$.  The $E,\t$
variables are a kind of polar coordinates as can be checked by drawing
the curves of constant $E$, ``energy levels'', in the plane $p,q$ in
the cases in \equ(5.4), see Fig. 1.

In the harmonic oscillator case all trajectories are periodic.
In the pendulum case all motions are periodic except the ones which
separate the oscillatory motions (the closed curves in the second
drawing) from the rotatory motions (the apparently open curves) which,
in fact, are on closed curves as well if the $q$ coordinate, \ie the
vertical coordinate in Fig. 1, is regarded as ``periodic'' with period
$2\p h$.  In the Kepler case only the negative energy trajectories are
periodic and a few of them are drawn in Fig. 1. The single dots
represent the equilibrium points in phase space.

\eqfig{250pt}{105pt}
{}{fig1}{(1)}

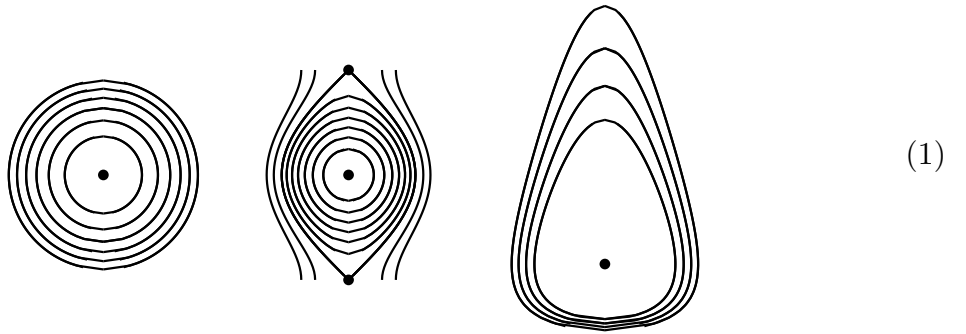
\0Fig. 1: {\nota The energy levels of the harmonic oscillator, the
pendulum, and the Kepler motion}

\0The region of phase space where motions are periodic is a set of
points $(p,q)$ with the topological structure of $\cup_{u\in
U}(\{u\}\times C_u)$ where $u$ is a coordinate varying in an open
interval $U$, for instance the set of values of the energy, and $C_u$
is a closed curve whose points $(p,q)$ are identified by a coordinate,
for instance by the time necessary to an arbitrarily fixed datum with
the same energy to evolve into $(p,q)$.

In the above cases, \equ(5.4), if the ``radial'' coordinate is chosen
to be the energy the set $U$ is the interval $(0,+\io)$ for the
harmonic oscillator, $(0,2 mg)$ or $(2mg,+\io)$ for the pendulum, and
$(-\fra12 \fra{mk^2 }{G^2},0)$ in the Kepler case.  The fixed datum
for the reference motion can be taken, in all cases, of the form
$(0,q_0)$ with the time coordinate $t_0$ given by \equ(5.3).

It is remarkable that the energy-time coordinates are canonical
coordinates: for instance in the vicinity of $(p_0,q_0)$ and if
$p_0>0$ this can be seen by setting

$$S(q,E)=\ig_{q_0}^q \sqrt{2 m(E-V(x))}\,dx\Eq(5.5)$$
and checking that $p=\dpr_q S(q,E), \,t=\dpr_E S(q,E)$ are identities
if $(p,q)$ and $(E,t)$ are coordinates for the same point so that the
criterion expressed by \equ(4.6) applies.

It is convenient to standardize the coordinates by replacing the time
variable by an angle$\a=\fra{2\p}{T(E)}t$; and instead of the energy
any invertible function of it can be used. 

It is natural to look for a coordinate $A=A(E)$ such that the map
$(p,q)\otto(A,\a)$ is a canonical map: this is easily done as the
function

$$\hat S(q,A)=\ig_{q_0}^q \sqrt{2 m(E(A)-V(x))}\,dx\Eq(5.6)$$
generates (locally) the correspondence between $p=\sqrt{2 m(E(A)-V(q))}$ and
$\a= E'(A) \ig_0^q$ $
\fra{dx}{\sqrt{2 m^{-1}(E(A)-V(x))}}$. Therefore, by the criterion
\equ(4.6),  if
$E'(A)=\fra{2\p}{T(E(A))}$, \ie if $A'(E)=\fra{T(E)}{2\p}$, the
coordinates $(A,\a)$ will be canonical coordinates. Hence, by
\equ(5.2),  $A(E)$  can be taken equal to

$$A=\fra1{2\p} 2\ig_{q_-(E)}^{q_+(E)}
\sqrt{2 m(E-V(q))}\,dq\=\fra1{2\p}\oint pdq\Eq(5.7)$$
where the last integral is extended to the closed curve of energy $E$,
see Fig. 1.  The {\it action--angle coordinates} $(A,\a)$ are defined
in open regions of phase space covered by periodic motions: in
action--angle coordinates such regions have the form $W=J\times \TTT$
of a product of an open interval $J$ and a one dimensional ``torus''
$\TTT=[0,2\p]$ (\ie a unit circle).  \*

\0{\it References:}  [LL68], [Ar68], [Ga83].
\*

\Section(6,Quasi periodicity and integrability)
\*

{\it A Hamiltonian is called {\it integrable} in an open region $W\subset
T^*(M)$ of phase space if

\0(1) there is an analytic and non singular (\ie with non zero Jacobian)
change of coordinates $(\V p,\V q)\otto(\V I,\Bf)$ mapping $W$ into a
set of the form $\II\times\TTT^\ell$ with $\II\subset \RRR^\ell$
(open) and {\it furthermore}

\0(2) the flow $t\to S_t(\V p,\V q)$ on phase space is
transformed into $(\V I,\Bf)\to (\V I,\Bf+\Bo(\V I)\,t)$ where $\Bo(\V
I)$ is a smooth function on $\II$.}
\*

This means that in suitable coordinates, that can be called 
``integrating coordinates'', the system appears as a set of
$\ell$ points with coordinates $\Bf=(\f_1,\ldots,\f_\ell)$ moving
on a unit circle at angular velocities $\Bo(\V I)=(\o_1(\V
I),\ldots,\o_{\ell}(\V I))$ depending on the actions of the initial
data.

A system integrable in a region $W$ which in integrating coordinates
$\V I,\Bf$ has the form $\II\times \TTT^\ell$ is said {\it
anisochronous} if $\det \dpr_{\V I}\Bo(\V I)\ne0$. It is said {\it
isochronous} if $\Bo(\V I)\=\Bo$ is {\it independent on $\V I$}.  The
motions of integrable systems are called {\it quasi periodic} with
frequency {\it spectrum} $\Bo(\V I)$, or with {\it frequencies}
$\Bo(\V I)/2\p$, in the coordinates $(\V I,\Bf)$.

Clearly an integrable system admits $\ell$ independent constants of
motion, the $\V I=(I_1,\ldots,I_\ell)$ and, for each choice of $\V I$,
the other coordinates vary on a ``standard'' $\ell$-dimensional torus
$\TTT^\ell$: hence it is possible to say that a phase space region of
integrability is {\it foliated into $\ell$-dimensional invariant tori}
$\TT(\V I)$ parameterized by the values of the constants of motion $\V
I\in \II$.

{\it If an integrable system is anisochronous then it is canonically
integrable}: \ie it is possible to define on $W$ a canonical change of
coordinates $(\V p,\V q)=\CC (\V A,\Ba)$ mapping $W$ onto
$J\times\TTT^{\ell}$ and such that $\HH(\CC(\V A,\Ba))=h(\V A)$ for a
suitable $h$: so that, if $\Bo(\V A)\defi \dpr_{\V A} h(\V A)$, the
equations of motion become

$$\dot{\V A}=\V0,\qquad\dot\Ba=\Bo(\V A)\Eq(6.1)$$
Given a system $(\V I,\Bf)$ of coordinates integrating an
anisochronous system the construction of action--angle coordinates can
be performed, {\it in principle}, via a classical procedure (under a few
extra assumptions).

Let $\g_1,\ldots,\g_\ell$ be $\ell$ {\it topologically independent}
circles on $\TTT^\ell$, for definiteness let $\g_i(\V
I)=\{\Bf\,|\,\f_1=\f_2=\ldots=\f_{i-1}=\f_{i+1}=\ldots=0,\
\f_i\in[0,2\p]\}$, and set

$$A_i(\V I)=\fra1{2\p}\oint_{\g_i(\V I)} \V p\cdot d\V q\Eq(6.2)$$
If the map $\V I\otto \V A(\V I)$ is analytically invertible as $\V
I=\V I(\V A)$ the function

$$S(\V A,\Bf)= (\l)\ig_{\V 0}^{\,\Bf} \V p\cdot d\V q \Eq(6.3)$$
is well defined if the integral is over any path $\l$ joining the
points $(p(\V I(\V A),\V0)$, $q(\V I(\V A),\V0))$ and $(p(\V I(\V A),$
$\Bf)),$ $q(\V I(\V A),\Bf)$ lying on the torus parameterized by $\V
I(\V A)$.

The key remark in the proof that \equ(6.3) really defines a function
of the only variables $\V A,\Bf$ is that {\it anisochrony implies the
vanishing of the Poisson brackets} (cf. \equ(4.4)): $\{I_i,I_j\}=0$
(hence also $\{ A_i,
A_j\}\=\sum_{h,k}\dpr_{I_k}\dpr_{I_h}\{I_k,I_h\}=0$).  And the
property $\{I_i,I_j\}=0$ can be checked to be precisely the
integrability condition for the differential form $\V p\cdot d\V q$
restricted to the surface obtained by varying $\V q$ while $\V p$ is
constrained so that $(\V p,\V q)$ stays on the surface $\V
I$=constant, \ie on the invariant torus of the points with fixed $\V
I$.

The latter property is necessary and sufficient in order that the
function $S(\V A,\Bf)$ be well defined (\ie be independent on the
integration path $\l$) up to an additive quantity of the form $\sum_i
2\p n_iA_i$ with $\V n=(n_1,\ldots,n_\ell)$ integers..

Then the action--angle variables are defined by the canonical change of
coordinates with $S(\V A,\Bf)$ as generating function, \ie by setting

$$\a_i=\dpr_{A_i} S(\V A,\Bf), \qquad I_i=\dpr_{\f_i} S(\V A,\Bf).\Eq(6.4)$$
and, since the computation of $S(\V A,\Bf)$ is ``reduced to
integrations'' which can be regarded as a natural extension of the
quadratures discussed in the one dimensional cases, also such systems
are called {\it integrable by quadratures}. The just described
construction is a version of the more general {\it Arnold-Liouville
theorem}.

In practice, however, the actual evaluation of the integrals in
\equ(6.2),\equ(6.3) can be difficult: its analysis in various cases
(even as ``elementary'' as the pendulum) has in fact led to key
progress in various domains, \eg in the theory of special functions
and in group theory.

In general any surface on phase space on which the restriction of the
differential form $\V p\cdot d\V q$ is locally integrable is called a
{\it Lagrangian manifold}: hence the invariant tori of an
anisochronous integrable system are Lagrangian manifolds.

If an integrable system is anisochronous it cannot admit more than
$\ell$ independent constants of motion; furthermore it does not admit
invariant tori of dimension $> \ell$. Hence $\ell$-dimensional
invariant tori are called {\it maximal}.

Of course invariant tori of dimension $<\ell$ can also exist: this
happens when the variables $\V I$ are such that the frequencies
$\Bo(\V I)$ admit nontrivial rational relations; \ie there is an
integer components vector $\Bn\in\ZZZ^\ell,
\,\Bn=(\n_1,\ldots,\n_\ell)\ne0$ such that

$$\Bo(\V I)\cdot\Bn=\sum_i \o_i(\V I)\,\n_i=\V0\Eq(6.5)$$
in this case the invariant torus $\TT(\V I)$ is called {\it resonant}.
If the system is anisochronous then $\det\dpr_{\V
I}\Bo(\V I)\ne0$ and, therefore, the resonant tori are associated with
values of the constants of motion $\V I$ which {\it form a set of
$0$-measure} in the space $\II$ but {\it which is not empty and dense}.

Examples of isochronous systems are the systems of harmonic
oscillators, \ie systems with Hamiltonian $\sum_{i=1}^\ell \fra1{2m_i}
p_i^2+ \fra12 \sum_{i,j}^{1,\ell} c_{ij}q_i q_j$ where the matrix $v$
is a positive definite matrix. This is an isochronous system with
frequencies $\Bo=(\o_1,\ldots,\o_\ell)$ whose squares are the
eigenvalues of the matrix $m_i^{-\fra12}c_{ij}m_j^{-\fra12}$. It is
integrable in the region $W$ of the data $\V x=(\V p,\V q)\in
\RRR^{2\ell}$ such that, setting $A_\b= \fra 1{2 \,
\o_\b}\, \big((\sum_{i=1}^\ell \fra{v_{\b,i} p_i}{\sqrt{m_i}})^2+
\o_\b^2 (\sum_{i=1}^\ell \fra{v_{\b,i}
q_i}{\sqrt{{m_i}^{-1}}})^2\big)$ for all eigenvectors $\V v_\b$,
$\b=1,\ldots,\ell$, of the above matrix, the vectors $\V A$ have all
components $>0$ .

Even though this system is isochronous it, nevertheless, admits a
system of canonical action--angle coordinates in which the Hamiltonian
takes the simplest form

$$h(\V A)=\sum_{\b=1}^\ell \o_\b A_\b\=\,\Bo\cdot\V A\Eq(6.6)$$
with $\a_\b=-\atan \fra{\sum_{i=1}^\ell \fra{v_{\b,i}
p_i}{\sqrt{m_i}}}{ \sum_{i=1}^\ell \sqrt{m_i}\o_\b{v_{\b,i}
q_i}} $ as conjugate angles.

An example of anisochronous system is the {\it free rotators} or {\it
  free wheels}: \ie $\ell$ noninteracting points on a circle of radius
  $R$ or $\ell$ noninteracting coaxial wheels of radius $R$. If
  $J_i=m_i R^2$ or, respectively, $J_i=\fra12 m_i R^2$ are the inertia
  moments and if the positions are determined by $\ell$ angles
  $\Ba=(\a_1,\ldots,\a_\ell)$ the angular velocities are constants
  related to the angular momenta $\V A=(A_1,\ldots,A_\ell)$ by
  $\o_i=A_i/J_i$. The Hamiltonian and the specrum are

$$h(\V A)=\sum_{i=1}^\ell\fra1{2J_i} A_i^2,\qquad \Bo(\V
  A)=\Big(\fra1{J_i} A_i\Big)_{i=1,\ldots,\ell}\Eq(6.7)$$

\0{\it References:}  [LL68], [Ar68], [Ga83], [Fa98].
\*

\Section(7,Multidimensional quadratures: central motion)
\*

Important mechanical systems with more than one degree of freedom are
integrable by canonical quadratures in vast regions of phase
space. This is checked by showing that there is a foliation into
invariant tori $\TT(\V I)$ of dimension equal to the number $\ell$ of
degrees of freedom parameterized by $\ell$ constants of motion $\V I$
{\it in involution}, \ie such that $\{I_i,I_j\}=0$. One then performs,
if possible, the construction of the action--angle variables by the
quadratures discussed in the previous section.

The above procedure is well exemplified by the theory of the planar
motion of a unit mass attracted by a coplanar center of force: the Lagrangian
is, in polar coordinates $(\r,\th)$, $\LL=\fra{m}{2}
(\dot\r^2+\r^2\dot\th^2)-V(\r)$. The planarity of the motion is not a
strong resttriction as central motion takes always place on a plane

Hence the equations of motion are
$\fra{d}{dt} m\r^2\dot\th=0$, \ie $m\r^2\dot\th=G$ is a constant of
motion (it is the angular momentum), and
$\ddot\r=-\dpr_\r V(\r)+\dpr_\r \fra{m}2\r^2\dot
\th^2=-\dpr_\r V(\r)+ \fra{G^2}{m\,\r^3}\defi-\dpr_\r V_G(\r)$. So that
energy conservation yields a second constant of motion $E$

$$\fra{m}2\dot\r^2 +\fra12 \fra{G^2}{m\r^2}+V(\r)=E=
\fra1{2m} p_\r^2+\fra1{2m}\fra{p_\th^2}{\r^2}+V(\r)\Eq(7.1)$$
The \rhs is the Hamiltonian for the system, derived from $\LL$, if
$p_\r,p_\th$ denote conjugate momenta of $\r,\th$: $p_\r=m\dot\r$ and
$p_\th=m\r^2\dot \th$ (note that $p_\th=G$).

Suppose $\r^2 V(\r)\tende{\r\to0}0$: then the singularity at the origin
cannot be reach\-ed by any motion starting with $\r>0$ if $G>0$. Assume
also that the function $V_G(\r)\defi
\fra12 \fra{G^2}{m\r^2}+V(\r)$ has only one minimum $E_0(G)$, no
maximum and no horizontal inflection and tends to a limit $E_\io(G)\le
\io$ when $\r\to\io$. Then the system is integrable in the domain
$W=\{(\V p,\V q)\,|\, E_0(G)<E<E_\io(G),\, G\ne0\}$.

This is checked by introducing a ``standard'' periodic solution $t\to
R(t)$ of $m\ddot\r=-\dpr_\r V_G(\r)$ with energy $E_0(G)<E<E_\io(G)$
and initial data $\r=\r_{E,-}(G)$, $\dot\r=0$ at time $t=0$, where
$\r_{E,\pm}(G)$ are the two solutions of $V_G(\r)=E$, see Section
\sec(5): this is a periodic analytic function of $t$ with period
$T(E,G)=2\ig_{\r_{E,-}(G)}^{\r_{E,+}(G)}\fra{d
x}{\sqrt{\fra2m(E-V_G(x))}}$.

The function $R(t)$ is given, for $0\le t\le \fra12T(E,G)$ or for $
\fra12T(E,G)\le t\le T(E,G)$, by the quadratures

$$\txt t=\ig_{\r_{E,-}(G)}^{R(t)}\fra{d
x}{\sqrt{\fra2m(E-V_G(x))}},\quad{\rm or}\quad
t=\fra{T(E,G)}2 -\ig_{\r_{E,+}(G)}^{R(t)}\fra{d
x}{\sqrt{\fra2m(E-V_G(x))}} \Eq(7.2)$$
respectively. The analytic regularity of $R(t)$ follows from
the general existence, uniqueness and regularity theorems applied to
the differential equation for $\ddot\r$.

Given an initial datum $\dot\r_0,\r_0,\dot\th_0,\th_0$ with energy $E$
and angular momentum $G$ define $t_0$ to be the time such that
$R(t_0)=\r_0, \dot R(t_0)=\dot\r_0$: then $\r(t)\= R(t+t_0)$ and $\th(t)$
can be computed as $\th(t)=\th_0+\ig_0^t \fra{G}{m R(t'+t_0)^2} dt'$: a
second quadrature.  Therefore we can use as coordinates for the motion
$E,G,t_0$, which determine $\dot\r_0,\r_0,\dot\th_0$ and a fourth
coordinate that determines $\th_0$ which could be $\th_0$ itself but
which is conveniently determined, via the second quadrature, as follows.

The function $G\, m^{-1}\,R(t)^{-2}$ is periodic with period $T(E,G)$,
hence it can be expressed in a Fourier series $\ch_0(E,G)+\sum_{k\ne0}
\ch_k(E,G) e^{\fra{2\p}{T(E,G)}\,i\, t\,k}$; the quadrature
for $\th(t)$ can be performed by integrating the
series terms. Setting $\lis\th(t_0)\defi
\fra{T(E,G)}{2\p}\sum_{k\ne0}\fra{\ch_k(E,G)}k
e^{\fra{2\p}{T(E,G)}\,i\,t_0\,k}$ and 
$\f_1(0)=\th_0-\lis\th(t_0)$ the $\th(t)=\th_0+\ig_0^t
\fra{G}{m R(t'+t_0)^2} dt'$ becomes

$$\f_1(t)=\f_1(0)+\ch_0(E,G)\,t\Eq(7.3)$$
Hence the system is integrable and the spectrum is
$\Bo(E,G)=(\o_0(E,G)$, $\o_1(E,G))$ $\=(\o_0,\o_1)$ with $\o_0\defi
\fra{2\p}{T(E,G)}$ and $\o_1\defi\ch_0(E,G)$, while $\V I=(E,G)$ are
constants of motion and the angles $\Bf=(\f_0,\f_1)$
can be taken $\f_0\defi \o_0t_0,\f_1\defi\th_0-\lis\th(t_0)$. 
At $E,G$ fixed the motion takes place
on a $2$-dimensional torus $\TT(E,G)$ with $\f_0,\f_1$ as angles.

In the aniso\-chronous cases, \ie when $\det\dpr_{E,G}\Bo(E,G)\ne0$,
canonical action--angle variables conjugated to $(p_\r,\r,p_\th,\th)$
can be constructed via \equ(6.2),\equ(6.3) by using two cycles
$\g_1,\g_2$ on the torus $\TT(E,G)$. It is convenient to choose 
\vskip1mm

\0(1) $\g_1$ as the cycle consisting of the points $\r=x,\th=0$ whose
first half (where $p_\r\ge0$) consists in the set $\r_{E,-}(G)\le x\le
\r_{E,+}(G)$, $p_\r=\sqrt{{2m}(E-V_G(x))}$ and $d\th=0$;

\0(2) $\g_2$ as the cycle $\r=const,\th\in[0,2\p]$ on which $d\r=0$ and
$p_\th=G$ obtaining

$$A_1=\fra2{2\p}\ig_{\r_{E,-}(G)}^{\r_{E,+}(G)}\sqrt{2m
(E-V_G(x))}dx,\qquad A_2=G\Eq(7.4)$$
\vskip-1mm

According to the general theory (cf. Sect. \sec(6)) a generating
function for the canonical change of coordinates from
$(p_\r,\r,p_\th,\th)$ to action--angle variables is (if, to fix
ideas, $p_\r>0$)

\vskip-1mm$$S(A_1,A_2,\r,\th)=G\th+\ig_{\r_{E,-}}^\r
\sqrt{2m(E-V_G(x))}dx\Eq(7.5)$$
In terms of the above $\o_0,\ch_0$  the Jacobian matrix
$\fra{\dpr(E,G)}{\dpr(A_1,A_2)}$ is computed from \equ(7.4),\equ(7.5)
to be $\pmatrix{\o_0&\ch_0\cr0&1\cr}$. It follows
$\dpr_E S=t, \,\dpr_G S=\th-\lis\th(t)-\ch_0t$ so that, see \equ(6.4),

\kern-1mm
$$\a_1\defi\dpr_{A_1} S=\o_0 t, \qquad\a_2=\defi\dpr_{A_2}
S=\th-\lis\th(t)
\Eq(7.6)$$
and $(A_1,\a_1),(A_2,\a_2)$ are the action--angle pairs.
\*

\0{\it References:}  [LL68], [Ga83].

\*
\Section(8, Newtonian potential and Kepler's laws)

\*
The anisochrony property, \ie
$\det\fra{\dpr(\o_0,\ch_0)}{\dpr(A_1,A_2)}\ne0$ or, equivalently,
$\det\fra{\dpr(\o_0,\ch_0)}{\dpr(E,G)}$ $\ne0$, {\it is not satisfied}
in the important cases of the harmonic potential and of the Newtonian
potential. Anisochrony being only a sufficient condition for canonical
integrability it is still possible (and true) that, nevertheless, in
both cases the canonical transformation generated by \equ(7.5)
integrates the system. This is expected since the two potentials are
limiting cases of anisochronous ones (\eg $|\V q|^{2+\e}$ and
$|q|^{-1-\e}$ with $\e\to0$).

The Newtonian potential $\HH(\V p,\V q)=\fra1{2m}\V p^2-\fra{k\,m}{|\V
q|}$ is integrable in the region $G\ne0, \,
E_0(G)=-\fra{k^2m^3}{2G^2}<E<0,
|G|<\sqrt{\fra{k^2m^3}{-2E}}$. Proceeding as in Section \sec(7) one
finds integrating coordinates and that the integrable motions develop
on ellipses with one focus on the center of attraction $S$ so that
motions are periodic hence not anisochronous: nevertheless the
construction of the canonical coordinates via
\equ(6.2),\equ(6.3),\equ(6.4) (hence \equ(7.5)), works and leads to
canonical coordinates $(L',\l',G',\g')$. To obtain action--angle
variables with a simple interpretation it is convenient to perform on
the variables $(L',\l',G',\g')$ (constructed by following the
procedure just indicated) a further trivial canonical tranformation by
setting $L=L'+G',\, G=G',\l=\l',\g=\g'-\l'$; then 
\vskip1mm

\halign{#\hfill&\ #\hfill\cr
$\l$& {\it average anomaly}, is the time necessary to the point $P$ to move
from the pericenter \cr& to its actual position in units of the period and
times $2\p$\cr
$L$& {\it action}, is essentially the energy
$E=-\fra{k^2m^3}{2L^2}$,\cr
$G$ & {\it angular momentum}, \cr
$\g$& {\it axis longitude}, is the angle between a fixed axis and
the major axis of the ellipse \cr& oriented from the center
of the ellipse $O$
to the center of attraction $S$.
\cr}

\kern5mm

\eqfig{330pt}{70pt}{}{fig2}{(2)}

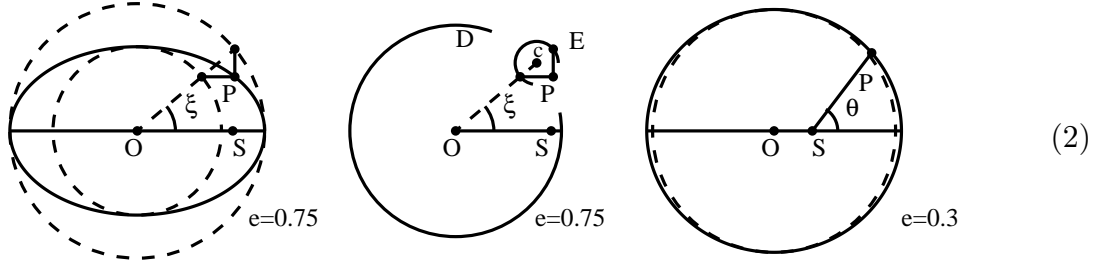
\0Fig. 2: {\nota Eccentric and true anomalies of $P$ which
moves on a small circle $E$ centered at a point $c$ moving on the
circle $D$ located half way between the two concentric circles
containing the Keplerian ellipse: the anomaly of $c$ with respect to
the axis $OS$ is $\x$. The circle $D$ is {\it eccentric} with respect
to $S$ and therefore $\x$ is, still today, called eccentric anomaly
while the circle $D$ is, in ancient terminology, the {\it deferent
circle} (eccentric circles were introduced in Astronomy by
Ptolemy). The small circle $E$ on which the point $P$ moves is, in
ancient terminology, an {\it epicycle}. The deferent and the
epicyclical motions are synchronous (\ie have the same period); Kepler
discovered that his key
\ap hypothesis of inverse proportionality between angular velocity on
the deferent and distance between $P$ and $S$ (\ie
$\r\dot\x=constant$) implied both synchrony and elliptical shape of
the orbit, with focus in $S$.  The latter law is equivalent to
$\r^2\dot\th=constant$ (because of the identity $a\dot\x=\r\dot\th$).
Small
eccentricity ellipses can be hardly distinguished from circles. }

The {\it eccentricity} of the ellipse is $e$ such that $G=\pm L
\sqrt{1-e^2}$. The ellipse equation is $\r= a\,(1-e \cos\x)$ where $\x$
is the {\it eccentric anomaly} (see Fig. 2), $a=\fra{L^2}{k m^2}$ is the major
semiaxis and $\r$ is the distance to the center of attraction S.

Finally the relations between
eccentric anoma\-ly $\x$, average anomaly $\l$, {\it true anomaly} $\th$
(the latter is the polar angle), and SP--distance $\r$ are
given by Kepler's equations

\kern-2truemm
$$\eqalign{
\l=&\x-e\sin \x,\qquad
(1-e\cos\x)(1+e\cos\th)=1-e^2,\cr
\l=&(1-e^2)^{\fra32}\ig_0^\th
\fra{d\th'}{(1+e\cos\th')^2},\qquad
\fra\r{a}=\fra{1-e^2}{1+e\cos\th}\cr}
\Eq(8.1)$$
and this relation between true anomaly and average anomaly can
be inverted in the form

$$\x=\l + g_\l, \qquad \th=\l+f_\l\ \tto\
\fra{\r}{a}=\fra{1-e^2}{1+e\cos(\l+f_\l)}
\Eq(8.2)$$
where $g_\l=g(e\sin\l,e\cos\l),\, f_\l=f(e\sin\l,e\cos\l)$ and
$g(x,y), f(x,y)$ are
suitable functions analytic for $|x|,|y|$ $<1$. Furthermore
$g(x,y)=x\,(1+y+\ldots), f(x,y)=2x\, (1+\fra54 y+\ldots)$ and
$\ldots$ denote terms of degree $2$ or higher in $x,y$,
containing only even powers of $x$.
\*
\0{\it References:}  [LL68], [Ga83].
\*

%\pagina
\Section(9,Rigid body)
\*

Another fundamental integrable system is the rigid body in absence of
gravity and with a fixed point $O$. It can be naturally described in terms
of the {\it Euler angles} $\tb,\fb,\psb$, see Fig. 3, and their
derivatives $\dot\tb,\dot\fb,\dot\psb$.

\eqfig{140pt}{105pt}
{}
{fig3}{(3)}

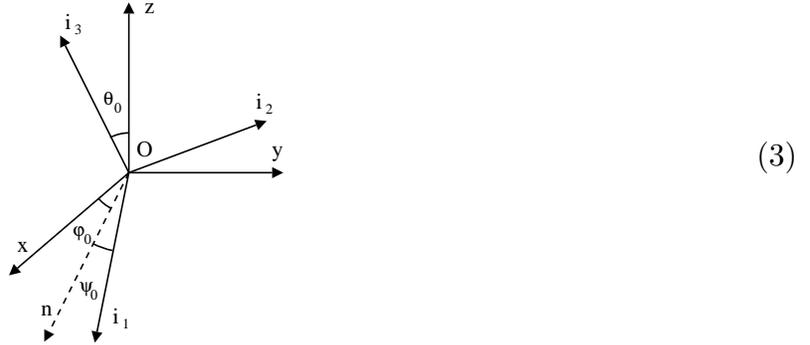
\0Fig. 3: {\nota The Euler angles of the comoving frame $\V i_1,\V i_2,
\V i_3$ with respect to a fixed frame $\V x,\V y, \V z$. The direction
$\V n$ is the ``node line, intersection between the planes ${\bf x,y}$
and ${\bf i_1, i_2}$.}

Let $I_1,I_2,I_3$ be the three {\it principal inertia moments} of the
body along the three {\it principal axes} with unit vectors $\V i_1,\V
i_2,\V i_3$. The inertia moments and the principal axes are the
eigenvalues and the associated unit eigenvectors of the $3\times3$
{\it inertia matrix} $\II$ which is defined by $\II_{hk}=\sum_{i=1}^n
m_i {({\V x_i})}_{h} {({\V x}_i)}_{k}$, where 
$h,k=1,2,3$ and $\V x_i$ is the position of
the $i$-th particle in a reference frame with origin at $O$ and in
which all particles are at rest: this {\it comoving frame} exists as a
consequence of the rigidity constraint. The principal axes form a
coordinate system which is comoving as well: \ie also in the frame
$(O; \V i_1,\V i_2,\V i_3)$ the particles are at rest.

The Lagrangian is simply the kinetic energy: \ie we imagine the
rigidity constraint to be ideal, \eg realized by internal central
forces in the limit of infinite rigidity, as mentioned in
Sec. \sec(3). The {\it angular velocity} of the rigid motion is
defined by

$$\Bo=\dot \tb\V n+\dot\fb \V z+\dot\psb \V i_3\Eq(9.1)$$
expressing that a generic infinitesimal motion must consist of a
variation of the three Euler angles and therefore it has to be a
rotation of speeds $\dot \tb,\dot\fb,\dot\psb$ around the axes $\V n,\V
z,\V i_3$ as shown in Fig. 3.

Let $(\o_1,\o_2,\o_3)$ be the components of $\Bo$ along the pricipal
axes $\V i_1,\V i_2,\V i_3$: for brevity the latter axes will often
be called $\V 1,\V 2,\V 3$. Then the angular momentum $\V M$, with
respect to the pivot point $O$, and the kinetic energy $K$ can be
checked to be
$$\V M=I_1\o_1\V i_1+I_2\o_2\V i_2+I_3\o_3\V i_3,\qquad
K=\fra12(I_1\o_1^2+I_2\o_2^2+I_2\o_3^2)\Eq(9.2)$$
and are constants of motion. From Fig.3 it follows that
$\o_1=\dot\tb\cos\psb+\dot\fb\sin\tb\sin\psb$,
$\o_2=-\dot\tb$ $\sin\psb+\dot\fb\sin\tb\cos\psb$ and
$\o_3=\dot\fb\cos\tb+\dot\psb$ so that the Lagrangian, uninspiring at
first, is

$$\eqalign{
\LL\defi&\fra12I_1(\dot\tb\cos\psb+\dot\fb\sin\tb\sin\psb)^2
+\cr&+\fra12I_2(-\dot\tb\sin\psb+\dot\fb\sin\tb\cos\psb)^2+
\fra12I_3(\dot\fb\cos\tb+\dot\psb)^2\cr}\Eq(9.3)$$

Angular momentum conservation does not imply that the components
$\o_j$ are constants because also $\V i_1,\V i_2,\V i_3$ change with
time according to $\fra{d}{dt}\V i_j=\Bo\wedge \V
i_j,\,j=1,2,3$. Hence $\dot {\V M}=\V0$ becomes, by the first of
\equ(9.2) and denoting $I\Bo=(I_1\o_1,I_2\o_2,I_3\o_3)$,
the {\it Euler equations} $I\dot\Bo+\Bo\wedge I\Bo=\V0$, or

$$I_1\dot\o_1=(I_2-I_3)\o_2\o_3,\ \
I_2\dot\o_2=(I_3-I_1)\o_3\o_1.\ \
I_3\dot\o_3=(I_1-I_2)\o_1\o_2\Eq(9.4)$$
which can be considered together with the conserved quantities, \equ(9.2).

\eqfig{125pt}{110pt}{}{fig4}{(4)}

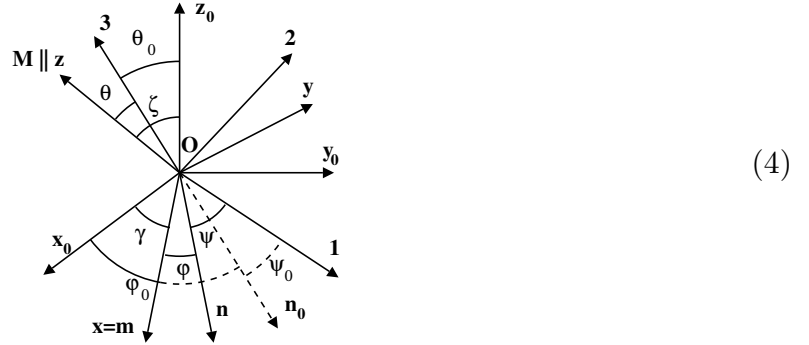
\0Fig. 4: {\nota The laboratory frame, the angular momentum frame, the
comoving frame (and the Deprit angles).}

Since angular momentum is conserved it is convenient to introduce the
{\it laboratory frame} $(O;\V x_0,\V y_0,\V z_0)$ with fixed axes
$\V x_0,\V y_0,\V z_0$ and (see Fig. 4):

\0(1) $(O;{\bf x,y,z})$: the {\it momentum frame} with fixed axes, but
    with ${\bf z}$-axis oriented as $\V M$, and ${\bf x}$-axis coinciding
    with the node (\ie the intersection) of the plane ${\bf x_0y_0}$
    and the plane ${\bf xy}$ (orthogonal to $\V M$).  Therefore ${\bf
    x,y,z}$ is determined by the {\it two} Euler angles $\z,\g$ of
    $(O;{\bf x,y,z})$ in $(O;{\bf x_0,y_0,z_0})$,

\0(2) $(O;{\bf 1,2,3})$: the {\it comoving frame}, \ie the frame fixed with the
body, and with unit vectors ${\V i_1,\V i_2,\V i_3}$ parallel to the
principal axes of the body. The frame is determined by three Euler
angles $\th_0,\f_0,\ps_0$

\0(3) The Euler angles of $(O;{\bf 1,2,3})$ with respect to $(O;{\bf
    x,y,z})$ which are denoted $\th,\f,\ps$

\0(4) $G$: the total angular momentum: $G^2=\sum_j I_j^2 \o_j^2$

\0(5) $M_3$: the angular momentum along the ${\bf z}_0$
    axis; $M_3=G \cos\z$

\0(6) $L$: the projection of $\V M$ on the axis ${\bf3}$, $L=G \cos \th$

\*
The quantities $G,M_3,L, \f,\g,\ps$ determine $\th_0,\f_0,\ps_0$ and
$\dot\th_0,\dot\f_0,\dot \ps_0$, or the $p_{\th_0},p_{\f_0},p_{\ps_0}$
variables conjugated to $\th_0,\f_0,\ps_0$ as shown by the following comment.

Considering Fig. 4, the angles $\z,\g$ determine location, in the
fixed frame $(O;{\bf x_0,\V y_0,\V z_0})$ of the direction of $\V M$
and the node line $\V m$, which are respectively the $\V z$-axis and
the $\V x$ axis of the fixed frame associated with the angular
momentum; the angles $\th,\f,\ps$ {\it then}  determine the position of the
comoving frame with respect to the fixed frame $(O;{\bf \V x,\V y,\V
z})$, hence its position with respect to $(O;{\bf x_0,y_0,z_0})$, \ie
$(\th_0,\f_0,\ps_0)$. From this and $G$ it is possible to determine
$\Bo$ because

$$\cos\th=\fra{I_3\o_3}{G},\ \tan\ps=\fra{I_2\o_2}{I_1\o_1},\quad
\o_2^2=I_2^{-2}(G^2-I_1^2\o_1^2-I_3^2\o_3^2)
%\dot\f=G\fra{I_1\o^2_1+I_2 \o_2^2}{I_1^2\o^2_1+I_2^2 \o_2^2}
\Eq(9.5)$$
and, from \equ(9.1), $\dot\th_0,\dot\f_0,\dot\ps_0$ are determined.

The Lagrangian \equ(9.3) gives immediately (after expressing $\Bo$,
\ie ${\bf n,\bf z,\bf i_3}$, in terms of the Euler angles
$\th_0,\f_0,\ps_0$) an expression for the variables
$p_{\th_0},p_{\f_0}$, $p_{\ps_0}$ conjugated to $\th_0,\f_0,\ps_0$

$$
p_{\th_0}={\bf M}\cdot{{\bf n_0}}\qquad
p_{\f_0}={\bf M}\cdot{\bf z_0}\qquad
p_{\ps_0}={\bf M}\cdot{\bf i_3}\Eq(9.6)$$
and in principle we could proceed to compute the Hamiltonian.

However
the computation {\it can be avoided} because of the very remarkable
property ({\cs Deprit}), which can be checked with some patience,
making use of \equ(9.6) and of elementary spherical trigonometry
identities,

$$M_3 d\g+Gd\f+L d\psi=p_{\f_0}d\f_0+p_{\ps_0}d\ps_0+p_{\th_0}d\th_0
\Eq(9.7)$$
which means that {\it the map $\big((M_3,\g),(L,\ps),(G,\f)\big)\otto
\big((p_{\th_0},\th_0),(p_{\f_0},\f_0),(p_{\ps_0},\ps_0)\big)$
is a canonical map}. And in the new coordinates the kinetic energy,
hence the Hamiltonian, takes the form

$$K={1\over 2}\Big[{L^2\over I_3}+(G^2-L^2)\Big(
{\sin^2\psi\over I_1}+{\cos^2\psi\over I_2}\Big)\Big]\Eq(9.8)$$
This again shows that $G,M_3$ are constants of motion, and the
$L,\ps$ variables are determined by a quadrature, because
the Hamilton equation for $\ps$ combined with the energy conservation
yields

$$\txt\dot\ps =
\pm (\fra1{I_3}-\fra{\sin^2\ps}{I_1}-\fra{\cos^2\ps}{I_2})\,
%L_E(\ps)\crL_E(\ps)&\defi
\sqrt{\fra{2E-G^2(\fra{\sin^2\ps}{I_1}+
\fra{\cos^2\ps}{I_2})}{\fra1{I_3}-\fra{\sin^2\ps}{I_1}-
\fra{\cos^2\ps}{I_2}}}
\Eq(9.9)$$
In the integrability region this motion is periodic with some period
$T_L(E,G)$. Once $\ps(t)$ is determined the Hamilton equation for $\f$
leads to the further quadrature

$$\dot \f=\big(\fra{\sin^2\ps(t)}{I_1}+\fra{\cos^2\ps(t)}{I_2}\big)\,G\Eq(9.10)$$
which determines a second periodic motion with period $T_G(E,G)$. The
$\g,M_3$ are constants and, therefore, the motion takes place on three
dimensional invariant tori $\TT_{E,G,M_3}$ in phase space each of
which is {\it always} foliated into two dimensional invariant tori
parameterized by the angle $\g$ which is constant by \equ(9.8)
(because $K$ is $M_3$-independent): the latter are in turn foliated by
one dimensional invariant tori, \ie by periodic orbits, with $E,G$
such that the value of $T_L(E,G)/T_G(E,G)$ is rational.

Note that if $I_1=I_2=I$ the above analysis is extremely
simplified. Furthermore if gravity $g$ acts on the system the Hamiltonian will
simply change by the addition of a potential $-m gz$ if $z$ is the
height of the center of mass. Then, see Fig. 4, if
the center of mass of the body is on the
axis $\V i_3$ and $z= h \cos\th_0$ and 
$h$ is the distance of the center of mass from $O$, since
$\cos\th_0=\cos\th\cos\z-\sin\th\sin\z\cos\f$, the Hamiltonian will
become $\HH=K-mgh\cos\th_0$ or

$$\txt\HH={G^2\over2I_3}+{G^2-L^2\over2I}
-m\,g\,h\,\left({M_3L\over G^2}-
\bigl(1-{M_3^2\over G^2}\bigr)^{1/2}\bigl(1-{L^2\over G^2}\bigr)^{1/2}
\cos\f\right)\Eq(9.11)$$
so that, again, the system is integrable by quadratures (with the
roles of $\ps$ and $\f$ ``interchanged'' with respect to the
previous case) in suitable regions of phase space. This is called the
{\it Lagrange's gyroscope}.

A less elementary integrable case is when the inertia moments are
related as $I_1=I_2=2I_3$ and the center of mass is in the plane $\V
i_1,\V i_2$ (rather than on the $\V i_3$ axis) and only gravity acts,
besides the constraint force on the pivot point $O$; this is called
the {\it Kowalevskaia's gyroscope}.
\*
\0{\it References:}  [Ga83].\*

\Section(10, Other quadratures)
\*

An interesting classical integrable motion is that of a point mass
attracted by two equal mass centers of gravitational attraction. Or a
point ideally constrained to move on the surface of a general ellipsoid.

New integrable systems have been discovered quite recently and have
generated a wealth of new developments ranging from group theory (as
integrable systems are closely related to symmetries) to partial
differential equations.

It is convenient to extend the notion of integrability by saying that a
system is integrable in a region $W$ of phase space if
\*
{\it
\0(1) there is a
change of coordinates $(\V p,\V q)\in W\otto \{\V A,\Ba,\V Y,\V y\}
\in (U\times\TTT^\ell)\times(V\times\RRR^m)$ where $U\subset
\RRR^\ell,\,V\subset \RRR^m$, with $\ell+m\ge1$,
are open sets and
\\
(2) the $\V A,\V Y$ are constants of motion while the
other coordinates vary ``linearly'':

$$(\Ba,\V y)\to (\Ba+\Bo(\V A,\V Y)t,\,\V y+\V v(\V A,\V Y) t)\Eq(10.1)$$
where $\Bo(\V A,\V Y),\V v(\V A,\V Y)$ are smooth functions.}
\*

In the new sense the systems studied in the previous sections are
integrable in much wider regions (essentially on the entire phase
space with the exception of a set of data which lie on lower
dimensional surfaces forming sets of zero volume). The notion is
convenient also because it allows us to say that even the systems of
free particles are integrable.

Two very remarkable systems integrable in the new sense are the
Hamiltonian systems, respectively called {\it Toda lattice} ({\cs
Kruskal, Zabusky}), and {\it Calogero lattice}, ({\cs Calogero,
Moser}); if $(p_i,q_i)\in \RRR^2$ they are

$$\eqalign{
\HH_T(\V p,\V q)=&\fra1{2m}\sum_{i=1}^n p_i^2+\sum_{i=1}^{n-1}
g \,e^{-\k(q_{i+1}-q_i)}\cr
\HH_C(\V p,\V q)=&\fra1{2m}\sum_{i=1}^n p_i^2+\sum_{i<j}^{n}
\fra{g}{(q_i-q_j)^2} +\fra12\sum_{i=1}^n m\o^2  q_i^2\cr}\Eq(10.2)$$
where $m>0$ and $\k,\o,g\ge 0$. They describe the motion of $n$
interacting particles on a line.

The integration method for the above systems is again to  find first
the constants of motion and later to look for quadratures, when
appropriate. The constants of motion can be found with the
method of the {\it Lax pairs}. One shows that there is a pair of
self adjoint  $n\times n$ matrices
$M(\V p,\V q), N(\V p,\V q)$ such that the equations of motion become

$$\txt\fra{d}{dt} M(\V p,\V q)\,
= \,i\, \big[M(\V p,\V q), N(\V p,\V q)\big],\qquad i=\sqrt{-1}\Eq(10.3)$$
which imply that $M(t)=U(t) M(0)U(t)^{-1}$, with $U(t)$ a unitary
matrix. When the equations can be written in the above form it is
clear that the $n$ eigenvalues of the matrix $M(0)=M(\V p_0,\V q_0)$
are constants of motion. When appropriate, \eg in the Calogero lattice
case with $\o>0$, it is possible to proceed to find canonical action--angle
coordinates: a task that is quite difficult due to the arbitrariness
of $n$, but which is possible.

The Lax pairs for the Calogero lattice (with $\o=0,g=m=1$) are

$$M_{hh}= p_h,\ \ N_{hh}=0, \ {\rm and}\
M_{hk}=\fra{i}{(q_h-q_k)},\ N_{hk}= \fra{1}{(q_h-q_k)^2}
\ h\ne k\Eq(10.4)$$
while for the Toda lattice (with $m=g=\fra12\k=1$) 
the non zero matrix elements of $M,N$
are

$$\eqalign{
M_{hh}&= p_h,\qquad
M_{h,h+1}=M_{h+1,h}=e^{-(q_h-q_{h+1})},\qquad\cr
N_{h,h+1}&= -N_{h+1,h}=\,i\,e^{-(q_h-q_{h+1})}\cr}
\Eq(10.5)$$
which are checked by first trying the case $n=2$.

Another integrable system ({\cs Sutherland}) is

$$\HH_S(\V p,\V q)=\fra1{2m}\sum_{i=k}^n p_k^2+\sum_{h<k}^{n}
\fra{g}{\sinh^2(q_h-q_k)} \Eq(10.6)$$
whose Lax pair is related to that of the Calogero lattice.

By taking suitable limits as $n\to\io$ and as the other parameters tend
to $0$ or $\io$ at suitable rates integrability of a
few differential equations, among which the {\it Korteweg-deVries equation}
or the {\it nonlinear Sch\"odinger equation}, can be derived.
\*

As mentioned in Section \sec(1) symmetry properties under continuous groups
imply existence of constants of motion. Hence it is natural to
think that integrability of a mechanical system reflects 
enough symmetry to imply the existence of as many constants of
motion, independent and in involution, as the number $n$ of degrees of
freedom.

This is in fact {\it always true}, and in some respects it is a
tautological statement in the anisochronous cases. Integrability in a
region $W$ implies existence of canonical action--angle coordinates
$(\V A,\Ba)$, see Section 6, and the Hamiltonian depends solely on the
$\V A$'s: therefore its restriction to $W$ is invariant with respect
to the action of the continuous commutative group $\TT^n$ of the
{\it translations of the angle variables}. The actions can be seen as
constants of motion whose existence follows from Noether's theorem, at
least in the anisochronous cases in which the Hamiltonian formulation
is equivalent to a Lagrangian formulation.

What is nontrivial is to recognize, prior to realizing integrability,
that a system admits this kind of symmetry: in most of the
interesting cases the systems either do not exhibit obvious symmetries
or they exhibit symmetries apparently unrelated to the group $\TT^n$,
which {\it nevertheless} imply existence of constants of motion in
number sufficient for integrability. Hence nontrivial integrable
systems possess a ``hidden'' symmetry under $\TT^n$: the rigid body is
an example.
\*

{\it However} very often the symmetries of a Hamiltonian $H$ which
imply integrability also imply partial isochrony, \ie imply that the
number of independent frequencies is smaller than $n$, see Section
\sec(6). Even in such cases often a map exists from the original
coordinates $(\V p,\V q)$ to the integrating variables $(\V A,\Ba)$ in
which $\V A$ are constants of motion and the $\Ba$ are uniformly
rotating angles (some of which are {\it also} constant) with spectrum
$\Bo(\V A)$ which is the gradient $\Dpr_{\V A}h(\V A)$ for some
function $h(\V A)$ depending only on a few of the $\V A$ coordinates.
However the map might fail to be {\it canonical}. The system is then
said to be {\it bihamiltonian}: in the sense that one can represent
motions in two systems of canonical coordinates, not related by a
canonical transformation, and by two Hamiltonian functions $H$ and
$H'\=h$ which generate the {\it same} motions in the respective
coordinates (the latter changes of variables are sometimes called
``canonical with respect to the pair $H,H'$'' while the
transformations considered in Section \sec(4) are called completely
canonical).  \*

\0{\it References:} [CD82].

\*

\Section(11, Generic nonintegrability)
\*

It is natural to try to prove that a system ``close'' to an integrable
one has motions with properties very close to quasi--periodic. This is
indeed the case but in a rather subtle way. That there is a problem is
easily seen in the case of a perturbation of an anisochronous
integrable system.

Assume that a system is integrable in a region $W$ of phase space
which, in the integrating action--angle variables $(\V A,\Ba)$, has the
standard form $U\times\TTT^\ell$ with a Hamiltonian $h(\V A)$ with
gradient $\Bo(\V A)=\dpr_{\V A} h(\V A)$. If the forces are perturbed
by a potential which is smooth then the new system will be described,
in the same coordinates, by a Hamiltonian like

$$\HH_\e(\V A,\Ba)=h(\V A)+\e f(\V A,\Ba)\Eq(11.1)$$
with $h,f$ analytic in the variables $\V A,\Ba$.

If the system really behaved like the unperturbed one it ought to have
$\ell$ constants of motion of the form $F_\e(\V A,\Ba)$ analytic in
$\e$ near $\e=0$ and {\it uniform}, \ie single valued (which is the
same as periodic) in the variables $\Ba$.  However the following
theorem ({\cs Poincar\'e}) shows that this is somewhat unlikely a
possibility.

\*
\0{\it If the matrix
$\Dpr^2_{\V A \V A} h(\V A)$ has rank $\ge2$ the Hamiltonian
\equ(11.1) ``generically'' 
(an intuitive notion precised below) cannot be integrated by a
canonical transformation $\CC_\e(\V A,\Ba)$ which
\\
(1) reduces to the identity as $\e\to0$, 
\\
(2) is
analytic in $\e$ near $\e=0$ and in $(\V A,\Ba)\in U'\times
\TTT^\ell$, with $U'\subset U$ open.
\\
Furthermore no uniform constants of motion $F_\e(\V A,\Ba)$, defined for $\e$
near $0$ and $(\V A,\Ba)$ in an open domain $U'\times\TTT^\ell$, exist
other than the functions of $\HH_\e$ itself.}  
\*

Integrability in the sense (1),(2) can be called {\it analytic
integrability} and it is the strongest (and most naive) sense
that can be given to the attribute.

The first part of the theorem, \ie (1),(2), holds simply because, if
integrability was assumed, a generating function of the integrating
map would have the form $\V A'\cdot\Ba+\F_\e(\V A',\Ba)$ with $\F$
admitting a power series expansion in $\e$ as
$\F_\e=\e\F^1+\e^2\F^2+\ldots$. Hence $\F^1$ would have to satisfy

$$\Bo(\V A')\cdot\Dpr_\Ba \F^1(\V A',\Ba)+ f(\V A',\Ba)=\lis f(\V
A')\Eq(11.2)$$
where $\lis f(\V A')$ depends only on $\V A'$ (hence integrating both
sides with $d\Ba$ it appears that $\lis f(\V A')$ must coincide with
the average of $f(\V A',\Ba)$ over $\Ba$).

This implies that the Fourier transform $f_{\Bn}(\V A)$,
$\Bn\in\ZZZ^\ell$, should satisfy

$$f_\Bn(\V A')=0\qquad {\rm if}\ \Bo(\V A')\cdot\Bn=0,\ \Bn\ne\V0\Eq(11.3)$$
which is equivalent to the existence of $\widetilde
f_\Bn(\V A')$ such that $f_\Bn(\V A)=\Bo(\V A')\cdot\Bn
\widetilde f_\Bn(\V A)$ for $\Bn\ne\V0$. But 
since there is no relation between $\Bo(\V A)$ and $f(\V A,\Ba)$ this
property ``generically'' will not hold in the sense that as close as
wished to a $f$ which satisfies the property \equ(11.3) there will be
another $f$ which does not satisfy it esentially no matter how
``closeness'' is defined, \eg with respect to the metric
$||f-g||=\sum_\Bn|f_\Bn(\V A)-g_\Bn(\V A)||$. This is so because the
rank of $\Dpr^2_{\V A \V A} h(\V A)$ is higher than $1$ and $\Bo(\V
A)$ varies at least on a two dimensional surface, so that
$\Bo\cdot\Bn=0$ becomes certainly possible for some $\Bn\ne\V0$ while
$f_{\Bn}(\V A)$ in general will not vanish, so that $\F^1$, hence
$\F_\e$, does not exist.

This means that close to a function $f$ there is a function $f'$
which violates \equ(11.3) for some $\Bn$. Of course this depends on
what is meant by ``close'': however here essentially any topology
introduced on the space of the functions $f$ will make the statement
correct. For instance if the distance between two functions is defined
by $\sum_\Bn \sup_{\V A\in U} |f_{\Bn}(\V A)-g_{\Bn}(\V A)|$ or by
$\sup_{\V A,\Ba} |f(\V A,\Ba)-g(\V A,\Ba)|$.

The idea behind the last of the theorem statement is in essence the
same: consider, for simplicity, the anisochronous case in which the
matrix $\Dpr^2_{\V A\V A}h(\V A)$ has maximal rank $\ell$, \ie the
determinant $\det\Dpr^2_{\V A\V A}h(\V A)$ does not
vanish. Anisochrony implies that $\Bo(\V A)\cdot\Bn\ne0$ for
all $\Bn\ne\V 0$ and for $\V A$ on a dense set and this property will
be used repeatedly in the following analysis.

Let $B(\e,\V A,\Ba)$ be a ``uniform'' constant of motion, meaning that it
is single--valued and analytic in the non simply connected region
$U\times\TTT^\ell$ and, for $\e$ small,

$$B(\e,\V A,\Ba)=B_0(\V A,\Ba)+\e B_1(\V A,\Ba)+\e^2 B_2(\V A,\Ba)
+\ldots\Eq(11.4)$$
The condition that $B$ is a constant of motion can be written order
by order in its expansion in $\e$: the first two orders are

$$\eqalign{
&\Bo(\V A)\cdot\dpr_{\Ba} B_0(\V A,\Ba)=0,
\cr
& \dpr_{\V A} f(\V A,\Ba)\cdot\dpr_{\Ba}B_0(\V A,\Ba)-
\dpr_{\Ba} f(\V A,\Ba)\cdot\dpr_{\V A}B_0(\V A,\Ba)+\cr&\kern3cm
+\Bo(\V
A)\cdot\dpr_{\Ba} B_1(\V A,\Ba)=0
\cr}\Eq(11.5)$$
Then the above two relations {\it and anisochrony} imply that $B_0$
must be a function of $\V A$ only and the second that $\Bo(\V
A)\cdot\Bn$ and $\dpr_{\V A} B_0(\V A)\cdot\Bn$ vanish simoultaneously
for all $\Bn$. Hence the gradient of $B_0$ {\it must} be proportional
to $\Bo(\V A)$, \ie to the gradient of $h(\V A)$: $\Dpr_{\V A} B_0(\V
A)= \l(\V A) \Dpr_{\V A}h(\V A)$. Therefore generically (because of
the anisochrony) it must be that $B_0$ depends on $\V A$ through $h(\V
A)$: $B_0(\V A)=F(h(\V A))$ for some $F$.

Looking again, with the new information, at the second of
\equ(11.5) it follows that at fixed $\V A$ the $\Ba$-derivative in the
direction $\Bo(\V A)$ of $B_1$ equals $F'(h(\V A))$ times the
$\Ba$-derivative of $f$, \ie
$B_1(\V A,\Ba)=f(\V A,\Ba)\,F'(h(\V A))+C_1(\V A)$.
% so that
%$\l(\V A)= F'(h(\V A))$ and $B_0(\V A)=F(h(\V A))$.

Summarizing: the constant of motion  $B$ has been written as
$B(\V A,\Ba)=F(h(\V A))+\e F'(h(\V A))\,f(\V A,\Ba)+\e C_1(\V A)+\e^2
B_2+\ldots$ which is equivalent to
$B(\V A,\Ba)$ $=F(\HH_\e)+\e(B'_0+\e B'_1+
\ldots)$ and therefore $B'_0+\e B'_1+\ldots$ is another analytic constant of
motion. Repeating the argument also $B'_0+\e B'_1+\ldots$ must have
the form $F_1(\HH_\e)+\e\,(B''_0+\e B''_1+\ldots)$; conclusion

$$B=F(\HH_\e)+\e\,F_1(\HH_\e)+\e^2 F_2(\HH_\e)+\ldots +\e^n F_n(\HH_\e)+
O(\e^{n+1})\Eq(11.6)$$
By analyticity $B=F_\e(\HH_\e(\V A,\Ba))$ for some $F_\e$: hence
generically all constants of motion are trivial.

Therefore a system close to integrable cannot behave as it would
naively be expected. The problem, however, was not manifest until {\cs
Poincar\'e}'s proof of the above results: because in most applications
the function $f$ has only finitely many Fourier components, or at
least is replaced by an approximation with this property, so that at
least \equ(11.3) and even a few of the higher order constraints like
\equ(11.5) become possible in open regions of action space. In fact it
may happen that the values of $\V A$ of interest are restricted so
that $\Bo(\V A)\cdot\Bn=0$ only for ``large'' values of $\Bn$ and
there $f_{\Bn}=0$.  {\it Nevertheless} the property that $f_\Bn(\V
A)=(\Bo(\V A)\cdot\Bn)\,\widetilde f_\Bn(\V A)$ (or the analogous
higher order conditions, \eg \equ(11.5)), which we have seen to be
necessary for analytic integrability of the perturbed system, can be
checked to fail in important problems, if no approximation is made on
$f$. Hence a conceptual problem arises.

\* 
\0{\it References:} [Po].
\*

\Section(12, Perturbing functions)
\*

To check, in a given problem, the nonexistence of nontrivial constants
of motion along the lines indicated in Sect. \sec(11) it is necessary
to express the potential, usually given in Cartesian coordinates as
$\e V(\V x)$, in terms of the action--angle variables of the
unperturbed, integrable, system. 

In particular the problem arises when trying to check nonexistence of
nontrivial constants of motion when the anisochrony assumption
(cf. Sect.\sec(11)) {\it is not satisfied}.  Usually it becomes
satisfied ``to second order'' (or higher): but to show this a more
detailed information on the structure of the perturbing function
expressed in action--angle variables is needed. For instance this is
often necessary even when the perturbation is approximated by a
trigonometric polynomial, as it is essentially always the case in
Celestial Mechanics.

Finding explicit expressions for the action--angle variables is in
itself a rather nontrivial task which leads to many problems of
intrinsic interest even in seemingly simple cases. For instance, in
the case of the planar gravitational central motion, the Kepler
equation $\l=\x-\e \sin\x$, see the first of \equ(8.1), must be solved
expressing $\x$ in terms of $\l$, see the first of \equ(8.2). It is
obvious that for small $\e$ the variable $\x$ can be expressed as an
analytic function of $\e$: {\it nevertheless} the actual construction
of this expression leads to several problems.  For small $\e$ an
interesting algorithm is the following.

Let $h(\l)=\x-\l$ so that the equation to solve (\ie the first of
\equ(8.1)) is

$$h(\l)= \e \sin(\l+h(\l))\=-\e \,\fra{\dpr c}{\dpr\l}(\l+h(\l))\Eq(12.1)$$
where $c(\l)=\cos\l$; the function $\l\to h(\l)$ should be periodic in
$\l$, with period $2\p$, and analytic in $\e,\l$ for $\e$ small and
$\l$ real. If $h(\l)=\e h^{(1)}+\e^2 h^{(2)}+\ldots$ the Fourier
transform of $h^{(k)}(\l)$ satisfies the recursion relation

$$h^{(k)}_\n=-\sum_{p=1}^\io\fra1{p!}\sum_{k_1+\ldots+k_p=k-1
\atop \n_0+\n_1+\ldots+\n_p=\n}
(i\n_0) c_{\n_0} (i\n_0)^p \prod h^{(k_j)}_{\n_j},\qquad k>1
\Eq(12.2)$$
with $c_\n$ the Fourier transform of the cosine ($c_{\pm1}=\fra12$,
$c_\n=0$ if $\n\ne\pm1$), and (of course) $h^{(1)}_\n=-i\n c_\n$.
Eq. \equ(12.2) is obtained by expanding the \rhs of \equ(12.1) in
powers of $h$ and then taking the Fourier transform of both sides
retaining only terms of order $k$ in $\e$.

Iterating the above relation imagine to draw all trees $\th$ with $k$
``branches'', or ``lines'', {\it distinguished by a label taking $k$
values}, and $k$ nodes and attach to each node $v$ a {\it harmonic}
label $\n_v=\pm1$ as in Fig. 5. The trees will assumed to start with a
{\it root line} $vr$ linking a point $r$ and the ``first node'' $v$,
see Fig. 5, and then bifurcate arbitrarily (such trees are sometimes
called ``rooted trees'').  \*

\eqfig{270pt}{70pt}{
\ins{54pt}{53pt}{$\st\Bn$}
\ins{87pt}{36pt}{$\st\Bn_0$}
\ins{137pt}{72pt}{$\st\Bn_1$}
\ins{186pt}{86pt}{$\st\Bn_4$}
\ins{136pt}{55pt}{$\st\Bn_2$}
\ins{137pt}{25pt}{$\st\Bn_{3}$}
\ins{187pt}{2pt}{$\st\Bn_{10}$}
\ins{187pt}{15pt}{$\st\Bn_{9}$}
\ins{187pt}{30pt}{$\st\Bn_8$}
\ins{187pt}{43pt}{$\st\Bn_7$}
\ins{187pt}{57pt}{$\st\Bn_6$}
\ins{187pt}{72pt}{$\st\Bn_5$}
}{fig5}{(5)}

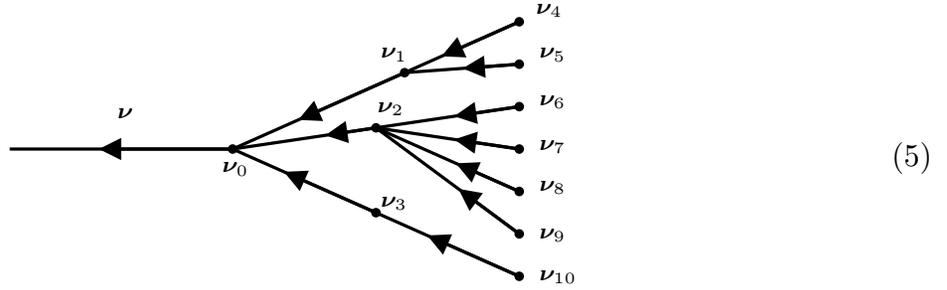
\0Fig. 5: {\nota An example of a tree graph and of its labels. It
contains only one simple node ($\st1$). Harmonics are indicated next
to their nodes. Labels distinguishing lines are not marked.}

Imagine the tree oriented from the endpoints towards the root $r$ (not
to be considered a node) and given a node $v$ call $v'$ the node immediately
following it. If $v$ is the first node before the root $r$ let $v'=r$
and $\n_{v'}=1$.  For each such decorated tree define its {\it
numerical value}

$$\Val(\th)= \fra{-i}{k!}
\prod_{lines\, l= \,v'v} (\n_{v'}\n_v)\prod_{nodes} c_{\n_v}\Eq(12.3)$$
and define a {\it current} $\n(l)$ on a line $l=v'v$ to be the sum of the
harmonics of the nodes preceding $v'$: $\n(l)=\sum_{w\le v}\n_v$.
Call $\n(\th)$ the current flowing in the root branch and call order
of $\th$ the number of nodes (or branches). Then

$$h^{(k)}_\n=\sum_{\th,\,\n(\th)=\n\atop
order(\th)=k} \Val (\th)\Eq(12.4)$$
provided trees are considered identical if they can be overlapped
(labels included) after suitably scaling the lengths of their branches
and pivoting them around the nodes out of which they emerge (the root
is always imagined fixed at the origin).

If the trees are stripped of the harmonic labels their number is
finite and it can be estimated $\le k! 4^k$ (because the labels which
distinguish the lines can be attached to an unlabeled tree in many
ways). The harmonic labels (\ie $\n_v=\pm1$) can be laid down in $2^k$
ways, and the value of each tree can be bounded by $\fra1{k!}
2^{-k}$ (because $c_{\pm1}=\fra12$).

Hence $\sum_{\n}| h^{(k)}_\n|\le 4^k$, which gives a (rough) estimate
of the radius of convergence of the expansion of $h$ in powers of
$\e$: namely $.25$ (easily improvable to $0.3678$ if $4^kk!$ is
replaced by $k^{k-1}$ using Cayley's formula for the enumeration of
rooted trees). A simple expression for $h^{(k)}(\ps)$ ({\cs Lagrange})
is $h^{(k)}(\ps)=\fra1{k!} \dpr_\ps^{k-1}\sin^k\ps$ (also readable
from the tree representation): the actual radius of convergence, first
determined by {\cs Laplace}, of the series for $h$ can be also
determined from the latter expression for $h$ ({\cs Rouch\'e}) or
directly from the tree representation: it is $\sim .6627$.

One can find better estimates or at least more efficient methods for
evaluating the sums in \equ(12.4): in fact in performing the sum in
\equ(12.4) important {\it cancellations} occurr. For instance the
harmonic labels can be subject to the further strong constraint
that {\it no line carries zero current} because the sum of the values
of the trees of fixed order and with at least one line carrying $0$
current vanishes.

The above expansion can also be simplified by ``partial
resummations''. For the purpose of an example, call {\it simple} the
nodes with one entering and one exiting line (see Fig. 5). Then all
tree graphs which on any line between two non simple nodes contain any
number of simple nodes can be eliminated. This is done by replacing,
in evaluating the (remaining) trees value, the factors $\n_{v'}\n_v$
in \equ(12.3) by $\n_{v'}\n_v/(1-\e\cos\ps )$: then the value of a
tree $\th$ becomes a function of $\ps$ and $\e$ to be denoted
$\Val(\th)_\ps$ {\it and} \equ(12.4) is replaced by

$$h(\ps)=\sum_{k=1}^\io {\sum_{\th,\,\n(\th)=\n\atop
order(\th)=k}}^{\kern-.3cm *}
\kern3mm\e^k\, e^{i\,\n\,\ps}\, \Val(\th)_\ps\Eq(12.5)$$
where the $*$ means that the trees are {\it subject to the further
restriction of not containing any simple node}. It should be noted
that the above graphical representation of the solution of the Kepler
equation is strongly reminiscent of the representations of quantities
in terms of graphs that occurr often in quantum field theory. Here the
trees correspond to ``Feynman graphs'', the factors associated with
the nodes are the ``couplings'', the factors associated with the lines
are the ``propagators'' and the resummations are analogous to the
``self-energy resummations'', while the mentioned cancellations can be
related to the class of identities called ``Ward identities''.  Not
only the analogy can be shown to be not superficial, but it turns out
to be even very helpful in key mechanical problems: see Appendix
\sec(A).

The existence of a vast number of identities relating the tree values
is shown already by the ``simple'' form of Lagrange's series and by
the even more remarkable resummation ({\cs
Levi-Civita}) leading to

$$h(\ps)=\sum_{k=1}^\io \fra{(\e\sin\ps)^k}{k!}
\Big(\fra1{1-\e\cos\ps}\dpr_\ps\Big)^k\,\ps\Eq(12.6)$$

It is even possible further collection of the series terms to
express it as series with much better convergence properties, for
instance its terms can be reorganized and collected ({\it resummed})
so that $h$ is expressed as a power series in the parameter

$$\h=\fra{\e\,e^{\sqrt{1-\e^2}}}{1+\sqrt{1-\e^2}}\Eq(12.7)$$
with radius of convergence $1$, which corresponds to $\e=1$ (via a
simple argument by {\cs Levi-Civita}). The analyticity domain for the
Lagrange series is $|\h|<1$. This also determines the Laplace radius
value: it is the point closest to the origin of the complex curve
$|\h(\e)|=1$: it is imaginary so that it is the root of the equation $\e
e^{\sqrt{1+\e^2}}/(1+\sqrt{1+\e^2})=1$.

The analysis provides an example, in a simple case of great interest
in applications, of the kind of computations actually necessary to
represent the perturbing function in terms of action--angle
variables. The property that the function $c(\l)$ in \equ(12.1) is the
cosine has been used only to limit the range of the label $\n$ to be
$\pm1$; hence the same method, with similar results, can be applied to
study the inversion of the relation between the average anomaly $\l$
and the true anomaly $\th$ and to get, for instance, quickly the
properties of $f,g$ in \equ(8.2).  \*

\0{\it References:} [LC].  
\*

\Section(13, Lindstedt and Birkhoff series. Divergences.)
\*

Nonexistence of constants of motion, rather than being the end of
the attempts to study by perturbation methods motions close to
integrable ones, marks the beginning of renewed efforts to understand
their nature.

Let $(\V A,\Ba)\in U\times \TTT^\ell$ be action--angle variables
defined in the integrability region for an analytic Hamiltonian and
let $h(\V A)$ be its value in the action--angle coordinates. Suppose
that $h(\V A)$ is {\it anisochronous} and let $f(\V A,\Ba)$ be an
analytic perturbing function. Consider, for $\e$ small, the Hamiltonian
$\HH_\e(\V A,\Ba)=\HH_0(\V A)+\e f(\V A,\Ba)$.

Let $\Bo_0=\Bo(\V A_0)\=\Dpr_{\V A} \HH_0(\V A)$ be the frequency
spectrum, see Section \sec(6), of one of the invariant tori of the
unperturbed system corresponding to an action $\V A_0$. Short of
integrability the question
to ask at this point is whether the perturbed system admits an
analytic invariant torus on which motion is quasi periodic with\\ (1)
the same spectrum $\Bo_0$ and
\\
(2) depends analytically
on $\e$ at least for $\e$ small and
\\
(3) reduces to the
``unperturbed torus'' $\{\V A_0\}\times\TTT^\ell$ as $\e\to0$.

More concretely the question is \*

{\it Are there functions $\V H_\e(\Bps),\V h_\e(\Bps)$ analytic in
$\Bps\in\TTT^\ell$ and in $\e$ near $0$, {\it vanishing as $\e\to0$}
and such that the torus with parametric equations

$$\V A=\V A_0+{\V H}_\e(\Bps),\qquad \Ba=\Bps+\V h_\e(\Bps)\qquad
\Bps\in\TTT^\ell
\Eq(13.1)$$
is invariant and, if $\Bo_0\defi\Bo(\V A_0)$, the motion on it is
simply $\Bps\to\Bps+\Bo_0 t$, \ie it is quasi periodic with spectum
$\Bo_0$?}
\*

In this context {\cs Poincar\'e}'s theorem of Sect.\sec(11) had
followed another key result, earlier developed in particular cases and
completed by him, which provides a {\it partial} answer to the
question.

Suppose that $\Bo_0=\Bo(\V A_0)\in\RRR^\ell$ satisfies a {\it Diophantine
property}, namely suppose that there exist
constants $C,\t>0$ such that

$$ |\Bo_0\cdot\Bn|\ge \fra1{C |\Bn|^\t},\qquad\hbox{for all} \ \V0\ne
\Bn\in\ZZZ^\ell\Eq(13.2)$$
which, for each $\t>\ell-1$ {\it fixed}, is a property enjoyed by all
$\Bo\in\RRR^\ell$ {\it but for a set of zero measure}. Then the
motions on the unperturbed torus run over trajectories that {\it fill
the torus densely} because of the ``irrationality'' of $\Bo_0$ implied
by \equ(13.2).  Writing Hamilton's equations,, $\dot\Ba=\dpr_{\V
A}\HH_0(\V A)+\e\Dpr_{\V A} f(\V A,\Ba)$, $\dot{\V
A}=-\e\Dpr_{\Ba}f(\V A,\Ba)$ with $\V A,\Ba$ given by \equ(13.1) with
$\Bps$ replaced by $\Bps+\Bo t$, and using the density of the
unperturbed trajectories implied by \equ(13.2), the condition that
\equ(13.1) are equations for an invariant torus on which the motion is
$\Bps\to\Bps+\Bo_0 t$ are

$$\eqalign{
&\Bo_0+(\Bo_0\cdot\Dpr_{\Bps}) \V h_\e(\Bps)= \Dpr_{\V A} \HH_0(\V
A_0+\V H_\e(\Bps))+\e\Dpr_{\V A} f(\V A_0+\V H_\e(\Bps), \Bps+\V
h_\e(\Bps))\cr
& (\Bo_0\cdot\Dpr_{\Bps}) \V H_\e(\Bps)=
-\e \Dpr_{\Ba} f(\V A_0+\V H_\e(\Bps), \Bps+\V
h_\e(\Bps))\cr}\Eq(13.3)$$
The theorem referred above ({\cs Poincar\'e}) is that
\*

\0{\it If the unperturbed system is anisochronous and $\Bo_0=\Bo(\V
A_0)$ satisfies \equ(13.2) for some $C,\t>0$ there exist two well
defined power series $\V h_\e(\Bps)=\sum_{k=1}^\io$ $ \e^k\,\V
h^{(k)}(\Bps)$ and $\V H_\e(\Bps)=\sum_{k=1}^\io \e^k\,\V
H^{(k)}(\Bps)$ which solve \equ(13.3) to all orders in $\e$. The
series for $\V H_\e$ is uniquely determined, and such is also the
series for $\V h_\e$ up to the addition of an arbitrary constant at
each order, so that it is unique if $\V h_\e$ is required, as
henceforth done with no loss of generality, to have zero average over
$\Bps$.}  \*

The algorithm for the construction is illustrated in a simple case in
Section \sec(14), \equ(14.4),\equ(14.5). Convergence of the above
series, called {\it Lindstedt series}, even for $\e$ small has been a
problem for a rather long time. {\cs Poincar\'e} proved existence of
the formal solution; but his other result, discussed in Sect.\sec(11),
casts doubts on convergence although {\it it does not exclude it} as
was immediately stressed by several authors (including {\cs
Poincar\'e} himself).  The result of Sect. 11 shows impossibility of
solving \equ(13.3) {\it for all} $\Bo_0$'s near a given spectrum,
analytically and uniformly, but it does not exclude the possibility of
solving it {\it for a single} $\Bo_0$.

The theorem admits several extensions or analogues: an interesting one
is to the case of isochronous unperturbed systems:
\*

\0{\it Given the Hamiltonian $\HH_\e(\V A,\Ba)=\Bo_0\cdot\V A+\e f(\V
A,\Ba)$, with $\Bo_0$ satisfying \equ(13.2) and $f$ analytic, there
exist power series $\CC_\e(\V A',\Ba'),\,u_\e(\V A')$ such that
$\HH_\e(\CC_\e(\V A',\Ba'))=\Bo_0\cdot\V A'+u_\e(\V A')$ holds as an
equality between formal power series (\ie order by order in $\e$) and
at the same time the $\CC_\e$ regarded as a map satisfies order by
order the condition (\ie {\rm \equ(4.3)}) that it is a canonical map.}
\*

This means that there is a generating function $\V
A'\cdot\Ba+\F_\e(\V A',\Ba)$ also defined by a formal power series $\F_\e(\V
A',\Ba)=\sum_{k=1}^\io \e^k\,
\F^{(k)}(\V A',\Ba)$, \ie such that if $\CC_\e(\V
A',\Ba')=(\V A,\Ba)$ then it is true, order by order in powers of
$\e$, that $\V A=\V A'+\Dpr_{\Ba}\F_\e(\V A',\Ba)$ and $\Ba'=\Ba+\Dpr_{\V
A'}\F_\e(\V A',\Ba)$. The series for $\F_\e,u_\e$ are called {\it
Birkhoff series}.

In this isochronous case {\it if Birkhoff series were convergent} for
small $\e$ and $(\V A',\Ba)$ in a region of the form $U\times
\TTT^\ell$, with $U\subset\RRR^\ell$ open and bounded, it would follow
that, for small $\e$, $\HH_\e$ would be integrable in a large
region of phase space (\ie where the generating function can be used
to build a canonical map: this would essentially be $U\times \TTT^\ell$
deprived of a small layer of points near the boundary of $U$). However
convergence for $\e$ small is false (in general) as shown by the
simple two dimensional example

$$\HH_\e(\V A,\Ba)=\Bo_0\cdot\V A+\e\,(A_2+f(\Ba)),\qquad (\V
A,\Ba)\in\RRR^2\times \TTT^2\Eq(13.4)$$
with $f(\Ba)$ an arbitrary analytic function with {\it all} Fourier
coefficients $f_\Bn$ positive for $\Bn\ne\V0$ and $f_{\V0}=0$.
In the latter case the solution is

$$u_\e(\V A')=\e A_2,\quad \F_\e(\V A',\Ba)=\sum_{k=1}^\io
\e^k \sum_{\V0\ne \Bn\in\zzz^2}
{f_\Bn\,e^{i\Ba\cdot\Bn}}\fra{(i\,\n_2)^k}
{(i(\o_{01}\n_1+\o_{02}\n_2))^{k+1}}\Eq(13.5)
$$
The series does not converge: in fact its convergence would imply
integrability and, consequently, bounded trajectories in phase space:
however the equations of motion for \equ(13.4) can be easily solved
{\it explicitly} and in any open region near a given initial data
there are other data which have unbounded trajectories if
$\o_{01}/(\o_{02}+\e)$ is rational.

Nevertheless even in this elementary case a formal sum of the series
yields

$$u(\V A')=\e A'_2,\quad \F_\e(\V A',\Ba)=\e \sum_{\V 0\ne\Bn\in\zzz^2}
\fra{f_\Bn\,e^{i\Ba\cdot\Bn}}{i(\o_{01}\n_1+(\o_{20}+\e)\n_2)}\Eq(13.6)$$
and the series in \equ(13.6) (no longer a power series in $\e$) is
really convergent if $\Bo=(\o_{01},\o_{02}+\e)$ is a Diophantine
vector (by \equ(13.2), because analyticity implies exponential decay of
$|f_\Bn|$). Remarkably for such values of $\e$ the Hamiltonian
$\HH_\e$ {\it is integrable and it is integrated} by the canonical map
generated by \equ(13.6), in spite of the fact that \equ(13.6) is
obtained, from \equ(13.5), via the {\it non rigorous sum rule}

$$\sum_{k=0}^\io z^k=\fra1{1-z} \qquad{\rm for}\ z\ne1\Eq(13.7)$$
(applied to cases with $|z|\ge1$, which are certainly realized for a
dense set of $\e$'s even if $\Bo$ is Diophantine because the $z$'s
have values $z=\fra{\n_2}{\Bo_0\cdot\Bn}$). In other words the
integration of the equations is elementary and once performed it
becomes apparent that, if $\Bo$ is diophantine, the solutions can be
rigorously found from \equ(13.6). Note that, for instance, this means
that relations like $\sum_{k=0}^\io 2^k=-1$ are really used to obtain
\equ(13.6) from \equ(13.5).

Another extension of Lindstedt series arises in a perturbation of an
an\-iso\-chro\-nous system when asking the question of what happens to the
unperturbed invariant tori $\TT_{\Bo_0}$ on which the spectrum {\it is
resonant}, \ie $\Bo_0\cdot\Bn=0$ for some $\Bn\ne\V0, \,
\Bn\in\ZZZ^\ell$.  The result is that even in such
case there is a formal power series solutions showing that {\it at least a few}
of the (infinitely many) invariant tori into which $\TT_{\Bo_0}$ is in
turn foliated in the unperturbed case can be formally continued at
$\e\ne0$, see Section \sec(15).
\*
\0{\it References:}  [Po].
\*

\Section(14, Quasi periodicity and KAM stability)
\*

To discuss more advanced results it is convenient to restrict
attention to a special (non trivial) paradigmatic case

$$\HH_\e(\V A,\Ba)=\fra12 \V A^2+\e\,f(\Ba)\Eq(14.1)$$
In this simple case (called {\it Thirring model}: representing $\ell$
particles on a circle interacting via a potential $\e f(\Ba)$) the equations for
the maximal tori \equ(13.3) reduce to equations for the only
functions $\V h_\e$:

$$(\Bo\cdot\Dpr_{\Bps})^2\V h_\e(\Bps)=-\e\,\Dpr_\Ba f(\Bps+\V
h_\e(\Bps)),\qquad \Bps\in\TTT^\ell\Eq(14.2)$$
as the second of \equ(13.3) simply becomes the definition of $\V H_\e$
because the \rhs does not involve $\V H_\e$.

The real problem is therefore whether the formal series considered in
Section \sec(13) converge
at least for small $\e$: and the example \equ(13.4) on the Birkhoff series
shows that sometimes {\it sum rules}
might be needed in order to give a meaning to the series. In fact
whenever a problem (of physical interest) admits a formal
power series solution which is not convergent,  or which it is not
known whether it is convergent, then one should look for sum rules for it.

The modern theory of perturbations starts with the proof of the
convergence for $\e$ small enough of the Lindstedt series
({\cs Kolmogorov}). The general ``KAM'' result is
\*

\0{\it Consider the Hamiltonian $\HH_\e(\V A,\Ba)=h(\V A)+\e f(\V
A,\Ba)$, defined in $U=V\times \TTT^\ell$ with $V\subset\RRR^\ell$
open and bounded and with $f(\V A,\Ba),h(\V A)$ analytic in the
closure $\lis V\times \TTT^\ell$ where $h(\V A)$ is also
anisochronous; let $\Bo_0\defi$ $ \Bo(\V A_0)=\dpr_{\V A}h(\V A_0)$
and assume that $\Bo_0$ satisfies \equ(13.2). Then 
\\ 
(1) there is $\e_{C,\t}>0$ such that the Lindstedt series converges
for $|\e|<\e_{C,\t}$,
\\ 
(2) its sum yields two function $\V H_\e(\Bps),\V
h_\e(\Bps)$ on $\TTT^\ell$ which parameterize an invariant torus
$\TT_{C,\t}(\V A_0,\e)$, 
\\ 
(3) on $\TT_{C,\t}(\V A_0,\e)$ the motion is $\Bps \to\Bps+\Bo_0t$,
see \equ(13.1).
\\ 
(4) the set of data in $U$ which belong to invariant tori
$\TT_{C,\t}(\V A_0,\e)$ with $\Bo(\V A_0)$ satisfying \equ(13.2) with
prefixed $C,\t$ has complement with volume $<\, const\, C^{-a}$ for a
suitable $a>0$ and with area also $<\, const\, C^{-a}$ on each
nontrivial surface of constant energy $\HH_\e=E$.  }
\*

In other words for $\e$ small the spectra of most unperturbed quasi
periodic motions can still be found as spectra of perturbed quasi
periodic motions developing on tori which are close
to the corresponding  unperturbed ones (\ie with the same spectrum).

This is a {\it stability result}: for instance in systems with two
degrees of freedom the invariant tori of dimension $2$ which lie on a
given energy surface, which has $3$ dimensions, will separate the
points on the energy surface into the set which is ``inside'' the
torus and the set which is ``outside'': hence an initial datum
starting (say) inside cannot reach the outside. Likewise a point
starting between two tori has to stay in between forever: and if the
two tori are close this means that motion will stay very localized in
action space, with a trajectory accessing only points close to the
tori and coming close to all such points, within a distance of the
order of the distance between the confining tori. The case of three or
more degrees of freedom is quite different, see Sect.\sec(17),\sec(19).

In the simple case of the rotators system \equ(14.1) the equations
for the parametric representation of the tori are the
\equ(14.2). The latter bear some analogy with the easier problem in
\equ(12.1): but the \equ(14.2) are $\ell$ equations instead of one and
they are differential equations rather than ordinary
equations. Furthermore the function $f(\Ba)$ which plays here the role
of $c(\l)$ in \equ(12.1) has Fourier coefficients $f_{\Bn}$ with no
restrictions on $\Bn$, while the Fourier coefficiens $c_\n$ for $c$ in
\equ(12.1) do not vanish only for $\n=\pm1$.

The above differences are, to some extent, ``minor'' and the power
series solution to \equ(14.2) can be constructed {\it by the same
algorithm} used in the case of \equ(12.1): namely one forms trees as
in Fig. 5 with the harmonic labels $\n_v\in\ZZZ$ replaced by
$\Bn_v\in\ZZZ^\ell$ (still to be thought of as possible harmonic
indices in the Fourier expansion of the perturbing function $f$). All
other labels affixed to the trees in Sec. 11 will be the same. In
particular the {\it current flowing on a branch} $l=v'v$ will be
defined as the sum of the harmonics of the nodes $w\le v$ preceding
$v$:

$$\Bn(l)\defi \sum_{w\le v}\Bn_w\Eq(14.3)$$
and we call $\Bn(\th)$ the current flowing in the root branch.

This time the value $\Val(\th)$ of a tree has to be defined
differently because the equation \equ(14.2) to be solved contains the
differential operator $(\Bo_0\cdot\Dpr_\Bps)^2$ which in Fourier
transform becomes multiplication of the Fourier component with
harmonic $\Bn$ by $(i\Bo\cdot\Bn)^2$.

The variation due to the presence of the operator
$(\Bo_0\cdot\Dpr_\Bps)^2$ and the necessity of its inversion in the
evaluation of $\V u\cdot\V h^{(k)}_{\Bn}$, \ie of the component of $\V
h_\Bn^{(k)}$ along an arbitrary unit vector $\V u$, is nevertheless
quite simple: the value of a tree graph $\th$ of order $k$ (\ie with
$k$ nodes and $k$ branches) has to be defined by (cf. \equ(12.3))

$$\Val(\th)\defi \fra{-i\,(-1)^k}{k!}
\Big(\prod_{lines\, l= \,v'v}
\fra{\Bn_{v'}\cdot \Bn_v}{\big(\Bo_0\cdot\Bn(l)\big)^2}\Big)
\Big(\prod_{nodes\,v} f_{\Bn_v}\Big)\Eq(14.4)$$
where the $\Bn_{v'}$ appearing in the factor relative to the root line
$rv$ from the first node $v$ to the root $r$, see Fig. 5, is
interpreted as an unit vector $\V u$ (it was interpreted as
$1$ in the ``one dimensional'' case \equ(12.1)).  The \equ(14.4) makes
sense only for trees in which no line carries $\V0$ current.  Then the
component along $\V u$ (the harmonic label attached to the root of a
tree) of $\V h^{(k)}$ is given, see also \equ(12.4), by

$$\V u\cdot \V h^{(k)}_\Bn=
\sum^*_{\th,\,\Bn(\th)=\Bn\atop
order(\th)=k} \Val (\th)\Eq(14.5)$$
where the $*$ means that the sum is only over trees in which a non
zero current $\Bn(l)$ flows on the lines $l\in \th$. The quantity $\V
u\cdot\V h^{(k)}_{\V0}$ will be $\V0$, see Section \sec(13).

In the case of \equ(12.1) zero current lines could appear: {\it but}
the contributions from tree graphs containing at least one zero
current line would cancel. In the present case the statement that the
above algorithm actually gives $\V h^{(k)}_\Bn$ {\it by simply ignoring}
trees with lines with $\V 0$ current is non trivial. It has been
{\cs Poincar\'e}'s contribution to the theory of Lindstedt series to show
that {\it even in the general case}, cf.\equ(13.3), the equations for the
invariant tori can be solved by a formal power series. The
\equ(14.5) is proved by induction on $k$ after checking it for the
first few orders.

The algorithm just described leading to \equ(14.4) can be extended to
the case of the general Hamiltonian considered in the KAM theorem.

The convergence proof is more delicate than the (elementary) one for
the equation \equ(12.1). In fact the values of trees of order $k$ can
give large contributions to $\V h^{(k)}_\Bn$: because the ``new''
factors $\big(\Bo_0\cdot\Bn(l)\big)^{2}$, although not zero, can be
quite small and their small size can overwhelm the smallness of the
factors $f_{\Bn}$ and $\e$: in fact even if $f$ is a trigonometric
polynomial (so that $f_{\Bn}$ vanishes identically for $|\Bn|$ large
enough) the currents flowing in the branches can be very large, of the
order of the number $k$ of nodes in the tree, see \equ(14.3).

This is called the {\it small divisors} problem. The key to its solution
goes back to a related work ({\cs Siegel}) which shows that

\*
\0{\it Consider the contribution to the sum in \equ(14.3) from graphs
$\th$ in which no pairs of lines which lie on the same path to the
root carry the same current and, furthermore, the node harmonics are
bounded by $|\Bn|\le N$ for some $N$. Then the number of lines $\ell$
in $\th$ with divisor $\Bo_0\cdot\Bn_\ell$ satisfying
$2^{-n}<C\,| \Bo_0\cdot\Bn_\ell|\le 2^{-n+1}$ does not exceed $4\, N \, k
2^{-n/\t}$.}  
\*

Hence, setting $F\defi \,C^2\,\max_{|\Bn|\le N}
|f_\Bn|$,  the corresponding $\Val(\th)$ can be bounded by

\kern-2mm
$$ \fra1{k!}F^k N^{2k} \prod_{n=0}^\io 2^{2n \,(4k 2^{-n/\t})}\defi
\fra1{k!}
B^k,\qquad
B=\, F N^2 2^{\sum_n 8n
2^{-n/\t}}
\Eq(14.6)$$
since the product is convergent. In the case in which $f$ is a
trigonometric polynomial of degree $N$ the above {\it restricted}
contributions to $\V u\cdot\V h^{(k)}_{\Bn}$ would generate a
convergent series for $\e$ small enough. In fact the number of trees
is bounded, as in Sec. \sec(12), by $k! 4^k (2N+1)^{\ell k}$ so that
the series $\sum _\Bn |\e|^k \,|\V u\cdot\V h^{(k)}_\Bn|$ would
converge for $\e$ small (\ie $|\e|< (B\cdot4 (2N+1)^\ell)^{-1}$).

Given this comment the analysis of the ``remaining contributions''
becomes the real problem and it requires new ideas because among the
excluded trees there are some simple $k$-th order trees whose value
alone, if considered separately from the other contributions, would
generate a factorially divergent power series in $\e$.

However the contributions of all large valued trees of order $k$ can
be shown to cancel: although not exactly (unlike the case of the
elementary problem of Sec. \sec(12), where the cancellation is not
necessary for the proof, in spite of its exact occurrence), but enough
so that in spite of the existence of exceedingly large values of
individual tree graphs their total sum can still be bounded by a
constant to the power $k$ so that the power series actually converges
for $\e$ small enough. The idea is discussed in Appendix \sec(A).  \*
\0{\it References:} [Po], [Ko55], [Mo62], [Ar68].  \*

\Section(15, Resonances and their stability)
\*

A quasi periodic motion with $r$ rationally independent frequencies is
called {\it resonant} if $r$ is strictly less than the number $\ell$
of degrees of freedom. The difference $s=\ell-r$ is the {\it
degree} of the resonance.

Of particular interest are the cases of a perturbation of an
integrable system in which resonant motions take place.

A typical example is the {\it n-body problem} which studies the mutual
perturbations of the motions of $n-1$ particles gravitating around a
more massive particle. If the particles masses can be considered
negligible the system will consist of $n-1$ central Keplerian motions:
it will therefore have $\ell=3(n-1)$ degrees of freedom. In general,
only one frequency per body occurs in absence of the perturbations
(the period of the Keplerian orbit): hence $r\le n-1$ and
$s\ge 2(n-1)$ (or in the planar case $s\ge (n-1)$) with equality
holding when the periods are rationally independent.

Another example is the rigid body with a fixed point perturbed by a
conservative force: in this case the unperturbed system has $3$
degrees of freedom but, in general, only two frequencies, see Section
\sec(9) after \equ(9.10).

Furthermore in the above examples there is the possibility that the
independent frequencies assume, for special initial data, values which
are rationally related, giving rise to resonances of even higher order
(\ie with smaller values of $r$).

In an integrable anisochronous system resonant motions will be dense
in phase space because the frequencies $\Bo(\V A)$ will vary as much
as the actions and therefore resonances of any order (\ie any
$r<\ell$) will be dense in phase space: in particular the periodic
motions (\ie the {\it highest order resonances}) will be dense.

Resonances, in integrable systems, can arise in {\it a priori
stable} integrable systems and in {\it \ap unstable} systems: the
first are systems whose Hamiltonian admits canonical action--angle
coordinates $(\V A,\Ba)\in U\times \TTT^\ell$ with $U\subset
\RRR^\ell$ open, while the second are systems whose Hamiltonian has,
in suitable {\it local} canonical coordinates, the form

$$\HH_0(\V A)+\sum_{i=1}^{s_1}\fra12(p_i^2-\l^2_i q_i^2)
+\sum_{j=1}^{s_2}\fra12(\p_j^2+\m_j^2 \k_j^2),\qquad \l_i,\m_j>0 \Eq(15.1)$$
where $(\V A,\Ba)\in U\times \TTT^r$, $U\in\RRR^r$, $(\V p,\V q)\in
V\subset \RRR^{2 s_1}$, $(\Bp,\Bk)\in V'\subset \RRR^{2 s_2}$ with
$V,V'$ neighborhoods of the origin and $\ell=r+s_1+s_2$,
$s_i\ge0,s_1+s_2>0$ and $\pm\sqrt{\l_j},\pm\sqrt{\m_j}$ are called
{\it Lyapunov coefficients} of the resonance. The perturbations
considered are supposed to have the form $\e f(\V A,\Ba,\V p,\V
q,\Bp,\Bk)$.  The denomination of \ap stable or unstable refers to the
properties of the ``\ap given unperturbed Hamiltonian''. The name of
\ap unstable is certainly appropriate if $s_1>0$: here also $s_1=0$ is
allowed for notational convenience implying that the Lyapunov
coefficients in \ap unstable cases are all of order $1$ (whether real,
$\l_j$ or imaginary $i\sqrt{\m_j}$).  In other words the \ap stable
case, $s_1=s_2=0$ in
\equ(15.1), is the only excluded case. Of course the stability
properties of the motions when a perturbation acts will depend on the
perturbation {\it in both cases}.

The {\it a priori stable} systems have usually a great variety of
resonances (\eg in the anisochro\-nous case resonances of any
dimension are dense). The {\it a priori unstable} systems have (among
possible other resonances) some {\it very special} $r$-dimensional
resonances occurring when the {\it unstable coordinates} $(\V p,\V q)$
and $(\Bp,\Bk)$ are zero and the frequencies of the $r$ action--angle
coordinates are rationally independent.

In the first case, \ap stable, the general question is whether the
resonant motions, which form invariant tori of dimension $r$ arranged
into families that fill $\ell$ dimensional invariant tori, continue to
exists, in presence of small enough perturbations $\e f(\V A,\Ba)$,
on slightly deformed invariant tori. Similar questions can be asked in
the \ap unstable cases. To examine more closely the matter consider the
formulation of the simplest problems.
\*

\0{\it A priori stable resonances}: More precisely let $\{\V
A_0\}\times\TTT^\ell$ be the unperturbed invariant torus $\TT_{\V
A_0}$ with spectrum $\Bo_0=\Bo(\V A_0)=\V\dpr_{\V A} \HH_0(\V A_0)$
with only $r$ rationally independent components. For simplicity
suppose that $\Bo_0=(\o_1,\ldots,\o_r,0,\ldots,0)\defi(\Bo,\V0)$ with
$\Bo\in\RRR^r$. The more general case in which $\Bo$ has only $r$
rationally independent components can be reduced to special case above
by a canonical linear change of coordinates at the price of changing
the $\HH_0$ to a new one, still quadratic in the actions but
containing mixed products $A_iB_j$: the proofs of the results that are
discussed here would not be really affected by such more general form
of $\HH$.

It is convenient to distinguish between the ``fast'' angles
$\a_1,\ldots,\a_r$ and the ``resonant'' angles
$\a_{r+1},\ldots,\a_\ell$ (also called ``slow'' or ``secular'') and
call $\Ba=(\Ba',\Bb)$ with $\Ba'\in\TTT^r$ and $\Bb\in
\TTT^s$. Likewise we distinguish the fast actions $\V
A'=(A_1,\ldots,A_r)$ and the resonant ones $A_{r+1},\ldots,A_{\ell}$
and set $\V A=(\V A',\V B)$ with $\V A'\in \RRR^r$ and $\V
B\in\RRR^s$.

Therefore the torus $\TT_{\V A_0}$, $\V A_0=(\V A'_0,\V B_0)$, is in
turn a {\it continuum of invariant tori} $\TT_{\V A_0,\Bb}$ with trivial
parametric equations: $\Bb$ fixed,  $\Ba'= \Bps$, $\Bps \in \TTT^r$, and
$\V A'=\V A'_0,\,\V B=\V B_0$. On each of them the motion is: $\V A',\V
B,\Bb$ constant and $\Ba'\to \Ba'+ \Bo\,t$, with rationally independent
$\Bo\in \RRR^r$.

Then the natural question is whether there exist functions $\V h_\e,\V
k_\e,\V H_\e,\V K_\e$ smooth in $\e$ near $\e=0$ and in
$\Bps\in\TTT^r$, vanishing for $\e=0$, and such that the torus
$\TT_{\V A_0,\Bb_0,\e}$ with parametric equations

$$\eqalign{
\V A'=&\V A'_0+\V H_\e(\Bps),\qquad \Ba'=\Bps+\V h_\e(\Bps),\cr
\V B=&\V B_0+\V K_\e(\Bps),\qquad
\Bb=\Bb_0+\V k_\e(\Bps),\cr}\qquad\Bps\in\TTT^r\Eq(15.2)$$
is invariant for the motions with Hamiltonian $\HH_\e(\V A,\Ba)=
\fra12{\V A'}^2+\fra12\V B^2+\e
f(\Ba',\Bb)$ and the motions on it are $\Bps\to \Bps+
\Bo\,t$. The above property, when satisfied, is summarized by saying
that the unperturbed resonant motions $\V A=(\V A'_0,\V B_0)$,
$\Ba=(\Ba'_0+\Bo' t,\Bb_0)$ can be {\it continued} in presence of
perturbation $\e f$, for small $\e$, to quasi periodic motions with the
same spectrum and on a slightly deformed torus $\TT_{\V
A'_0,\Bb_0,\e}$.
\*

\0{\it A priori unstable resonances}: here the question is whether the
{\it special} invariant tori continue to exist in presence of small
enough perturbations, of course slightly deformed. This means asking
whether, given $\V A_0$ such that $\Bo(\V A_0)=\V\dpr_{\V A}\HH_0(\V
A_0)$ has rationally independent components, there are functions $(\V
H_\e(\Bps),$ $\V h_\e(\Bps))$, $(\V P_\e(\Bps),\V Q_\e(\Bps))$ and
$(\BP_\e(\Bps), \BK_\e (\Bps))$ smooth in $\e$ near $\e=0$, vanishing
for $\e=0$, analytic in $\Bps\in\TTT^r$ and such that the
$r$-dimensional surface

$$\eqalign{
\V A=&\V A_0+\V H_\e(\Bps),\qquad \Ba=\Bps+\V h_\e(\Bps)\cr
\V p=&\V P_\e(\Bps),\kern1.7cm \V q=\V Q_\e(\Bps)\cr
\Bp=&\BP_\e(\Bps),\kern1.8cm \Bk=\BK_\e(\Bps)\cr}
\qquad \Bps\in \TTT^r\Eq(15.3)$$
is an invariant torus $\TT_{\V A_0,\e}$ on which the motion is
$\Bps\to\Bps+\Bo(\V A_0)\,t$. Again the above property is summarized
by saying that the unperturbed special resonant motions can be {\it
continued} in presence of perturbation $\e f$ for small $\e$ to quasi
periodic motions with the same spectrum and on a slightly deformed
torus $\TT_{\V A_0,\e}$.

Some answers to the above questions are presented in the following
section.
\*

\0{\it References:} [GBG04].
\*
\Section(16, Resonances and Lindstedt series)
\*

We discuss the equations \equ(15.2) in the paradigmatic case in which
the Hamiltonian $\HH_0(\V A)$ is $\fra12\V A^2$ (cf. \equ(14.1)).  It
will be $\Bo(\V A')\=\V A'$ so that $\V A_0=\Bo,\V B_0=\V0$ and the
perturbation $f(\Ba)$ can be considered as a function of $\Ba=(\Ba',\Bb)$: let
$\lis f(\Bb)$ be defined as its average over $\Ba'$. The determination
of the invariant torus of dimension $r$ which can be continued in the
sense discussed in Sect. \sec(15) is easily understood in this case.

A resonant invariant torus which, among the tori $\TT_{\V A_0,\Bb}$, has
parameric equations that can be continued as a {\it formal powers
series} in $\e$ is the torus $\TT_{\V A_0,\Bb_0}$ with $\Bb_0$ a
stationarity point for $\lis f(\Bb)$, \ie an {\it equilibrium point
for the average perturbation}: $\dpr_\Bb\lis f(\Bb_0)=0$. In fact the
following theorem holds

\kern3mm 
\0{\it If $\Bo\in \RRR^r$ satisfies a Diophantine property
and if $\Bb_0$ is a nondegenerate stationarity point for the ``fast
angle average'' $\lis f(\Bb)$ (\ie such that $\det \dpr^2_{\Bb\,\Bb}\lis
f(\Bb_0)\ne0$), then the equations for the functions $\V h_\e,\V
k_\e$:

$$\eqalign{
(\Bo\cdot\dpr_\Bps)^2 {\V h}_\e(\Bps)=&-\e\dpr_{\Ba'} f(\Bps+\V
h_\e(\Bps),\Bb_0+\V k_\e(\Bps))\cr
(\Bo\cdot\dpr_\Bps)^2 {\V k}_\e(\Bps)=&-\e\dpr_{\Bb} f(\Bps+\V
h_\e(\Bps)+\V k_\e(\Bps))\cr}\Eq(16.1)$$
can be formally solved in powers of $\e$.}

\kern3mm
Given the simplicity of the Hamiltonian that we are considering,
\ie \equ(14.1), it is not necessary to discuss the functions $\V H_\e,\V
K_\e$ because the equations that they should obey reduce to their
definitions as in the case of Sec. \sec(14), and for the same reason.

In other words also the resonant tori admit a Lindstedt series
representation.  {\it It is however very unlikely that the series are,
in general, convergent}. 

Physically this {\it new} aspect is due to the fact that the
linearization of the motion near the torus $\TT_{\V A_0,\Bb_0}$
introduces {\it oscillatory motions} around $\TT_{\V A'_0,\Bb_0}$ with
frequencies proportional to the square roots of the {\it positive}
eigenvalues of the matrix $\e\dpr^2_{\Bb\,\Bb}\lis f(\Bb_0)$:
therefore it is naively expected that it has to be necessary that a
Diophantine property be required on the vector $(\Bo,\sqrt{\e
\m_1},\ldots)$ where $\e\m_j$ are the positive eigenvalues. Hence some
values of $\e$, those for which $(\Bo,\sqrt{\e \m_1},\ldots)$ is not a
Diophantine vector or is too close to a non Diophantine vector, should be
excluded or at least should be expected to generate difficulties. Note
that the problem arises no matter what is supposed on the non
degenerate matrix $\dpr^2_{\Bb\,\Bb}\lis f(\Bb_0)$ since $\e$ can have
either sign; and no matter how $|\e|$ is supposed small. But we can
expect that if the matrix $\dpr^2_{\Bb\Bb}\lis f(\Bb_0)$ is (say)
positive definite (\ie $\Bb_0$ is a minimum point for $\lis f(\Bb)$)
then the problem should be easier for $\e<0$ and viceversa if $\Bb_0$
is a maximum it should be easier for $\e>0$ (\ie in the cases in which
the eigenvalues of $\e \dpr^2_{\Bb\Bb}\lis f(\Bb_0)$ are negative and
their roots do not have the interpretation of frequencies).

Technically the sums of the formal series can be given (so far) a
meaning only via summation rules involving divergent series: typically
one has to identify in the formal expressions (denumerably many)
geometric series which although divergent can be given a meaning by
applying the rule \equ(13.7).  Since the rule can only be applied if
$z\ne1$ this {\it leads to conditions on the parameter $\e$}, in order
to exclude that the various $z$ that have to be considered are too
close to $1$. Hence this stability result turns out to be {\it rather
different} from the KAM result for the maximal tori. Namely the
series can be given a meaning via summation rules provided $f$ and
$\Bb_0$ satisfy certain additional conditions and provided certain
values of $\e$ are excluded.  An example of a theorem is the
following: \*

\0{\it Given the Hamiltonian \equ(14.1) and a resonant torus $\TT_{\V
A'_0,\Bb_0}$ with $\Bo=\V A'_0\in\RRR^r$ satisfying a Diophantine
property let $\Bb_0$ be a non degenerate maximum point for the
average potential $\lis f(\Bb)\defi(2\p)^{-r}\ig_{\tttt^r}
f(\Ba',\Bb)d^{\,r}\Ba'$.  Consider the Lindstedt series solution for
the equations \equ(16.1) of the perturbed resonant torus with spectrum
$(\Bo,\V0)$. It is possible to express the single $n$-th order term of the
series as a sum of many terms and then rearrange the series thus obtained
so that the resummed series converges for $\e$ in a domain $\EE$
which contains a segment $[0,\e_0]$ and also contains a subset of
$[-\e_0,0]$ which, although with {\it open dense complement}, is so
large that it has $0$ as a Lebesgue density point. Furthermore the
resummed series for $\V h_\e,\V k_\e$ define an invariant $r$
dimensional analytic torus with spectrum $\Bo$.}
\*

More generally if $\Bb_0$ is only a nondegenerate stationarity point
for $\lis f(\Bb)$ the domain of definition of the resummed series is a
set $\EE\subset [-\e_0,\e_0]$ which on {\it both sides} of the origin has an
open dense complement although it has $0$ as a Lebesgue density point.

The above theorem can be naturally extended to the general case
in which the Hamiltonian is the most general perturbation of an
anisochronous integrable system $\HH_\e(\V A,\Ba)=h(\V A)+\e f (\V
A,\Ba)$ if $\dpr^2_{\V A \V A}h$ is a non singular matrix and the
resonance arises from a spectrum $\Bo(\V A_0)$ which has $r$
independent components (while the remaining are not necessarily
$\V0$).

We see that the convergence is a delicate problem for the
Lindstedt series for nearly integrable resonant motions. They might
even be divergent (mathematically a proof of divergence is an open
problem but it is a very resonable conjecture in view of the above
physical interpretation), {\it nevertheless} the above theorem shows
that sum rules can be given that ``sometimes'', \ie for $\e$ in a
large set near $\e=0$, yield a true solution
to the problem.

This is reminiscent of the phenomenon met in discussing
perturbations of isochro\-nous systems in
\equ(13.4), but it is a much more complex situation. And it leaves
many open problems: {\it foremost of them is the question of
uniqueness}. The sum rules of divergent series always contain some
arbitrary choices: which lead to doubt about the uniqueness of the
functions parameterizing the invariant tori constructed in this
way. It might even be that the convergence set $\EE$ may depend upon
the arbitrary choices, and that considering several of them no $\e$ with
$|\e|<\e_0$ is left out.

The case of \ap unstable systems has also been widely studied: in this
case too resonances with Diophantine $r$-dimensional spectrum $\Bo$
are considered. However in the case $s_2=0$ (called \ap unstable {\it
hyperbolic resonance}) the Lindstedt series can be shown to be
convergent while in the case $s_1=0$ (called \ap unstable {\it
elliptic resonance}) or in the {\it mixed} cases $s_1,s_2>0$ extra
conditions are needed. They involve $\Bo$ and
$\Bm=(\m_1,\ldots,\m_{s_2})$, cf. \equ(15.1), and properties of the
perturbations as well.  It is also possible to study a slightly
different problem: namely to look for conditions on $\Bo,\Bm,f$ which
imply that for small $\e$ invariant tori with spectrum $\e$-dependent
but {\it close}, in a suitable sense, to $\Bo$ exist.

The literature is vast but it seems fair to say that, given the above
comments, {\it particularly those concerning uniqueness and analyticity},
the situation is still quite unsatisfactory.
\*

\0{\it References:}  [GBG04].
%\ifnum\tipo=2\pagina\fi
\*

\Section(17, Diffusion in phase space)
\*

The KAM theorem implies that a perturbation of an analytic
anisochro\-nous integrable system, \ie with an analytic Hamiltonian
$\HH_\e(\V A,\Ba)=\HH_0(\V A)$ $+\e f(\V A,\Ba)$ and non degenerate
Hessian matrix $\dpr^2_{\V A\V A}h(\V A)$, generates large families of
maximal invariant tori. Such tori lie on the energy surfaces but do
not have codimension $1$ on them, \ie do not split the
$(2\ell-1)$--dimensional energy surfaces into disconnected
regions {\it except}, of course, in the case of $2$--degrees of
freedom systems, see Sect.\sec(14)..

Therefore there might exist trajectories with initial data close in
action space to $\V A^i$ which reach phase space points close in
action space to $\V A^f\ne \V A^i$ for $\e\ne0$, {\it no matter how
small}. Such {\it diffusion} phenomenon would occurr in spite of the
fact that the corresponding trajectory has to move in a space in which
very close to each $\{\V A\}\times \TTT^\ell$ there is an invariant
surface on which points move keeping $\V A$ constant within
$O(\e)$, which for $\e$ small can be $\ll|\V A^f-\V A^i|$.

In \ap unstable systems (cf. Sect. 15) with $s_1=1,s_2=0$ it is not
difficult to see that the corresponding phenomenon can actually
happen: the paradigmatic example ({\cs Arnold}) is the \ap unstable
system
$$\HH_\e=\fra{A_1^2}2+A_2+\fra{p^2}2+ g\, (\cos q-1)+\e\, (\cos
\a_1+\sin\a_2)(\cos q-1)
\Eq(17.1)$$
This is a system describing a motion of a ``pendulum'' ($(p,q)$
coordinates) interacting with a ``rotating wheel'' ($(A_1,\a_1)$
coordinates) and a ``clock'' ($(A_2,\a_2)$ coordinates) \ap unstable
near the points $p=0,q=0,2\p$, ($s_1=1,s_2=0$, $\l_1=\sqrt{g}$,
cf. \equ(15.1)). And it can be proved that on the energy surface of
energy $E$ and for each $\e\ne0$ small enough (no matter how small)
{\it there are initial data with action coordinates close to $\V
A^i=(A^i_1,A^i_2)$ with $\fra12 {A^{i\,2}_1}+ {A^i_2}$ close to $E$
eventually evolving to a datum $\V A'=(A'_1,A'_2)$} with $A'_1$ at
distance from $A^f_1$ smaller than an {\it arbitrarily} prefixed
distance (of course with energy $E$). {\it Furthermore} during the
whole process the pendulum energy stays close to $0$ within $o(\e)$
(\ie the pendulum swings following closely the unperturbed
separatrices).

In other words \equ(17.1) describes a machine (the pendulum) which,
working approximately in a cycle, extracts energy from a reservoir
(the clock) to transfer it to a mechanical device (the wheel).  The
statement that diffusion is possible means that the machine can work
as soon as $\e\ne0$, if the initial actions and the initial phases
(\ie $\a_1,\a_2,p,q$) are suitably tuned (as functions of $\e$).

The peculiarity of the system \equ(17.1) is that the unperturbed
pendulum fixed points $P_\pm$ (\ie the equilibria $p=0,q=0,2\p$) {\it
remain unstable equilibria even when $\e\ne0$}: and this is an
important simplifying feature.

\eqfig{240pt}{100pt}{}{fig6}{(6)}

\kern-4truemm
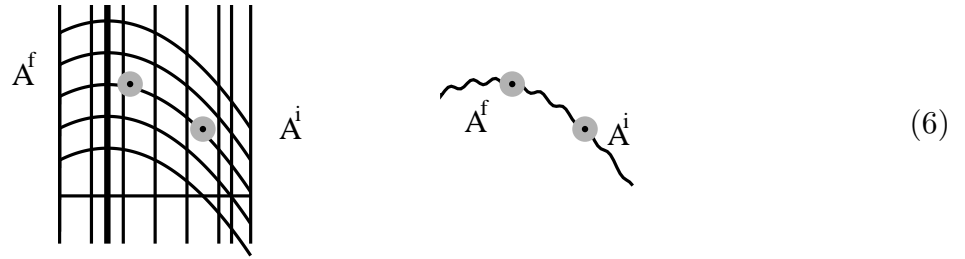
\0Fig. 6: {\nota The first drawing represents the $\e=0$ geometry: the
``partial energy'' lines are parabolae, $\fra12 A_1^2+A_2=const$. The
vertical lines are the resonances $A_1=rational$ (\ie $\n_1
A_1+\n_2=0$). The disks are neighborhoods of the points $\V A^i$ and
$\V A^f$ (the dots at their centers). The second drawing ($\e\ne0$) is
an artist rendering of a trajectory in $\V A$ space, driven by the
pendulum swings to accelerate the wheel from $A^i_1$ to $A^f_1$ at the
expenses of the clock energy, sneaking through invariant tori (not
represented and approximately) located ``away'' from the intersections
between resonances and partial energy lines (a dense set,
however). The pendulum coordinates are not shown: its energy stays
close to $0$ within a power of $\e$. Hence the pendulum swings staying
close to the separatrix. The oscillations symbolize the wiggly
behavior of the partial energy $\fra12 A_1^2+A_2$ in the process of
sneaking between invariant tori which, because of their invariance,
would be impossible without the pendulum. The energy $\fra12A^2_1$ of
the wheel increases slightly at each pendulum swing: accurate
estimates yield an increase of the wheel speed $A_1$ of the order of
$\e/\hbox{\nota log}\, \e^{-1}$ at each swing of the pendulum implying
a transition time of the order of $g^{-\fra12}\e^{-1}\hbox{\nota
log}\,\e^{-1}$.\vfil}

It permits bypassing the obstacle, arising in the analysis
of more general cases, represented by the resonance surfaces
consisting in the $\V A$'s with $A_1\n_1+\n_2=0$: the latter
correspond to harmonics $(\n_1,\n_2)$ present in the perturbing
function, \ie the harmonics which would lead to division by zero in an
attempt to construct (as necessary in the proof by Arnold's method) the
parametric equations of the perturbed invariant tori with action close
such $\V A$'s. In the case of \equ(17.1) the problem arises only on
the resonance marked in Fig.6 by a heavy line \ie $A_1=0$ corresponding to
$\cos \a_1$ in \equ(17.1).

If $\e=0$ the points $P_-$ with $p=0,q=0$ and the point $P_+$ with
$p=0,q=2\p$ are both unstable equilibria (and they are of course the
same point, if $q$ is an angular variable). The unstable manifold
(it is a curve) of $P_+$ coincides with the stable manifold of $P_-$
and viceversa. So that the unperturbed system admits non trivial
motions leading from $P_+$ to $P_-$ and from $P_-$ to $P_+$, both in a
biinfinite time interval $(-\io,\io)$: the $p,q$ variables describe a
pendulum and $P_\pm$ are its unstable equilibria which are connected
by the separatrices (which constitute the $0$-energy surfaces for the
pendulum).

The latter property remains true for more general \ap unstable
Hamiltonians 

$$\HH_\e=\HH_0(\V A)+\HH_u(p,q)+\e\, f(\V A,\Ba,p,q), \qquad {\rm in}\
(U\times\TTT^\ell)\times(\RRR^2)\Eq(17.2)$$
where $\HH_u$ is a one dimensional Hamiltonian which has two
unstable equilibrium points $P_+$ and $P_-$
linearly repulsive in one direction and linearly attractive in
another which are connected by two {\it heteroclinic} trajectories
which, as time tends to $\pm\io$, approach $P_-$ and $P_+$ and
viceversa.

Actually the points need not be different but, if coinciding, the
trajectories linking them must be nontrivial: in the case \equ(17.1)
the variable $q$ can be considered an angle and then $P_+$ and $P_-$
would coincide (but are connected by nontrivial trajectories, \ie by
trajectories that visit points different from $P_\pm$). Such
trajectories are called {\it heteroclinic} if $P_+\ne P_-$ and {\it
homoclinc} if $P_+=P_-$.

In the general case besides the homoclinicity (or heteroclinicity)
condition certain weak genericity conditions, automatically satisfied
in the example \equ(17.1), have to be imposed in order to show that
given $\V A^i$ and $\V A^f$ with the same unperturbed energy $E$ one
can find, for all $\e$ small enough but not equal to $0$, initial data
($\e$-dependent) with actions arbitrarily close to $\V A^i$ which
evolve to data with actions arbitrarily close to $\V A^f$. This is a
phenomenon called {\it Arnold diffusion}.  Simple sufficient
conditions for a transition from near $\V A^i$ to near $\V A^f$ are
expressed by the following result

\vskip2mm
\0{\it Given the Hamiltonian \equ(17.2) with $\HH_u$ admitting two
hyperbolic fixed points $P_\pm$ with heteroclinic connections, $t\to
(p_a(t),q_a(t)),\,a=1,2$, suppose that
\\
\0(1) On the unperturbed energy surface of energy $E=\HH(\V
A^i)+\HH_u(P_\pm)$ there is a regular curve $\g\,:\,s\to \V A(s)$
joining $\V A^i$ to $\V A^f$ such that the unperturbed tori $\{\V
A(s)\}\times \TTT^\ell$ can be continued at $\e\ne0$ into invariant
tori $\TT_{\V A(s),\e}$ for a set of values
of $s$ which fills the curve $\g$ leaving only gaps of size of order
$o(\e)$.

\0(2) The $\ell\times\ell$ matrix $D_{ij}$ of the second derivatives of the
integral of $f$ over the heteroclinic motions is not degenerate, \ie

\vskip-1.5mm
$$|\det D|=\big|\det \,\Big(\ig_{-\io}^\io dt \, \dpr_{\a_i\a_j} f(\V A,\Ba
+\Bo(\V A)\,t, p_a(t),q_a(t))\Big)\big|>c>0\Eq(17.3)$$
\vskip-1mm
\0for all $\V A$'s on the curve $\g$ and all $\Ba\in\TTT^2$. 
\\ 
Given arbitrarily $\r>0$, for $\e\ne0$ small enough there are initial
data with action and energy closer than $\r$ to $\V A^i$ and $E$,
respectively, which after a long enough time acquire an action closer
than $\r$ to $\V A^f$ (keeping the initial energy).}  
\vskip2mm

The above two conditions can be shown to hold generically for many
pairs $\V A^i\ne \V A^f$ (and many choices of the curves $\g$
connecting them) if the number of degree of freedom is $\ge3$.
Thus the result, obtained by a simple extension of the argument
originally outlined by Arnold to discuss the paradigmatic example
\equ(17.1), proves the existence of diffusion in {\it a priori unstable}
system. The integral in \equ(17.3) is called {\it Melnikov integral}.

The real difficulty is to estimate the time needed for the transition:
it is a time that obviously has to diverge as $\e\to0$. Assuming $g$
fixed (\ie $\e$--independent) a naive approach easily leads to
estimates which can even be worse than $O(e^{a\e^{-b}})$ with some
$a,b>0$.  It has finally been shown that in such cases the minimum
time can be, for rather general perturbations $\e f(\Ba,q)$, estimated
above by $O(\e^{-1}\log\e^{-1})$, which is the best that can be hoped
for under generic assumptions.  \*

\0{\it References:} [Ar68], [CV00].
\*

\Section(18, Long time stability of quasi periodic motions)
\*

A more difficult problem is whether the same phenomenon of migration
in action space occurs in \ap stable systems. The root of the
difficulty is a remarkable stability property of quasi periodic
motions. Consider Hamiltonians $\HH_\e(\V A,\Ba)=
h(\V A)+\e f(\V A,\Ba)$ with $\HH_0(\V
A)=h(\V A)$ {\it strictly convex, analytic and  anisochronous} on the
closure $\lis U$ of an open {\it bounded} region
$U\subset\RRR^\ell$, and a perturbation $\e f(\V A,\Ba)$ analytic in 
$\lis U\times\TTT^\ell$.

Then \ap bounds are available on how long it can possibly take to
migrate from an action close to $\V A_1$ to one close to $\V A_2$: and
the bound is of ``exponential type'' as $\e\to0$ (\ie it admits a
lower bound which behaves as the exponential of an inverse power of
$\e$). The simplest theorem is ({\cs Nekhorossev}):
\vskip2mm

\0{\it There are constants $0<a,b,d,g,\t$ such that any
initial datum $(\V A,\Ba)$ evolves so that the $\V A$ will not change by more
than $a\e^{g}$ before a long time bounded below by $\t e^{b\e^{-d}}$.}
\vskip2mm

Thus this puts an exponential bound, \ie a bound exponential in an
inverse power of $\e$, to the diffusion time: {\it before a time $\t
e^{b\e^{-d}}$ actions can only change by $O(\e^{g})$} so that their
variation cannot be large no matter how small $\e\ne0$ is
chosen. This places a (long) lower bound to the time of diffusion in
\ap stable systems. 

The proof of the theorem provides, actually, an interesting and
detailed picture of the actions variations showing that some actions
ay vary slower than others.  

The theorem is constructive, \ie all constants
$0<a,b,d,\t$ can be explicitly chosen and depend on
$\ell,\HH_0,f$ although some of them can be fixed to depend only on
$\ell$ and on the minimum curvature of the convex graph of
$\HH_0$. Its proof can be adapted to cover many cases which do not
fall in the class of systems with strictly convex unperturbed
Hamiltonian. And even to cases with a resonant unperturbed
Hamiltonian. 

However in important problems, \eg in the $3$ body problems met in
Celestial Mechanics, there is empirical evidence that diffusion takes
place at a fast pace (\ie not exponentialy slow in the above sense)
while the above results would forbid a rapid migration in phase space
{\it if they applied}: however in such problems the assumptions of the
theorem are not satisfied, because the unperturbed system is strongly
resonant (as in the celestial mechanics problems where the number of
independent frequencies is a fraction of the number of degrees of
freedom and $h(\V A)$ is far from strictly convex), leaving wide open
the possibility of observing rapid diffusion.

And changing the assumptions can dramatically change the results. For
instance rapid diffusion can sometimes be proved even though it might
be feared that it should require exponentially long times:
an example that has been proposed is the case of a {\it three
time scales system}, with Hamiltonian

$$\o_1 A_1+\o_2 A_2+\fra{p^2}2+g (1+\cos q)+ \e f(\a_1,\a_2,p,q)\Eq(18.1)$$
with $\Bo_\e\defi (\o_1,\o_2)$ with $
\o_1=\e^{-\fra12} \lis\o,\,\o_2=\e^{\fra12}\widetilde\o$ and
$\lis\o, \widetilde\o>0$ constants. The three scales are
$\o_1^{-1},\sqrt{g^{-1}},\o_2^{-1}$. In this case there are many
(although by no means all) pairs $\V A_1,\V A_2$ which can be
connected within a time that can be estimated to be of order
$O(\e^{-1}\log \e^{-1})$.

This is a rapid diffusion case in a \ap unstable system in which
condition \equ(17.3) is {\it not} satisfied: because
the $\e$--dependence of
$\Bo(\V A)$ implies that the lower bound $c$ in \equ(17.3) must depend
on $\e$ (and be exponentiallly small with an inverse power of $\e$ as
$\e\to0$).

The unperturbed system in \equ(18.1) is non resonant in the part
$\HH_0$ for $\e>0$ outside a set of zero measure (\ie where the vector
$\Bo_\e$ satisfies a suitable Diophantine property) and, furthermore,
it is \ap unstable: cases met in applications can be \ap stable and
resonant (and often not anisochronous) in the part $\HH_0$. And in
such system not only the speed of diffusion is not understood but
proposals to prove its existence, if present (as expected), have so
far not given really satisfactory results.  
\*

\0{\it References:} [Ne77].
\*

\Section(19, The three bodies problem)
\*

Mechanics and the three bodies problem can be almost identified in the
sense that the motion of three gravitating masses has been since the
beginning a key astronomical problem and at the same time the source
of inspiration for many techniques: foremost among them the theory of
perturbations.

As an introduction consider a special case. Let three masses
$m_S=m_0,m_J=m_1,m_M=m_2$ interact via gravity, \ie with interaction
potential $-k m_i m_j |\V x_i-\V x_j|^{-1}$: the simplest problem arises
when the third body has a neglegible mass compared to the two others
and the latter are supposed to be on a circular orbit; furthermore the mass
$m_J$ is $\e m_S$ with $\e$ small and the mass $m_M$ moves in the plane of
the circular orbit. This will be called the {\it circular
restricted three body problem}.

In a reference system with center $S$ and rotating at the angular
speed of $J$ around $S$ inertial forces (centrifugal and Coriolis')
act. Supposing that the body $J$ is located on the axis with unit
vector $\V i$ at distance $R$ from the origin $S$, the acceleration of
the point $M$ is $\ddot\Br={\V F}+\o^2_0(\Br-\fra{\e R}{1+\e}\V
i)-2\Bo_0\wedge\dot\Br$ if $\V F$ is the force of attraction and
$\Bo_0\wedge\dot\Br\=\o_0\dot\Br^\perp$ where $\Bo_0$ is a vector with
$|\Bo_0|=\o_0$ and perpendicular to the orbital plane and 
$\Br^\perp\defi (-\r_2,\r_1)$
if $\Br=(\r_1,\r_2)$. Here, taking into account that the origin $S$
rotates around the fixed center of mass, $\o^2_0(\Br-\fra{\e
  R}{1+\e}\V i)$ is the centrifugal force while $-2\Bo_0\wedge\dot\Br$ is
the Coriolis force.  The equations of motion can therefore be derived
from a Lagrangian

$$\LL=\fra12\dot{\Br}^2-W
+ \o_0\Br^\perp\cdot\dot{\Br}+\fra12\o_0^2{\Br}^2-
\o_0^2\fra{\e \,R}{1+\e}\Br\cdot\V i\Eq(19.1)$$
where $\o_0^2R^3=k m_S (1+\e)\defi g_0$ and $W=-\fra{k m_S}{|\Br|}
-\fra{k\,m_S\,\e}{|\Br-R\V i|}$ if $k$ is the gravitational constant, $R$
the distance between $S$ and $J$ and finally the last three terms come from the
Coriolis force (the first) and from the centripetal force (the other
two, taking into account that the origin $S$ rotates around the fixed
center of mass).

Setting $g=g_0/(1+\e)\= k m_S$, the Hamiltonian of
the system is

$$\txt\HH=\fra12(\V p-\o_0 \Br^\perp)^2
-\fra{g}{|\Br|}-\fra12\o_0^2\Br^2-\e\fra{g}{R}
\big({|\fra\Br{R}-\V i|^{-1}}
-\fra{\Br}R\cdot\V i\big)\Eq(19.2)
$$

The first part can be expressed immediately in the action--angle
coordinates for the two body problem, cf. Sect. \sec(8). Calling such
coordinates $(L_0,\l_0,G_0,\g_0)$ and $\th_0$ the polar angle of $M$
with respect to the ellipse major axis and $\l_0$ the mean anomaly of
$M$ on its ellipse, the Hamiltonian becomes, taking into account that
for $\e=0$ the ellipse axis rotates at speed $-\o_0$,

$$\txt
\HH=-\fra{g^2}{2L_0^2}-\o_0G_0 -\e\fra{g}{R}
\big({|\fra\Br{R}-\V i|^{-1}}
-\fra{\Br}R\cdot\V i\big)\Eq(19.3)$$
which is convenient if we study the {\it interior problem}, \ie $|\Br|<
R$. This can be expressed in the action--angle coordinates via
\equ(8.1), \equ(8.2):

$$\eqalign{\txt \th_0&\txt=\l_0+f_{\l_0},\kern2cm \th_0+\g_0=\l_0+\g_0+
f_{\l_0},\cr
\txt e&\txt=\big(1-\fra{G^2_0}{L^2_0}\big)^{\fra12},\kern1cm\fra{|\Br|}{R}=
\fra{ G_0^2}{gR}\fra1{1+e\cos(\l_0+f_{\l_0})},\cr}\Eq(19.4)$$
where, see \equ(8.2), $f_\l=(f(e \sin\l, e\cos\l)$ and
$f(x,y)=2x(1+\fra54y+\ldots)$ with the $\ldots$ denoting higher orders
in $x,y$ even in $x$. The Hamiltonian takes the form, if $\o^2=g R^{-3}$,

$$\HH_\e=-\fra{g^2}{2 L_0^2}-\o G_0+ \e\fra{g}{R}
F(G_0,L_0,\l_0,\l_0+\g_0)\Eq(19.5)$$
where the only important feature (for our purposes) is that
$F(L,G,\a,\b)$ is an {\it analytic function} of $L,G,\a,\b$ near a datum
with $|G|<L$ (\ie $e>0$) and $|\Br|< R$. {\it However} the domain of
analyticity in $G$ is rather small as it is constrained by $|G|<L$
excluding in particular the circular orbit case $G=\pm L$.

Note that apparently the KAM theorem fails to be applicable to
\equ(19.5) because the matrix of the second derivatives of
$\HH_0(L,G)$ has $0$ determinant. Nevertheless the proof of the
theorem goes through also in this case, with minor changes. This can
be checked by studying the proof or, following a remark by {\cs
Poincar\'e}, by simply remarking that the ``squared'' Hamiltonian
$\HH'_\e\defi (\HH_\e)^2$ has the form

$$\HH'_\e=(-\fra{g^2}{2 L_0^2}-\o\, G_0)^2+\e
F'(G_0,L_0,\l_0,\l_0+\g_0)\Eq(19.6)$$
with $F'$ still analytic. {\it But} this time
$\det\fra{\dpr^2\HH'_0}{\dpr(G_0,L_0)}=-6 g^2 L_0^{-4}\o_0^2 h\ne0$
if $h=-g^2\,L_0^{-2}-2\o\, G_0\ne0$.

Therefore {\it the KAM theorem applies to $\HH'_\e$} and the key
observation is that the orbits generated by the Hamiltonian
$(\HH_\e)^2$ are {\it geometrically the same} as those generated by the
Hamiltonian $\HH_\e$: they are  only run at a different speed because
of the need of a time rescaling by the {\it constant} factor $2\HH_\e$.

This shows that, given an unperturbed ellipse of parameters
$(L_0,G_0)$ such that $\Bo=(\fra{g^2}{L_0^3},-\o)$, $G_0>0$, with
$\o_1/\o_2$ Diophantine, then the perturbed system admits a motion
which is quasi periodic with spectrum proportional to $\Bo$ and takes
place on an orbit which wraps around a torus remaining {\it forever
close} to the unperturbed torus (which can be visualized as described
by a point moving, according to the area law on an ellipse rotating at
rate $-\o_0$) with actions $(L_0,G_0)$, provided $\e$ is small enough.
Hence
\*

\0{\it The KAM theorem answers, at least conceptually, the
classical question: can a solution of the three body problem remain
close to an unperturbed one forever? \ie is it possible that a solar
system is stable forever?}  \*

Assuming $e, |\Br|/R\ll 1$ and retaining only the lowest orders in
$e$ and $|\Br|/R\ll 1$ the Hamiltonian \equ(19.5) simplifies into

\vskip-3mm
$$\eqalign{
\txt\HH=&\txt-\fra{g^2}{2L_0^2}-
\o G_0 +\d_\e(G_0)-\fra{\e g}{2R}
\fra{G_0^4}{g^2R^2} \big(\txt3 \cos2(\l_0+\g_0)-\cr
&\txt- e\,
\cos\l_0-\fra92 e \cos(\l_0+2\g_0)+\fra32 e \cos(3\l_0+2\g_0)
 %%%???-3e\sin\l_0\sin 2(\l_0+\g_0)
\big)\cr}\Eq(19.7)$$
where $\d_\e(G_0)=-((1+\e)^{\fra12}-1)\,\o \,G_0-\fra{\e\,g}{2R}
\fra{G_0^4}{g^2 R^2}$ and $e=(1-G_0^2/L_0^2)^{\fra12}$.

It is an interesting exercise to estimate, assuming as model the
\equ(19.7) and following the proof of the KAM theorem, how small has
$\e$ to be if a planet with the data of Mercury can be stable forever
on a (slowly precessing) orbit with actions close to the present day
values under the influence of a mass $\e$ times the solar mass
orbiting on a circle, at a distance from the Sun equal to that of
Jupiter. It is possible to follow either the above reduction to the
ordinary KAM theorem or to apply directly to \equ(19.7) the Lindstedt
series expansion, proceeding along the lines of Sect. \sec(14). The
first approach is easy but the second is more efficient: unless
the estimates are done in a particularly careful manner the
value found for $\e m_S$ does not have astronomical interest.
 \*
\0{\it References:} [Ar68]. \*

\Section(20, Rationalization and regularization of singularities)
\*

Often integrable systems have interesting data which lie on the
boundary of the integrability domain. For instance the central motion
when $L=G$ (circular orbits) or the rigid body in a rotation around
one of the principal axes or the two body problem when $G=0$
(collisional data). In such cases perturbation theory cannot be
applied as discussed above. Typically the perturbation depends on
quantities like $\sqrt{L-G}$ and is {\it not analytic} at
$L=G$. Nevertheless it is sometimes possible to enlarge phase space
and introduce new coordinates in the vicinity of the data which in the
initial phase space are singular.

A notable example is the failure of the analysis of the circular
restricted three body problem: it apparently fails when the orbit that
we want to perturb is circular.

It is convenient to introduce the canonical coordinates $L,\l$ and
$G,\g$

$$L=L_0,\ G=L_0-G_0,\ \l=\l_0+\g_0,\ \g=-\g_0\Eq(20.1)$$
so that $e=\sqrt{2G L^{-1}}\sqrt{1-G(2L)^{-1}}$ and $\l_0=\l+\g$ and
$\th_0=\l_0+f_{\l_0}$ where $f_{\l_0}$ is defined in \equ(8.2) (see
also \equ(19.4)). Hence

\kern-3mm
$$\eqalign{\txt \th_0&\txt=\l+\g+f_{\l+\g},\qquad \th_0+\g_0=\l+
f_{\l+\g},\cr
\txt e=&\txt\sqrt{2G}
\sqrt{\fra{1}{L}\big(1-\fra{G}{2L}\big)},\qquad
\fra{|\Br|}{R}=
\fra{L^2 (1-e^2)}{gR}\fra1{1+e\cos(\l+\g+f_{\l+\g})},\cr}\Eq(20.2) $$
and the Hamiltonian \equ(19.7) takes the form

$$\HH_\e=-\fra{g^2}{2 L^2}-\o L+\o G+\e\fra{g}{R}
F(L-G,L,\l+\g,\l)\Eq(20.3)$$
In the coordinates $L,G$ of \equ(20.1) the unperturbed circular case
corresponds to $G=0$ and the \equ(19.3) once expressed in the
action--angle variables $G,L,\g,\l$ is analytic in a domain whose size
is controlled by $\sqrt{G}$. Nevertheless very often problems of
perturbation theory can be ``{\it regularized}''.

This is done by ``enlarging the integrability'' domain by
adding to it points (one or more) around the singularity (a boundary
point of the domain of the coordinates)
and introduce new coordinates to describe simultaneously the data
close to the singularity and the newly added points: in many
interesting cases the equations of motion are no longer singular, \ie
become analytic, in the new coordinates and therefore apt to describe
the motions that reach the singularity in a finite time. {\it One can
say that the singul;arity was only apparent}.

Perhaps this is best illustrated precisely in the above circular
restricted three body problem. There the singularity is where $G=0$,
\ie at a circular unperturbed orbit. If we describe the points with
$G$ small in a new system of coordinates obtained from the one in
\equ(20.1) by letting alone $L, \l$ and setting

\vskip-2mm
$$p=\sqrt{2G}\cos\g,\qquad q=\sqrt{2G}\sin\g\Eq(20.4)$$
then $p,q$ vary in a neighborhood of the origin
with the origin itself {\it excluded}.

Adding the origin of the $p,q$ plane then in a full neighborhood of
the origin the Hamiltonian \equ(19.3) is {\it analytic} in
$L,\l,p,q$. This is because it is analytic,
cf. \equ(19.3),\equ(19.4), as a function of $L,\l$ and $e\cos\th_0$
and of $\cos(\l_0+\th_0)$. 
Since $\th_0=\l+\g+f_{\l+\g}$ and $\th_0+\l_0=\l+ f_{\l+\g}$ by
\equ(19.4), the Hamiltonian \equ(19.3)
is analytic in $L,\l, e\cos(\l+\g+f_{\l+\g}),\,
\cos(\l+f_{\l+\g})$ for $e$ small (\ie for $G$ small) and, by
\equ(8.2), $f_{\l+\g}$ is analytic in $e\sin(\l+\g)$ and $e\cos(\l+\g)$.
Hence the trigonometric identities

\vskip-3mm
$$
\txt e\sin (\l+\g)\,=
\,\fra
{p \sin\l+q\cos\l}{\sqrt L}\,\sqrt{1-\fra{G}{2L}},
\qquad
e\cos(\l+\g)\,=
\,\fra
{p \cos\l-q\sin\l}{\sqrt L}\,\sqrt{1-\fra{G}{2L}} 
\Eq(20.5)$$
together with $G=\fra12(p^2+q^2)$ imply that \equ(20.3) is analytic
near $p=q=0$ and $L>0,\l\in[0,2\p]$. The Hamiltonian becomes
analytic and the new coordinates are suitable to describe motions
crossing the origin: \eg setting
$C\defi\fra12(1-\fra{p^2+q^2}{4L})\,L^{-\fra12}$
\equ(19.7) becomes

\vskip-3truemm
$$\eqalign{
\txt\HH=&\txt-\fra{g^2}{2L^2}-\o L+\o
  \fra12(p^2+q^2)+\d_\e(\fra12(p^2+q^2))-\fra{\e g}{2R}
\fra{(L-\fra12(p^2+q^2))^4}{g^2 R^2}\cdot\cr
&\txt\cdot
\Big(3\cos2\l
-\big( (-11\cos\l +3\cos3\l)\,p-\,(7\sin\l+3\sin3\l)\,q\big)\, C\Big)
\cr}\Eq(20.6)$$

The KAM theorem does not apply in the form discussed above to
``cartesian coordinates'' \ie when, as in \equ(20.6), the unperturbed
system is not assigned in action--angle variables: however there are
versions of the theorem (actually corollaries of it) which do apply
and therefore it becomes possible to obtain some results even for the
perturbations of circular motions by the techniques that have been
illustrated here.  

\vskip1mm
Likewise the Hamiltonian of the rigid body with a fixed point $O$ and
subject to analytic external forces becomes singular, if expressed in
the action--angle coordinates of Deprit, when the body motion nears a
rotation around a principal axis or more generally nears a
configuration in which any two of the axes $\V i_3$, $\V z$, $\V z_0$
coincide (\ie any two among the principal axis, the angular momentum
axis and the inertial $z$-axis coincide, see Section
\sec(9)). Nevertheless by imitating the procedure just described 
in the simpler cases of the circular three body problem, it is possible
to enlarge phase space so that in the new coordinates the Hamiltonian
is analytic near the singular configurations.

\vskip1mm
A regularization also arises when considering collisional orbits in
the unrestricted planar three body problem. In this respect a very
remarkable result is the regularization of collisional orbits in the
planar three body problem. After proving that if the total angular
momentum does not vanish simoultaneous collisions of the three masses
cannot happen within any finite time interval the question is reduced
to the regularization of two bodies collisions, under the assumption
that the total angular momentum does not vanish.

The {\it local} change of coordinates which changes the relative
position coordinates $(x,y)$ of two colliding bodies as
$(x,y)\to(\x,\h)$ with $x+i y=(\x+i\h)^2$ is not one to one, hence it
has to be regarded as an enlargement of the positions space, if points
with different $(\x,\h)$ are considered different. However the
equations of motion written in the variables $\x,\h$ have no
singularity at $\x,\h=0$, ({\cs Levi-Civita}).
\vskip1mm

Another celebrated regularization is the regularization of the
{\it Schwar\-tz\-schild metric}, \ie of the general relativity version of the
two body problem: it is however somewhat out of the scope of this
review ({\cs Synge, Kruskal}).
\*

\0{\it References:} [LC].

\*
\appendix(A, KAM resummation scheme)
\*

The idea to control the ``remaining contributions'' is to reduce the
problem to the case in which there are no pairs of lines that follow
each other in the tree order and which have the {\it same current}.
Mark by a {\it scale label} ``$0$'' the lines of a tree whose divisors
are $>1$: these are lines which give no problems in the
estimates. Then mark by a scale label ``$\ge1$'' the lines with 
current $\Bn(l)$ such that
$|\Bo_0\cdot\Bn(l)|\le 2^{-n+1}$ for $n=1$ (\ie the remaining lines).

The lines labeled $0$ are said to be {\it on scale $0$}, while those
labeled $\ge1$ are said to be {\it on scale $\ge1$}. A {\it cluster of
scale $0$} will be a {\it maximal} collection of lines of scale $0$
forming a connected subgraph of a tree $\th$.

Consider only trees $\th_0\in \Th_0$ of the family $\Th_0$ of trees
{\it containing no clusters of lines with scale label $0$ which have
only one line entering the cluster and one exiting it with equal
current}.

It is useful to introduce the notion of a line $\ell_1$ situated
``{\it between}'' two lines $\ell,\ell'$ with $\ell'>\ell$: this will mean
that $\ell_1$ precedes $\ell'$ but not $\ell$.

All trees $\th$ in which there are some pairs $l'> l$ of consecutive
lines of scale label $\ge1$ which have equal current and such that all
lines between them bear scale label $0$ are obtained by ``inserting''
on the lines of trees in $\Th_0$ with label $\ge1$ any number of
clusters of lines and nodes, with lines of scale $0$ and with the
property that the sum of the harmonics of the nodes inserted {\it
vanishes}.

Consider a line $l_0\in \th_0\in\Th_0$ linking nodes $v_1<v_2$ and
labeled $\ge1$ and imagine inserting on it a cluster $\g$ of lines of
scale $0$ {\it with sum of the node harmonics vanishing and out of
which emerges one line connecting a node $v_{out}$ in $\g$ to $v_2$
and into which enters one line linking $v_1$ to a node
$v_{in}\in\g$}. The insertion of a $k$--lines, $|\g|=(k+1)$-nodes,
cluster changes the tree value by replacing the line factor, that will
be briefly called ``value of the cluster $\g$,

$$\fra{\Bn_{v_1}\cdot
\Bn_{v_2}}{\Bo_0\cdot\Bn(l_0)^2}\ \to\
\fra{(\Bn_{v_1}\cdot M(\g;\Bn(l_0))\,
\Bn_{v_2})}{\Bo_0\cdot\Bn(l_0)^2}
\fra1{\Bo_0\cdot\Bn(l_0)^2}\Eqa(A1.1)$$
where $M$ is a $\ell\times \ell$ matrix
$M_{rs}(\g,\Bn(l_0))=\fra{ \e^{|\g|}}{k!}\n_{out,r}\n_{in,s}
\prod_{v\in\g} (-f_{\Bn_v}) \prod_{l\in\g} \fra{\Bn_v\cdot
\Bn_{v'}}{\Bo_0\cdot\Bn(l)^2}$ if $\ell=v'v$ denotes a line linking
$v'$ and $v$. Therefore if all possible connected clusters are
inserted and the resulting values are added up the result can be taken
into account by attributing to the original line $l_0$ a factor like
\equ(A1.1) with $M^{(0)}(\Bn(l_0))\defi \sum_\g 
M(\g;\Bn(l_0))$ replacing $M(\g;\Bn(l_0))$.

If several connected clusters $\g$ are inserted on the same line and
their values are summed the result is a modification of the factor
associated with the line $l_0$ into

\vskip-2mm
$$\txt\sum_{k=0}^\io
\Bn_{v_1}\cdot\Big(\fra{M^{(0)}(\Bn(l_0))}{\Bo_0\cdot\Bn(l_0)^2}
\Big)^k\Bn_{v_2}
\fra1{\Bo_0\cdot\Bn(l_0)^2}=
(\Bn_{v_1}\cdot
\fra1{\Bo_0\cdot\Bn(l_0)^2 - M^{(0)}(\Bn(l_0))}\Bn_{v_2})\Eqa(A1.2)$$
The series defining $M^{(0)}$ involves, {\it by construction}, only
trees with lines of scale $0$, hence with large divisors so that it
converges to a matrix of small size of order $\e$ (actually $\e^2$,
looking more carefully) if $\e$ is small enough.

Convergence {\it can} be established by simply remarking that the
series defining $M^{(1)}$ is built with lines with values $>\fra12$ of
the propagator, so that it certainly converges for $\e$ small enough
(by the estimates in Section \sec(12) where the propagators were
identically $1$) and the sum is of order $\e$ (actually $\e^2$), hence
$<1$. {\it However} such an argument cannot be repeated when dealing
with lines with smaller propagators (which still have to be
discussed). Therefore a method not relying on so trivial a remark on
the size of the propagators has eventually to be used when considering
lines of scale higher than $1$, as it will soon become necessary.

The advantage of the collection of terms achieved with \equ(A1.2) is
that we can represent $\V h$ as a sum of values of trees which are
{\it simpler} because they contain no pair of lines of scale $\ge1$
with in between lines of scale $0$ with total sum of the node
harmonics vanishing. The price is that the divisors are now more
involved and we even have a problem  due to the fact that we have not
proved that the series in \equ(A1.2) converges. In fact it is a
geometric series whose value is the \rhs of \equ(A1.2) obtained by the
sum rule \equ(13.7) {\it unless we can prove that the ratio of the
geometric series is $<1$}. This is trivial in this case by the
previous remark: but it is better to remark that there is anothr
reason for convergence, whose use is not really necesary here but it
will become essential later.

The property that the ratio of the geometric series is $<1$ can be
regarded as due to the consequence of the {\it cancellation} mentioned
in Section \sec(14) which can be shown to imply that the ratio is $<1$
because $M^{(0)}(\Bn)=\e^2 \,(\Bo_0\cdot\Bn)^2 m^{(0)}(\Bn)$ with
$|m^{(0)}(\Bn)|< D_0$ for some $D_0>0$ and for all $|\e|<\e_0$ for
some $\e_0$: so that for small $\e$ the divisor in \equ(A1.2) is
essentially still what it was before starting the resummation.

At this point an induction can be started. Consider trees evaluated
with the new rule and place a scale lavel ``$\ge2$'' on the lines with
$|\Bo_0\cdot\Bn(l)|\le 2^{-n+1}$ for $n=2$: leave the label ``$0$'' on
the lines already marked so and label by ``$1$'' the other lines. The
lines of scale ``$1$'' will satisfy $2^{-n}< |\Bo_0\cdot\Bn(l)|\le 2^{-n+1}$
for $n=1$. And the graphs will now possibly contain lines {\it of
scale $0,1$ or $\ge2$ while lines with label ``$\ge1$'' no longer can
appear}, by construction.

A {\it cluster of scale $1$} will be a maximal collection of lines of
scales $0,1$ forming a connected subgraph of a tree $\th$ and
containing at least one line of scale $1$.

The construction carried considering clusters of scale $0$ can be
repeated by considering trees $\th_1\in\Th_1$ with $\Th_1$ the
collection of trees with lines marked $0,1$ or $\ge2$ and in which no
pairs of lines with equal momentum appear to follow each other if
between them there are only lines marked $0$ or $1$.

Insertion of connected clusters $\g$ of such lines on a line $l_0$
of $\th_1$ leads to define a matrix $M^{(1)}$ formed by summing tree
values of clusters $\g$ with lines of scales $0$ or $1$ evaluated with the
line factors defined in \equ(A1.1) and with the restriction that {\it
in $\g$ there are no pairs of lines $\ell<\ell'$ with the same current
and which follow each other while any line between them has lower scale
(\ie $0$}), here between means preceding $l'$ but not preceding $l$, as
above.

Therefore a scale independent method has to devised to check
convergence for $M^{(1)}$ and for the matrices to be introduced
later to deal with even smaller propagators. This is achieved by the
following extension of Siegel's theorem mentioned in Section \sec(14):
\vskip2mm

\0{\it Let $\Bo_0$ satisfy \equ(13.2) and set $\Bo= C\Bo_0$.
Consider the contribution to the sum in \equ(14.3) from graphs
$\th$ in which 
\\
(1) no pairs $\ell'>\ell$ of lines which lie on the same path to the
root carry the same current $\Bn$ if all lines $\ell_1$ between
them have current $\Bn(\ell_1)$ such that $|\Bo\cdot\Bn(\ell_1)|> 2
|\Bo\cdot\Bn|$.
\\
(2) the node harmonics are bounded by $|\Bn|\le N$ for some $N$.
\\
Then the number of lines $\ell$
in $\th$ with divisor $\Bo\cdot\Bn_\ell$ satisfying
$2^{-n}<|\Bo\cdot\Bn_\ell|\le 2^{-n+1}$ does not exceed $4\, N \, k\,
2^{-n/\t}$, $n=1,2,\ldots$.} 
\vskip2mm%\*

This implies, by the same estimates in \equ(14.6), that the series
defining $M^{(1)}$ converges. Again it must be checked that there
are cancellations implying that $M^{(1)}(\Bn)=\e^2 \,(\Bo_0\cdot\Bn)^2
m^{(1)}(\Bn)$ with $|m^{(1)}(\Bn)|< D_0$ for the {\it same} $D_0>0$ and
the same $\e_0$.

At this point one deals with trees containing only lines carrying
labels $0,1,\ge2$ and the line factors for the lines $\ell=v'v$ of
scale $0$ are $\Bn_{v'}\cdot\Bn_{v}/(\Bo_0\cdot\Bn(\ell))^2$, those of
the lines $\ell=v'v$ of scale $1$ have line factors
$\Bn_{v'}\cdot(\Bo_0\cdot\Bn(\ell)^2-M^{(0)}(\Bn(\ell)))^{-1} \Bn_v$
and those of the lines $\ell=v'v$ of scale $\ge2$ have line factors
$\Bn_{v'}\cdot(\Bo_0\cdot\Bn(\ell)^2-M^{(1)}(\Bn(\ell)))^{-1} \Bn_v$.
{\it Furthermore no pair of lines of scale ``$1$'' or of scale ``$\ge
2$'' with the same momentum and with only lines of lower scale (\ie of
scale ``$0$'' in the first case or of scale ``$0$'',''$1$'' in the
second) between then can follow each other.}

And so on until, after infinitely many steps, the problem is reduced
to the evaluation of tree values in which each line carries a scale
label $n$ and there are no pairs of lines which follow each other and
which have only lines of lower scale in between. Then the Siegel argument
applies once more and the series so resummed is an absolutely
convergent series of functions analytic in $\e$: hence the original
series is convergent.

Although at each step there is a lower bound on the denominators it
would not be possible to avoid using Siegel's theorem.  In fact the
lower bound would becomes worse and worse as the scale increases. In
order to check the estimates of the constants $D_0,\e_0$ which
control the scale independence of the convergence of the various
series it is necessary to take advantage of the theorem, and of
the absence at each step of the necessity of considering trees with
pairs of consecutive lines with equal momentum and intermediate lines
of higher scale.
 
One could also perform the analysis by bounding $h^{(k)}$ order by
order with no resummations (\ie without changing the line factors) and
exhibiting the necessary cancellations. Or the paths
that {\cs Kolmogorov}, {\cs Arnold} and {\cs Moser} used to prove the
first three (somewhat different) versions of the theorem, by
successive approximations of th equations for the tori, can be followed.

The invariant tori are {\it Lagrangian manifolds} just as the unperturbed
ones (cf. comments after \equ(6.4)) and, in the case of the Hamiltonian
\equ(14.1) the generating function $\V A\cdot\Bps+\F(\V A,\Bps)$ can
be expressed in terms of their parametric equations
\pagina
%\vskip-2mm
$$\eqalign{
&\F(\V A,\Bps)=G(\Bps)+\V a\cdot\Bps+
\V h(\Bps)\cdot(\V A-\Bo-\D \V h(\Bps))\cr
&
\Dpr_\Bps G(\Bps)\defi-\D \V h(\Bps)+\T h(\Bps)\Dpr_\Bps \D \T
h(\Bps)-\V a\cr
&
\V a\defi \ig (-\D \V h(\Bps)+\T h(\Bps)\Dpr_\Bps \D \T
h(\Bps))\fra{d\Bps}{(2\p)^\ell}=\ig \T h(\Bps)\Dpr_\Bps \D \T
h(\Bps)\fra{d\Bps}{(2\p)^\ell}
\cr
}\Eqa(A1.3)
$$
where $\D=(\Bo\cdot \Dpr_\Bps)$ and the invariant torus corresponds to
$\V A'=\Bo$ in the map $\Ba=\Bps+\Dpr_{\V A}\F(\V A,\Bps)$ and $\V
A'=\V A+\Dpr_\Bps\F(\V A,\Bps)$. In fact by \equ(A1.3) the latter
becomes
$\V A'=\V A-\D\V h$ and, from the second of \equ(13.3) written for $f$
depending only on he angles $\Ba$, it is $\V A=\Bo+\D\V h$ when $\V
A,\Ba$ are on the ivariant torus.  

Note that if $\V a$ exists it is necessarily determined by the third
relation but the check that the second equation in
\equ(A1.3) is soluble (\ie that the \rhs is an exact gradient up to a
constant) is nontrivial. The canonical map generated by $\V A\cdot\Bps
+\F(\V A,\Bps)$ is {\it also} defined for $\V A'$ close to $\Bo$ and
foliates the neighborhood of the invariant torus with other
tori: of course for $\V A'\ne \Bo$ the tori defined in this way are, in
general, not invariant.
\*

\0{\it References:} [GBG94].
\*

\appendix(B, Coriolis and Lorentz forces. Larmor precession)

{\it Larmor precession} is the part of the motion of an electrically
charged particle caused by the action of a magnetic field $\V H$ (in an
inertial frame of reference). It is due to the {\it Lorentz force}
which, on a unit mass with unit charge, produces an acceleration
$\ddot\Br=\V v\wedge \V H$ if the speed of light is $c=1$.

Therefore if $\V H=H \V k$ is directed along the $\V k$
axis the acceleration it produces is the same that the Coriolis force would
impress on a unit mass located in a reference frame which rotates with
angular velocity $\o_0\V k$ around the $\V k$ axis if $\V H=2\o_0 \V
k$.

The above remarks imply that a homogeneous sphere homogeneoulsy
electrically charged with a unit charge and freely pivoting about its
center in a constant magnetic field $H$ directed along the $\V k$ axis
undergoes the same motion it would follow if not subject to the
magnetic field but seen in a non inertial reference frame rotating at
constant angular velocity $\o_0$ around the $\V k$ axis if $H$ and
$\o_0$ are related by $H=2\o_0$: in this frame the Coriolis force
is interpreted as a magnetic field.

This holds, however, only if the centrifugal force has zero moment
with rerspect to the center: true in the spherical symmetry case only.
In spherically non symmetric cases the centrifugal forces have in
general non zero moment so the equivalence between Coriolis forces and
magnetic fields is only approximate.

The {\it Larmor theorem} makes this more precise. It gives a
quantitative estimate of the difference between the motion of a
general system of particles of mass $m$ in a magnetic field and the
motion of the same particles in a rotating frame of reference but in
absence of a magnetic field. The approximation is estimated in terms
of the size of the {\it Larmor frequency} $eH/2mc\,$: which should be
small compared to the other characteristic frequencies of the motion
of the system: the physical meaning is that the centrifugal force
should be small compared to the other forces.

The vector potential $\V A$ for a constant magnetic field in the $\V
k$-direction $\V H=2\o_0 \V k$ is $\V A= 2\o_0\V k\wedge\Br\=
2\o_0\Br^\perp$. Therefore, from the treatment of the Coriolis force
in Section \sec(19), see \equ(19.2), the motion of a charge $e$ with
mass $m$ in a magnetic field $\V H$ with vector potential $\V A$ and
subject to other forces with potential $W$ can be described, in an
inertial frame and in generic units in which the speed of light is
$c$, by a Hamiltonian

$$\HH =\fra1{2m}(\V p-\fra{e}c\V A)^2+ W(\Br)\Eqa(A2.1)$$
where $\V p=m\dot\Br+\fra{e}c \V A$ and $\Br$ are canonically conjugated.

%\ifnum\tipo=0\pagina\fi
\*

\def\*{\vskip.6mm}
\0{\bf References}
\*
\nota

\0{\bf [Ar68] Arnold, V.I.:} {\it Mathematical methods of
classical mechanics}, Sprin\-ger-Verlag, 1989.
\*

\0{\bf [CD82] Calogero, F., Degasperis, A.:} {\it Spectral transform
  and solitons}, North Holland, 1982.
\*

\0{\bf [CV00] Chierchia, L., Valdinoci, E.:}
{\it A note on the construction of Hamiltonian trajectories along
heteroclinic chains}, Forum Mathematicum, {\bf12}, 247-255, 2000.
\*

\0{\bf [Fa98] Fass\`o, F.}: {\it Quasi-periodicity of motions and
complete integrability of Hamiltonian systems}, Ergodic Theory and
Dynamical Systems, {\bf 18}, 1349-1362, 1998.
\*

\0{\bf [Ga83] Gallavotti, G.:} {\it The elements of mechanics},
Sringer Verlag, New York, 1983.
\*

\0{\bf [GBG04] Gallavotti, G., Bonetto, F., Gentile, G.:}
{\it Aspects of the ergodic, qualitative and statistical properties
of motion}, Springer--Verlag, Berlin, 2004.
\*

\0{\bf [Ko55] N. Kolmogorov:} {\it On the preservation of conditionally
periodic motions}, Doklady Akade\-mia Nauk SSSR, {\bf 96}, 527-- 530, 1954.
\*

\0{\bf [LL68] Landau, L.D., Lifshitz, E.M.}:
{\it Mechanics}, Pergamon Press, 1976.
\*

\0{\bf [LC] Levi-Civita, T.} {\it Opere Matematiche}, Accademia
Nazionale dei Lincei, Zani\-chelli, Bologna, 1956.
\*

\0{\bf [Mo62] J. Moser:} {\it On invariant curves of an area
preserving mapping of the annulus}, Nachricten Akadenie Wissenschaften
G\"ottingen, {\bf 11}, 1--20, 1962.
\*

\0{\bf [Ne77] Nekhorossev, V.}
{\it An exponential estimate of the time of stability
of nearly integrable Hamiltonian systems}, Russian Mathematical
Surveys, 32 (6), 1 {\it65}, 1977.
\*

\0{\bf [Po] Poincar\'e, H.:} {\it
M\'ethodes nouvelles de la m\'ecanique cel\`este}, Vol. I,
Gauthier-Villars, reprinted by Gabay, Paris, 1987.

\end{document}